\documentclass[a4paper, 12pt, titlepage, draft=false]{scrartcl} 
\makeatletter
\DeclareOldFontCommand{\rm}{\normalfont\rmfamily}{\mathrm}
\DeclareOldFontCommand{\sf}{\normalfont\sffamily}{\mathsf}
\DeclareOldFontCommand{\tt}{\normalfont\ttfamily}{\mathtt}
\DeclareOldFontCommand{\bf}{\normalfont\bfseries}{\mathbf}
\DeclareOldFontCommand{\it}{\normalfont\itshape}{\mathit}
\DeclareOldFontCommand{\sl}{\normalfont\slshape}{\@nomath\sl}
\DeclareOldFontCommand{\sc}{\normalfont\scshape}{\@nomath\sc}
\makeatother

\usepackage{ifdraft} 

\ifdraft{
}

\usepackage{color}
\usepackage[usenames,dvipsnames]{xcolor}
\usepackage{verbatim}
\usepackage{comment}

\makeatletter
\def\@xobeysp{\mbox{}\space}
\def\verbatim@font{\normalfont\ttfamily\raggedright\leftskip\@totalleftmargin}
\makeatother

\usepackage{csvsimple}
\usepackage{amsmath}
\usepackage{amssymb}
\usepackage{amsfonts}
\usepackage{amsthm}
\usepackage{todonotes}
\usepackage{import}
\usepackage{booktabs}
\usepackage{tabularx}
\usepackage{longtable}[=v4.13]
\usepackage{cellspace}
\usepackage{rotating}
\usepackage{siunitx}
\ExplSyntaxOn
\cs_new_eq:NN \siunitx_table_collect_begin:Nn \__siunitx_table_collect_begin:Nn
\ExplSyntaxOff
\usepackage{tabu}
\usepackage{hhline}
\usepackage{float}
\usepackage{lscape}
\usepackage[export]{adjustbox}
\usepackage{changepage}

\usepackage[utf8]{inputenc}
\usepackage[T1]{fontenc}
\usepackage[ngerman,english]{babel}

\usepackage[nomain,acronym,nonumberlist,nopostdot,automake]{glossaries}
\usepackage{glossary-mcols}

\makeglossaries
\newacronym{www}{WWW}{World Wide Web}
\newacronym{tls}{TLS}{Transport Layer Security}
\newacronym{ssl}{SSL}{Secure Sockets Layer}
\newacronym{mitm}{MITM}{Man-In-The-Middle}
\newacronym{html}{HTML}{Hyper Text Markup Language}
\newacronym{gdpr}{GDPR}{General Data Protection Regulation}
\newacronym{eu}{EU}{European Union}
\newacronym{w3c}{W3C}{World Wide Web Consortium}
\newacronym{css}{CSS}{Cascading Style Sheets}
\newacronym{pc}{PC}{Personal Computer}
\newacronym{gpu}{GPU}{Graphical Processing Unit}
\newacronym{gps}{GPS}{Global Positioning System}
\newacronym{cpu}{CPU}{Central Processing Unit}
\newacronym{http}{HTTP}{Hypertext Transfer Protocol}
\newacronym{api}{API}{Application Programming Interface}
\newacronym{osi}{OSI}{Open Systems Interconnection}
\newacronym{mtu}{MTU}{Maximum Transmission Unit}
\newacronym{tcp}{TCP}{Transmission Control Protocol}
\newacronym{ip}{IP}{Internet Protocol}
\newacronym{os}{OS}{Operating System}
\newacronym{dnt}{DNT}{Do Not Track}
\newacronym{ajax}{AJAX}{Asynchronous JavaScript And XML}
\newacronym{sop}{SOP}{Same-Origin Policy}
\newacronym{xhr}{XHR}{XMLHttpRequest}
\newacronym{dom}{DOM}{Document Object Model}
\newacronym{xss}{XSS}{Cross-Site Scripting}
\newacronym{cdn}{CDN}{Content Delivery Network}
\newacronym{sql}{SQL}{Structured Query Language}
\newacronym{json}{JSON}{JavaScript Object Notation}
\newacronym{csv}{CSV}{Comma-Separated Values}
\newacronym{vnc}{VNC}{Virtual Network Computing}
\newacronym{orm}{ORM}{Object Relational Mapper}
\newacronym{erd}{ERD}{Entity Relationship Diagram}
\newacronym{url}{URL}{Uniform Resource Locator}
\newacronym{tld}{TLD}{Top-Level Domain}
\newacronym{io}{I/O}{Input / Output}
\newacronym{ram}{RAM}{Random Access Memory}
\newacronym{hpkp}{HPKP}{HTTP Public Key Pinning}
\newacronym{nat}{NAT}{Network Address Translation}
\newacronym{tcf}{TCF}{Transparency and Consent Framework}
\usepackage{geometry}

\usepackage{listings}

\usepackage{inconsolata}

\usepackage{color}
\definecolor{pblue}{rgb}{0.13,0.13,1}
\definecolor{pgreen}{rgb}{0,0.5,0}
\definecolor{pred}{rgb}{0.9,0,0}
\definecolor{pgrey}{rgb}{0.46,0.45,0.48}
\definecolor{lightgray}{rgb}{.9,.9,.9}
\definecolor{darkgray}{rgb}{.4,.4,.4}
\definecolor{purple}{rgb}{0.65, 0.12, 0.82}
\lstdefinelanguage{JavaScript}{
	keywords={break, const, let, case, catch, continue, debugger, default, delete, do, else, false, finally, for, function, if, in, instanceof, new, null, return, switch, this, throw, true, try, typeof, var, void, while, with},
	morecomment=[l]{//},
	morecomment=[s]{/*}{*/},
	morestring=[b]',
	morestring=[b]",
	ndkeywords={class, export, boolean, throw, implements, import, this},
	keywordstyle=\color{blue}\bfseries,
	ndkeywordstyle=\color{darkgray}\bfseries,
	identifierstyle=\color{black},
	commentstyle=\color{purple}\ttfamily,
	stringstyle=\color{red}\ttfamily,
	sensitive=true
}
\usepackage[labelfont=sc]{caption}

\usepackage{algorithm}
\usepackage{algorithmicx}
\usepackage[noend]{algpseudocode}

\usepackage{graphicx}
\graphicspath{{}{./figures/}}

\usepackage{caption}
\usepackage{subcaption}
\usepackage[hyphenbreaks]{breakurl}
\usepackage[hyphens]{url}
\usepackage{hyperref}
\usepackage{etoolbox}
\hypersetup{colorlinks=true}
\hypersetup{
	urlcolor=black,
	citecolor=black,
	linkcolor=black,
}
\makeatletter
\def\@footnotecolor{black}
\define@key{Hyp}{footnotecolor}{%
	\HyColor@HyperrefColor{#1}\@footnotecolor%
}
\patchcmd{\@footnotemark}{\hyper@linkstart{link}}{\hyper@linkstart{footnote}}{}{}
\makeatother

\usepackage[autostyle=true]{csquotes}
\NewDocumentCommand{\qnameref}{sm}{``\IfBooleanTF{#1}{\nameref*{#2}}{\nameref{#2}}''}

\usepackage[multiple]{footmisc}

\usepackage{indentfirst}
\usepackage{soul}

\usepackage{enumitem}
\usepackage[sort&compress,nameinlink,noabbrev]{cleveref} %
\usepackage{nameref} 
\usepackage[backend=bibtex,style=numeric,
hyperref=true,natbib=true,
backref=false,backrefstyle=three,
sortcites=false,
maxbibnames=50,
maxcitenames=3,
firstinits=true, 
isbn=true,url=true,doi=true]
{biblatex} 
\patchcmd{\thebibliography}{\clubpenalty4000}{\clubpenalty10000}{}{}
\patchcmd{\thebibliography}{\widowpenalty4000}{\widowpenalty10000}{}{}
\patchcmd{\bibsetup}{\interlinepenalty=5000}{\interlinepenalty=10000}{}{}

\bibliography{bibliography}

\patchcmd{\url}{\clubpenalty4000}{\clubpenalty10000}{}{}
\patchcmd{\url}{\widowpenalty4000}{\widowpenalty10000}{}{}
\patchcmd{\verb}{\clubpenalty4000}{\clubpenalty10000}{}{}
\patchcmd{\verb}{\widowpenalty4000}{\widowpenalty10000}{}{}

\usepackage{array}
\newcolumntype{x}[1]{>{\centering\arraybackslash\hspace{0pt}}m{#1}}

\theoremstyle{plain}

\usepackage{tikz}
\usepackage{tikz-qtree}
\usepackage{pgfplots}
\usepackage{pgfplotstable}
\usepgfplotslibrary{groupplots}
\pgfplotsset{compat=1.14}
\usepackage{etoolbox}

\newenvironment{smallitemize}
{ \begin{itemize}
		\setlength{\itemsep}{0pt}
		\setlength{\parskip}{0pt}
		\setlength{\parsep}{0pt}     
	}
	{ \end{itemize}                  } 

\newenvironment{smallenumerate}
{ \begin{enumerate}
		\setlength{\itemsep}{0pt}
		\setlength{\parskip}{0pt}
		\setlength{\parsep}{0pt}     
	}
	{ \end{enumerate}                  } 

\usepackage{fancyhdr}

\makeatletter
\newcommand{\rightorleftmark}{%
	\begingroup\protected@edef\x{\rightmark}%
	\ifx\x\@empty
	\endgroup\leftmark
	\else
	\endgroup\rightmark
	\fi}
\makeatother
\usepackage{setspace}
\usepackage{array}
\usepackage{ragged2e}
\newcolumntype{L}{>{\RaggedRight\hspace{0pt}}X}

\usepackage{units}
\usepackage[a4paper]{anysize}
\marginsize{3.5cm}{2.75cm}{2cm}{3cm} %
\usepackage[many]{tcolorbox}
\newtcolorbox{myboxi}[1][]{
	breakable,
	title=#1,
	colback=white,
	colbacktitle=white,
	coltitle=black,
	fonttitle=\bfseries,
	bottomrule=1pt,
	toprule=1pt,
	leftrule=1pt,
	rightrule=1pt,
	titlerule=0pt,
	arc=2pt,
	outer arc=2pt,
	colframe=black,
	enlarge top by=12pt,
}

\usepackage[section]{placeins}
\makeatletter
\AtBeginDocument{%
	\expandafter\renewcommand\expandafter\subsection\expandafter{%
		\expandafter\@fb@secFB\subsection
	}%
}
\makeatother

\newcommand{\FPNET}{\texttt{FPNET}}
\newcommand{\FPMON}{\texttt{FPMON}}
\newcommand{\FPCRAWL}{\texttt{FPCRAWL}}

\begin{document}
\pagenumbering{Roman}

\newgeometry{left=3cm,bottom=2cm,right=3cm,top=3cm}
\begin{titlepage}
\centering

\vspace*{-0.8em}
{\LARGE \textbf{Technische Universität Berlin}}

\vspace{0.5cm}

{\large Institut für Softwaretechnik und Theoretische Informatik\\[1mm]}
{\large Security in Telecommunications\\[5mm]}

Fakultät IV\\
Ernst-Reuter-Platz 7\\
10587 Berlin\\
\href{https://www.isti.tu-berlin.de/}{www.isti.tu-berlin.de}\\

\vspace*{1cm}

\includegraphics[width=5cm]{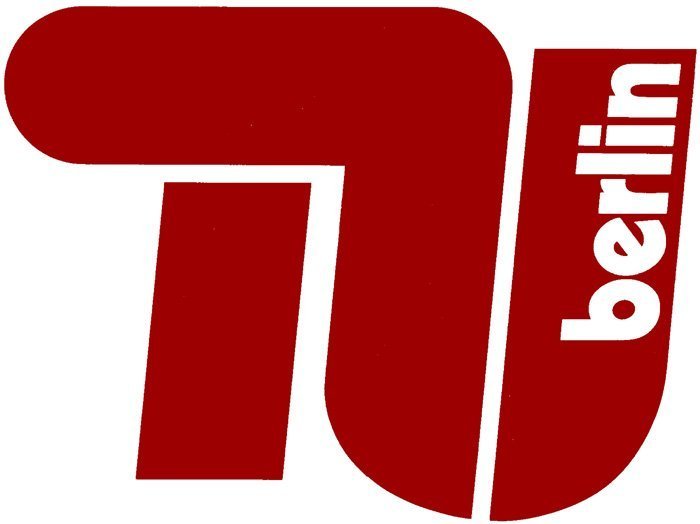}\\

\vspace*{1.0cm}

{\LARGE Master Thesis}\\

\vspace{0.5cm}
{\Large\textbf{Uncovering Fingerprinting Networks.\\An Analysis of In-Browser Tracking using a Behavior-based Approach}}\\
\vspace*{1cm}
{\LARGE Sebastian Neef}
\\
\vspace*{0.3cm}
Matriculation Number: 350692\\
s.neef@campus.tu-berlin.de\\
\vspace*{0.3cm}
29.03.2021 \\ 
\vspace*{0.3cm}
Supervised by\\
\vspace*{0.2mm}
Prof. Dr. Jean-Pierre Seifert\\
\vspace*{0.3cm}
Second Supervisor\\
\vspace*{0.2mm}
Prof. Dr. Florian Tschorsch\\
\vspace*{0.3cm}
Assistant Supervisor\\
\vspace*{0.2mm}
\hspace{2mm}M.Sc. Julian Fietkau

\end{titlepage}

\restoregeometry

\clearpage
\thispagestyle{empty}
\mbox{}
\newpage

\section*{Declaration}

I hereby declare that the thesis submitted is my own, unaided work, completed without any unpermitted external help. Only the sources and resources listed were used.
\\
The independent and unaided completion of the thesis is affirmed by affidavit:
\begin{flushleft}
	Berlin, the 29th of March, 2021
	
	\ \\
	\ \\
	\ \\
	Sebastian Neef
	
\end{flushleft}
\ \\
Due to the collaboration with SecT's research assistant Julian Fietkau and master's degree candidate Felix Kybranz on the paper \textit{The Elephant in the Background. A Quantitative Approach to Empower Users Against Web Browser Fingerprinting} \cite{fp_paper} prior to the thesis, this work references their goals, methods and tools. Therefore, it is closely related to the paper and Kybranz's thesis \cite{fp_paper,kybranz}, but comprises the essential ideas, implementation, and results of Sebastian Neef's work. 
\newpage

\section*{Acknowledgements} \label{sec:acknowledgements}
My profound gratitude goes to the whole SecT department for making me feel at home. Foremost to Prof. Seifert for supervising my thesis and Julian Fietkau for assisting with insightful ideas, productive discussions, and helpful guidance. Furthermore, I sincerely appreciate countless tips and constructive feedback from all members of AG Rechnersicherheit. Finally, I must thank my whole family and friends for always motivating and supporting me throughout my life. Thank you all!

\section*{Abstract} \label{sec:abstract}
\selectlanguage{english}
Throughout recent years, the importance of internet-privacy has continuously risen. The General Data Protection Regulation by the European Union fundamentally changed digital data processing by requiring explicit consent for processing personally identifiable information. In combination with the cookie law, users can opt-out of being profiled by advertisers or other entities. Browser fingerprinting is a technique that does not require cookies or persistent identifiers. It derives a sufficiently unique identifier from the various browser or device properties. Academic work has covered offensive and defensive fingerprinting methods for almost a decade, observing a rise in popularity.

This thesis explores the current state of browser fingerprinting on the internet. For that, we implement \FPNET~- a scalable \& reliable tool based on \FPMON, to identify fingerprinting scripts on large sets of websites by observing their behavior. By scanning the Alexa Top 10,000 websites, we spot several hundred networks of equally behaving scripts. For each network, we determine the actor behind it. We track down companies like Google, Yandex, Maxmind, Sift, or FingerprintJS, to name a few. 

In three complementary studies, we further investigate the uncovered networks with regards to I) randomization of filenames or domains, II) behavior changes, III) security. Two consecutive scans reveal that only less than 12.5\% of the pages do not change script files. With our behavior-based approach, we successfully re-identify almost 9,000 scripts whose filename or domain changed, and over 86\% of the scripts without URL changes. The security analysis shows an adoption of TLS/SSL to over 98\% and specific web security headers set for over 30\% of the scripts.

Finally, we voice concerns about the unavoidability of modern fingerprinting and its implications for internet users' privacy since we believe that many users are unaware of being fingerprinted or have insufficient possibilities to protect against it.

\pagebreak
\section*{Zusammenfassung} \label{sec:zusammenfassung}
\selectlanguage{ngerman}
Über die letzten Jahre ist Privatsphäre zu einem immer wichtigeren Thema des Internets geworden. Die Datenschutzgrundverordnung hat die digitale Datenverarbeitung fundamental verändert, bspw. indem der Verarbeitung von personenbezogenen Daten explizit zugestimmt werden muss. In Kombination mit dem Cookie-Gesetz haben Internetnutzer die Möglichkeit der Profilbildung von Werbetreibenden oder anderen Unternehmen zu wiedersprechen. Browser Fingerprinting ist eine Technik die keine Speicherung von Cookies oder Identifikatoren voraussetzt. Stattdessen wird ein hinreichend eindeutiger Identifikator aus verschiedenen Browser- oder Geräteeigenschaften abgeleitet. In der Wissenschaft werden offensive und defensive Fingerprinting-Methoden seit fast einem Jahrzehnt erforscht, wobei ein Anstieg in der Popularität verzeichnet wird. 

Diese Arbeit untersucht die aktuelle Landschaft des Browser-Fingerprintings im Internet. Dafür implementieren wir \FPNET~- ein skalierbares \& zuverlässiges Programm, welches auf \FPMON~basiert und Fingerprinting-Skripte anhand ihres Verhaltens auf großen Mengen von Webseiten anhand deren Verhaltens identifizieren kann. Mit einer Untersuchung der Alexa Top 10.000 Webseiten, erkennen wir mehrere Hundert Netzwerke bestehend aus verhaltensähnlichen Skripten. Für jedes Netzwerk bestimmen wir den Akteur und entdecken unter anderem Unternehmen wie Google, Yandex, Maxmind, Sift oder FingerprintJS. 

In drei weiteren Studien erforschen wir die Netzwerke mit Hinblick auf I) Randomisierung der Dateinamen oder Domains, II) Verhaltensänderungen, III) Sicherheit. Zwei aufeinanderfolgende Untersuchungen zeigen weniger als 12,5\% der Webseiten ohne Skriptänderungen. Unser verhaltensbasierte Ansatz kann erfolgreich fast 9,000 Scripte wiedererkennen, trotz Änderung des Dateinamens oder der Domain. Skripte mit unveränderten URLs sind zu über 86\% re-identifizierbar. Die Netzwerkanalyse zeigt einen Anstieg der TLS/SSL-Nutzung auf über 98\%, sowie bestimmte Websicherheits-Header auf 30\% der Skripte.

Schließlich äußern wir Bedenken an der Unvermeidbarkeit des modernen Fingerprintings und dessen Auswirkungen auf die Privatsphäre der Internetnutzer, weil wir glauben, dass viele Nutzer unwissend gefingerprintet werden oder nicht genügend Möglichkeiten besitzen, sich dagegen zu schützen.
\selectlanguage{english}

\newpage
\setcounter{tocdepth}{3}
\tableofcontents
\newpage

\listoffigures
\newpage
\listoftables
\newpage
\renewcommand{\lstlistlistingname}{List of \lstlistingname s}
\lstlistoflistings
\newpage
\renewcommand{\glsnamefont}[1]{\textbf{#1}}
\glsaddall
\printglossary[type=\acronymtype,title=List of Abbreviations,style=super,nogroupskip=true]

\pagestyle{fancy}
\renewcommand{\sectionmark}[1]{\markboth{}{\thesection~#1}}
\renewcommand{\subsectionmark}[1]{}

\pagenumbering{arabic}

\section{Introduction} \label{sec:introduction} 
\subsection{Motivation} \label{sec:introduction-motivation}

In 2013, Edward Snowden leaked classified documents about mass surveillance programs run by various secret services which enable tracking of individuals on the internet \cite{snowden_nsa_lucas_2014}.  Since those revelations, internet privacy and security have increased in relevance. New \gls{gdpr} laws were passed in the \gls{eu} a few years later, redefining how personal data can be processed and empowering users to request that information \cite{gdpr_paper}. After the formation of Let's Encrypt in 2014 \cite{letsencrypt_website}, the use of secure, encrypted connections on the \gls{www} reached over 90\% \cite{google_transparency, letsencrypt_stats}, as \gls{tls} certificates could now be obtained free of cost.

However, thwarting network-based attacks, such as \gls{mitm}, is insufficient to evade online tracking, since application-level techniques are agnostic to the network layer. For example, \citeauthor{panopticlick} showed that browsers can be fingerprinted and identified \cite{panopticlick}. With the continuous evolution of \gls{html}, \gls{css}, JavaScript or other technology, more functionality is introduced into internet browsers. Every new feature could become a piece of information used to identify a browser and the user uniquely. Academia has covered defensive and offensive techniques in the past, as discussed in Chapter \ref{sec:background-related-work}.

Online identification and tracking techniques are not limited to sophisticated actors. Businesses rely on those techniques for risk assessment, fraud detection, or customer scoring, as the The New York Times Company reports \cite{nytimes_sift}. Website operators might not even be aware of their partners' (e.g., advertisement companies) tracking practices. These practices leave users at the risk of being tracked and identified throughout the internet. Depending on the technique, a user's or device's information is collected in the background, completely oblivious to the user.

Felix Kybranz's and assistant supervisor Julian Fietkau's research on browser fingerprinting \cite{fp_paper,kybranz} motivates further research in this area.

\subsection{Research Goals} \label{sec:introduction-goal}
The thesis' goal is to identify and analyze existing fingerprinting networks on the internet. To achieve this goal, a behavior-based approach will be explored that tries to identify fingerprinting scripts by observing their behavior, e.g., function calls, function names, return values, or other properties. For this, \FPMON~\cite{kybranz} will be extended and modified.

After collecting the behavior information, the next goal is to group similar behaving scripts into \emph{fingerprinting networks} and identify the actors behind them. To validate the expectation of finding businesses in the advertisement, security, or similar industries, we manual review and classify the networks and actors.

In order to discover and estimate the real-world impact of such fingerprinting networks, we apply the approach to a large set of popular, public websites sorted by their Alexa Traffic rank \cite{alexa_topsites}. This approach requires a suitable scanning system. It might pose a challenge to automate headless browsers properly and simulate real ones to capture a fingerprinter's full behavior. Furthermore, scalability and performance will be considered, but its evaluation is not part of this work.

The final goal is to assess the privacy implications of being exposed to fingerprinting networks. We hope the final results will enable further research into novel defensive techniques.

The thesis aims to answer the following questions in particular:
\begin{smallitemize}
	\item How relevant is JavaScript for modern websites?
	\item How common and privacy-invasive are browser fingerprinting scripts?
	\item What fingerprinting scripts can be identified? 
	\item What are the resulting networks and actors? What size and who are they?
	\item What defensive techniques are utilized? 
	\item Is randomization or obfuscation common?
	\item Do fingerprinting scripts change or differ?
	\item Is transport security used for fingerprinting scripts? 
	\item Are web security headers in place?
\end{smallitemize}

\newpage
\subsection{Outline} \label{sec:introduction-structure}
The thesis consists of multiple chapters for better readability and clarity. 

Chapter \ref{sec:background-related-work} aims to provide the reader with sufficient background knowledge to follow the thesis. First, browser fingerprinting is introduced, followed a review of relevant related work in this research field. 

Chapter \ref{sec:design} and \ref{sec:implementation} elaborate on the design and implementation of the tools to better research browser fingerprinting on a large scale using a behavior-based approach. They introduce our scanning system \FPNET~with its design decisions, problems and solutions as well as the necessary modifications to the existing \FPMON~extension. Furthermore, the general methodology for the upcoming chapter will be outlined.

Chapter \ref{sec:evaluation} presents the five complementary studies that were be conducted, including each study's methodology and results. All studies are related and lead this research's path to uncover fingerprinting networks. In the first study, the \gls{www}'s dependency on JavaScript is examined. The second study focuses on the identification of fingerprinting networks using the behavior-based approach. Randomization is evaluated in Study 3, followed by a file-content approach to detect and block fingerprinting networks in study 5. Finally, study 5 analyzes the fingerprinting scripts' web security.

Chapter \ref{sec:discussion} discusses the previously obtained results. It reviews the uncovered fingerprinting networks and the actors behind them. Additionally, the strengths and weaknesses of the chosen approach as well as the privacy implications for internet users will be examined. The chapter ends with ideas and thoughts for future work on this topic.  

The final Chapter \ref{sec:conclusion} summarizes this thesis' work and reflects on the research goals, research questions, and their answers.

\section{Background \& Related Work} \label{sec:background-related-work}
This chapter provides the reader with necessary terminology, basic concepts, and fundamental knowledge to improve the next chapters' understanding. Past and current related academic work will be introduced to position and differentiate the thesis' approach in this research area. 

\subsection{Browser-based Fingerprinting}
\subsubsection{The History of Browsers}
Since the invention and public release of the \gls{www} by \citeauthor{www_berners_lee} in the early 1990s, hypertext documents were suggested to be viewed using graphical interfaces, called \emph{browsers} \cite{www_berners_lee}.  
Jim Clark and Marc Andreessen founded Netscape in 1994 to develop the browser \emph{Navigator}, reaching an 87\% market share by 1996 \cite{Spinello2005}.
The company Microsoft, known for its operating system \emph{Windows}, used ``questionable tactics'' (i.e., bundling its browser with Windows) to establish the browser \emph{Internet Explorer} \cite{Spinello2005}. By the end of 1999, it dominated 80\% of the market, leading to ``one of the most significant antitrust cases of the last century: \textit{The United States of America versus Microsoft}'' \cite{Spinello2005}. The case was settled in 2002, and Microsoft was no longer allowed to ``restrict the freedom [...] to install non-Microsoft software [...]'' or ``[...] selling competing software programs bundled with Windows'' \cite{Spinello2005}. 

Since then, the browser market has diversified as new browsers emerged: Chrome 65.9\%, Safari 16.6\%, Firefox 4.3\%, Opera 2.1\% and Edge 1.9\% \cite{browser_marketshare}.
The \gls{w3c}, lead by Tim Berners-Lee, ``develop[s] protocols and guidelines that ensure long-term growth for the Web'' \cite{w3c_help}.
This includes specifications for \gls{html}, \gls{css}, JavaScript, Graphics, Audio and Video, Accessibility, Internalization and other important technology that drives modern websites \cite{w3c_standards}.
As not to lose market share, web browsers need to support as much functionality as possible. At the same time, this trend poses higher demands to a device's hardware and its integration. Rendering high-quality video streams or animations requires a \gls{gpu}; positioning or location features benefit from a \gls{gps} module; web applications excel in performance and responsiveness with a powerful \gls{cpu}. A device that usually fulfills those requirements is a \gls{pc}, laptop, or smartphone.

Similar to the constant evolution of web browsers and their functionality, fingerprinting technology evolves, too, resulting in more sophisticated techniques to extract specific bits of information about a device. 

\subsubsection{Fingerprinting Purposes} \label{sec:background-related-work:fp-purposes}
\begin{figure}[h]
	\begin{subfigure}{0.4\textwidth}
		\includegraphics[width=\textwidth]{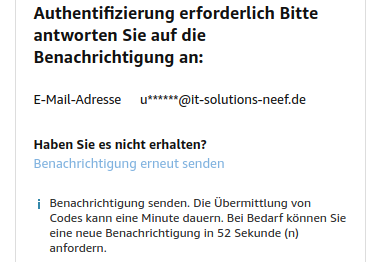}
		\caption{An additional authentication step asking for a confirmation code sent via E-Mail.}
		\label{fig:amazon_login:login}
	\end{subfigure}
	\begin{subfigure}{0.6\textwidth}
		\includegraphics[width=\textwidth]{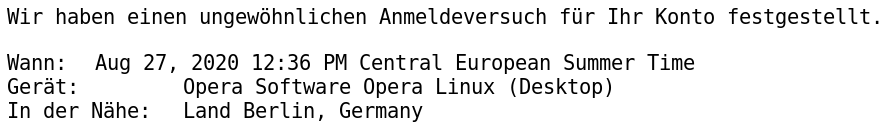}
		\caption{E-Mail notification, including browser details.}
		\label{fig:amazon_login:email}
	\end{subfigure}
	\caption{Fingerprinting used as an additional security layer during login to detect anomalies like changed browsers.}
	\label{fig:amazon_login}
\end{figure}

Browser fingerprinting can serve multiple purposes. To better understand the topic's controversy, two examples are highlighted where fingerprinting brings benefits on one side but drawbacks on the other. 
\paragraph{Security}
Not all purposes are necessarily evil or dubious in terms of privacy. Fingerprinting can be part of a website's security practices. For example, a popular e-commerce website notices a login attempt with Opera instead of Firefox and promptly asks the user to perform an additional authentication step. Figure \ref{fig:amazon_login} shows the prompt (\ref{fig:amazon_login:login}) and the corresponding notification email (\ref{fig:amazon_login:email}). The latter includes detailed information about the login date, device, and approximate location. Thus, a malicious attacker is thwarted from gaining access to that account while notifying the victim of a login anomaly. Security breaches resulting in exposed credentials are not uncommon. For example, \cite{haveibeenpwned} lists over 10 billion ``pwned accounts'' and almost 500 hacked websites, including reputable businesses such as Adobe, Avast, Comcast, Yandex and others.

By tying already authenticated sessions with browser fingerprinting identifiers, authentication-related attacks such as session hijacking can be detected and prevented. Despite an attacker obtaining a valid session token (e.g., through \gls{xss}), they might not be able to exactly replicate the browser or hardware properties, thus failing to hijack the session.

Not only do users benefit from additional account security, but businesses gain a reduced risk of fraudulent activity. Anti-fraud services also rely on fingerprinting \cite{iovation_fraud}. Similarly, companies offer device identification techniques to detect non-human bot traffic \cite{datadome_botdetection}.

\paragraph{Analytics \& Advertising}
Unfortunately, there is another side to the coin. Browser fingerprinting facilitates user identification across websites. Industries profiting from this are analytics and advertising-related. While tracking-information stored in browser cookies can easily be cleared by a user \cite{clear_cookies}, device or browser properties cannot trivially, although some browsers try to change that as elaborated in section \qnameref{sec:background:relatedwork:defensive}. When a user completely disables cookies, device fingerprinting can serve as an alternative to track her nonetheless. If cookies are enabled, the fingerprint can improve the tracking accuracy by linking the data. After all, precisely identifying a returning visitor or customer can enhance a service's tracking quality. 

\subsubsection{Fingerprinting Process}
Browser fingerprinting is a broad topic that has several facets. \emph{Passive} and \emph{active} fingerprinting define how information is obtained. To further discern the necessity of persistent client-side storage, the terms \emph{stateful} and \emph{stateless} fingerprinting supplement the categorization. While this thesis' work primarily focuses on stateless and active fingerprinting, all four categories can work together and are therefore introduced.
\newpage
\paragraph{Browser Fingerprint}
A \emph{device fingerprint} or \emph{browser fingerprint} usually consists of information gathered from the hardware, operating system or the browser's configuration \cite{Laperdrix2020}. The thesis follows \cite{Laperdrix2020} limitations and focuses only on fingerprinting information collected through the browser. IP addresses or additional properties can be used on mobile devices \cite{Kurtz2016}, but are not considered in this work.

\paragraph{Passive Fingerprinting}
Although older literature \cite{Boda2012} uses a different interpretation of passive fingerprinting, the \gls{w3c} defines the term as follows:
\begin{quote}
``Passive fingerprinting is browser fingerprinting based on characteristics observable in the contents of Web requests, without the use of any code executed on the client.'' \cite{w3c_fp}
\end{quote}
When a browser requests a document, it includes information about itself in the \gls{http} headers. Table \ref{tab:http-headers} lists typical headers and values. \texttt{Accept}, \texttt{Accept-Encoding} and \texttt{Accept-Language} provide information about a browser's processing capabilities as well as localization settings. The \gls{dnt} header's presence can be part of a fingerprint, too.

\begin{table*}[h]
	{
		\footnotesize
	\begin{tabularx}{\linewidth}{lX}
		\toprule
		Header & Value \\
		\midrule\midrule
		User-Agent & Mozilla/5.0 (X11; Linux x86\_64; rv:79.0) Gecko/20100101 Firefox/79.0 \\
		\midrule
		Accept & text/html,application/xhtml+xml,application/xml; q=0.9,image/webp,*/*;q=0.8 \\
		\midrule
		Accept-Language & en-US,en;q=0.5 \\
		\midrule
		Accept-Encoding & gzip, deflate, br \\
		\midrule
		Referer & https://browserleaks.com/ \\
		\midrule
		DNT & 1 \\
		\midrule
		Upgrade-Insecure-Requests & 1 \\
		\bottomrule
	\end{tabularx}
	\caption{Typical headers in a \gls{http}-request.}
	\label{tab:http-headers}
	}
\end{table*}

\begin{figure}
	\includegraphics[width=\textwidth]{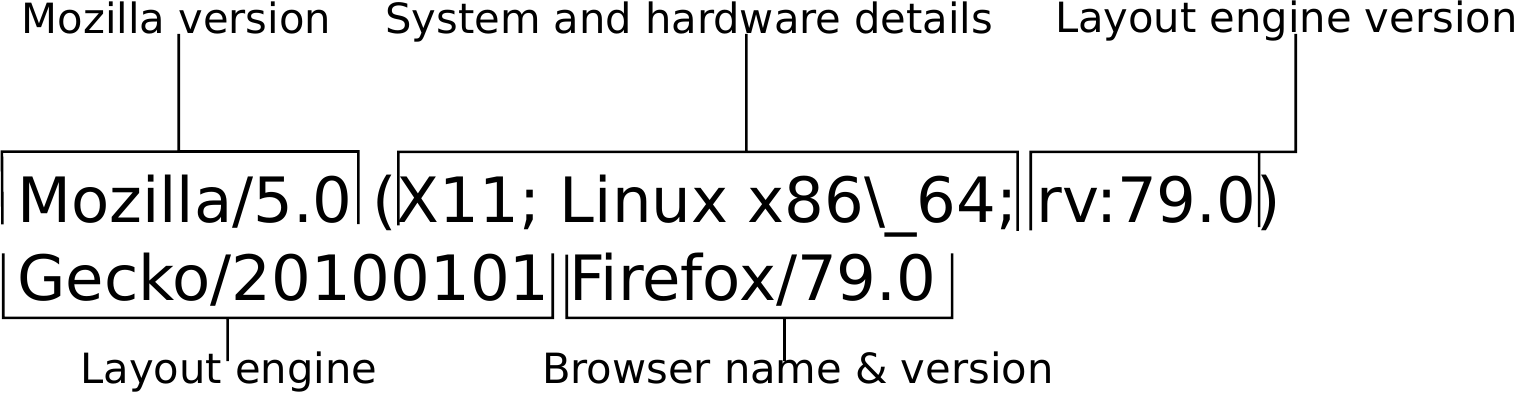}
	\caption{Information disclosed by the \texttt{User-Agent} header.}
	\label{fig:useragent}
\end{figure}

The \texttt{User-Agent} string discloses various types of information, including the system's software and hardware details, version numbers, and more as illustrated in Figure \ref{fig:useragent}.

Additionally, information about the device can be derived by inspecting lower \gls{osi} layers. For example, the operating system, \gls{mtu}, approximate distance in hops are detectable through \gls{tcp}/\gls{ip} fingerprinting \cite{Fried03passiveoperating,browserleaks}. Table \ref{tab:tcp-fingerprint} provides an example obtained from the \gls{ip} address tool \cite{browserleaks}.

As no code is executed on the client, all information is processed server-side, rendering the process highly opaque. A user cannot tell whether or what passive fingerprinting techniques are performed on a server.
\begin{table*}[h]
	\centering
	\footnotesize
		\begin{tabularx}{\textwidth}{XX}
			\toprule
			Fingerprint type & Value \\
			\midrule\midrule
			\gls{os} & Linux (2.2.x-3.x) \\
			\midrule
			\gls{mtu} & 1452 \\
			\midrule
			Distance & 13 Hops\\
			\bottomrule
		\end{tabularx}
		\caption{\gls{tcp}/\gls{ip} fingerprinting example}
		\label{tab:tcp-fingerprint}
\end{table*}
\vspace*{-\baselineskip}
\paragraph{Active Fingerprinting}
In contrast to passive fingerprinting, the execution of code to retrieve information is considered \emph{active fingerprinting} in the specification to mitigate browser fingerprinting by the \gls{w3c}:
\begin{quote}
	``For active fingerprinting, we also consider techniques where a site runs JavaScript or other code on the local client to observe additional characteristics about the browser, user, device or other context.'' \cite{w3c_fp}
\end{quote}
JavaScript is a client-side browser programming language.  With \gls{ajax}, the web application model changed from synchronous to asynchronous, making websites more responsive to user interaction \cite{garrett2005ajax}. The scripting language's adoption on the internet reaches more than 95\% according to \cite{w3techs_js}. Alternatives, for example Java, Flash, and Silverlight, are rarely used \cite{w3techs_js}. 

JavaScript's dominance might be an indicator for its use in active fingerprinting techniques. Table \ref{tab:browser-js-props} provides an overview of the properties it can access. A more detailed breakdown of all the properties and their functions is provided in the appendix (see  \qnameref{sec:appendix:featuregroups-ratings}).

\begin{table*}[h]
	\centering
	\footnotesize
	\begin{tabularx}{\linewidth}{lX}
		\toprule
		Feature & Description \\
		\midrule\midrule
		Browser & Information about the browser's User-Agent, installed plugins, or various settings such as language, timezone, location, Do Not Track, and more. \\
		\midrule
		Hardware & Information about a device's memory, \gls{cpu}, concurrency, screen, or battery are obtainable. \\
		\midrule
		Storage & Read/write access to the client side storage APIs, such as WebKit, session or local storage, or IndexedDB.\\
		\midrule
		Media & Enumerating audio and video formats, or observing media processing behavior.\\
		\midrule
		Graphics & Monitoring slight variances in Canvas and WebGL that depend on the hardware (e.g. \gls{gpu}).\\
		\midrule
		Fonts & The list of supported fonts can vary per device.\\
		\midrule
		... & ...\\
		\bottomrule
	\end{tabularx}
	\caption{Examples of information accessible by JavaScript.}
	\label{tab:browser-js-props}
\end{table*}

History stealing is a technique to extract a browser's history \cite{Wondracek2010}. It also classifies as active fingerprinting, since it executes code to exploit known or unknown bugs in browsers to derive information about the history. Although these bugs are almost two decades old \cite{bugzilla_57351,bugzilla_147777,bugzilla_224954}, \cite{browser_history_revisited} discovered similar issues in \citeyear{browser_history_revisited}. According to \cite{browser_histories_study}, up to 80\% of users can be re-identified using the browser history. 
\paragraph{Stateful Fingerprinting}
Some fingerprinting techniques require persistence. A web server can request the browser to save specific key-value pairs using the \texttt{Set-Cookie} header. Due to security restrictions from the \gls{sop}, such cookies are only sent to the originating website. Nonetheless, they can contain unique identifies or previously collected fingerprinting information. \cite{Englehardt2015} found that users can be surveilled by observing third-party tracking cookies. Since browsers allow the user to remove cookies at any time \cite{clear_cookies}, other persisting techniques were invented. \cite{Ayenson2011} discussed Flash and HTML5 local storage and discovered it on popular websites, whereas \cite{Acar2014} analyzed so-called \texttt{evercookies} that try to survive and restore cookie removals. Passive fingerprinting benefits from persisted information because unique identifiers can be set and retrieved by the web server. 

\paragraph{Stateless Fingerprinting}
When no data is stored on the client-side, the fingerprinting technique is considered stateless. Information about the browser or device can still be collected (i.e., using active fingerprinting) but it is not persisted. Therefore, the obtained information or unique identifiers need to be transferred back to a server. When using JavaScript, this can be achieved through at least two methods:

\begin{smallenumerate}
	\item Deliberately instructing the browser to send an \gls{ajax}-based \gls{xhr} containing the information.
	\item Indirectly instructing the browser to send a \gls{http} request by creating a new \gls{html} element in the \gls{dom}, e.g. \texttt{<img src="https://tracker.tld/track?identifier=[data]">}  
\end{smallenumerate}
In theory, the fingerprinted properties will not often change, and therefore, the same or highly similar fingerprint can be calculated, forming a unique, stateless identifier.

\begin{figure}[h]
	\centering
	\includegraphics[width=0.8\textwidth]{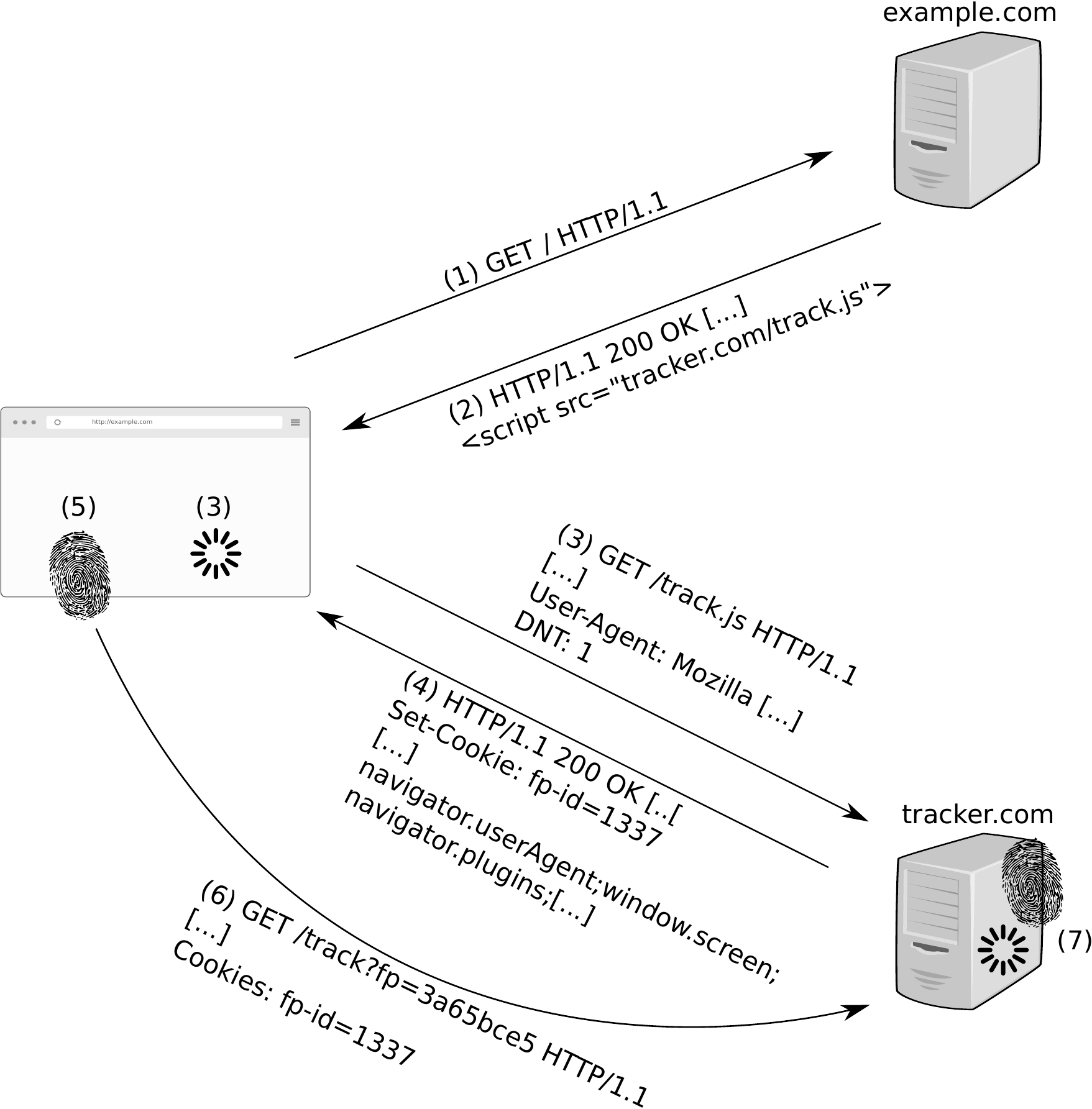}
	\caption{Illustration of passive and active fingerprinting mechanisms}
	\label{fig:fp-mechanism}
\end{figure}

\paragraph{Fingerprinting Mechanism}
Figure \ref{fig:fp-mechanism} illustrates a fictional interaction with a website that includes a fingerprinting script. It tries to serve as an example featuring all four fingerprinting classes. 
\begin{enumerate}[label=(\arabic*)]
	\setlength{\itemsep}{0pt}
	\setlength{\parskip}{0pt}
	\setlength{\parsep}{0pt}  
	\item First, the browser requests a document from the web server \texttt{example.com}.
	\item The server processes the \gls{http} request. A document referencing a 3rd-party fingerprinting script from \texttt{tracker.com} is returned.
	\item The browser receives the \texttt{example.com} page and processes it. When encountering the 3rd-party script, it sends a request to fetch it from \texttt{tracker.com}. The \gls{http} request contains headers with information about the browser.
	\item The web server at \texttt{tracker.com} processes the request. It can use the provided header information to perform passive fingerprinting. At the same time, it generates a unique token to be used in a cookie for stateful fingerprinting. The returned JavaScript code contains active fingerprinting functionality.
	\item Once the 3rd-party script is received, it is executed in the document's context.
It gathers more information about the browser and device using the JavaScript \gls{api}. 
	\item The gathered data is optionally hashed and then sent to \texttt{tracker.com} without persisting data on the client, thus performing stateless fingerprinting.
	\item \texttt{tracker.com} cross-references and optionally validates the data obtained from passive and active fingerprinting.
\end{enumerate}
On subsequent requests to the website or other websites that include a script from \texttt{tracker.com}, the cookie set in (4) would be added. \texttt{tracker.com} could therefore cross-reference the data received from step (6). Discrepancies between values extracted from passive and active fingerprinting could be detected (7), e.g., the \texttt{User-Agent} header being different from \texttt{navigator.userAgent}.

\subsubsection{Privacy Considerations}

Having control over how one's personal information is processed on the internet is referred to as online privacy \cite{Pavlou2011}. Personal information usually needs to be provided in order to use an online service. For example, the email address, name, payment details (e.g., credit card information), and billing address might be required to sign up for a paid service. Although this information can help with revenue, customer relationship management, enhancing the service, or be required by law (e.g., tax reasons), it could be leaked \cite{haveibeenpwned} or otherwise misused by companies. \citeauthor{Baruh2017} analyzed more than 150 papers on online privacy from 1990 to 2016 and found that ``privacy concerns predict the extent to which individuals use online services and engage in privacy management [...]'' \cite{Baruh2017}. 

\begin{figure}[h]
	\centering
	\includegraphics[width=\textwidth]{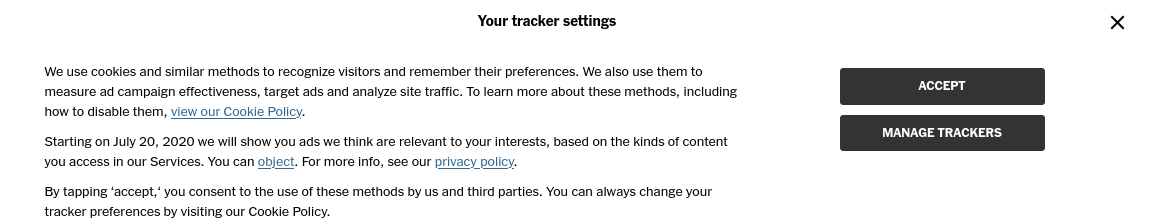}
	\caption{``Cookie Bar'' asking for consent to set cookies.}
	\label{fig:cookie_banner}
\end{figure}

In an attempt to create awareness and strengthen online users' privacy, the \gls{eu} began to regulate the use of tracking technologies since 2002 \cite{eu_eprivacy}. Most cookies require explicit consent by a user, thus limiting the effectiveness of stateful fingerprinting. However, a study conducted by \citeauthor{Trevisan2019} in \citeyear{Trevisan2019} shows that a large percentage of websites sets tracking cookies before getting consent through the ``Cookie Bar'' (e.g., in Figure \ref{fig:cookie_banner}), thereby violating the \gls{gdpr} since it entered into force in 2018 \cite{Trevisan2019}. 

As stateful fingerprinting requires consent, which a privacy-conscious person will probably withhold, stateless fingerprinting technology is likely to gain more traction. According to \cite{gdpr_paper}, the \gls{gdpr} covers 
\begin{quote}
	``Personal data is information that, directly or indirectly, can identify an individual, and specifically includes online identifiers such as IP addresses, cookies and digital fingerprinting, and location data that could identify individuals.'' \cite{gdpr_paper}
\end{quote}
Even single browser properties that are used for fingerprinting purposes might fall under the \gls{gdpr} regulation \cite{piwik_gdpr}. Before such data can be lawfully processed, consent must be provided. However, one could argue that since some browser properties (e.g, screen size, fonts, canvas, and more) need to be accessed and used on the client-side to properly layout a website, play videos, or perform other legitimate actions, it is hard to identify and discern between legitimate and illegitimate usage. As \cite{Trevisan2019} suggests, some websites might not obtain prior consent, or users might consent without understanding all consequences (i.e., assuming cookie tracking instead of fingerprinting). \cite{192371} showed that only every other user was aware of online behavioral advertising, while only a fraction was familiar with privacy-protection tools. 42\% found advertising beneficial, and 23\% did not mind targeted ads. What remains is the question of whether tracking pursuits legitimate or vicious intent. A user cannot easily determine the intent since the collected fingerprints or browser properties are often transferred to a server. Once the information is there, the user cannot reason about its use. Both security or tracking-related processing is plausible, as well as sending the data to other business partners through backend communication.

\subsection{Overview of Related Work}\label{sec:background-related-work:related-work}

Browser fingerprinting is not a new research topic. Past and current related academic work will be outlined to the best of the author's knowledge. 

This thesis is the continuation of recent research from the chair of Security in Telecommunication at the Technical University of Berlin. Fietkau and Kybranz built \FPMON~ and focused on identifying fingerprinting on a large set of popular websites \cite{kybranz, fp_paper}. Extending on the existing tooling, this thesis examines the individual fingerprinting scripts and the actors controlling them. The initial, promising results are part of the publication \cite{fp_paper}.

Furthermore, there exists extensive research by others as well. According to \cite{Laperdrix2020}, it began in \citeyear{mayer2009any} when \cite{mayer2009any} uniquely identified over 95\% out of 1,328 clients using less than 5 properties. \cite{panopticlick} is often cited for the first broad study on browser fingerprinting using the Panopticlick project about a year later. From more than 470 thousand collected fingerprints, 83.6\% were unique when only three \gls{http} headers, five values from JavaScript and Flash or Java were used as sources of information. The Panopticlick website \cite{panopticlick_website} is online to this date, and users are encouraged to learn about their browser's uniqueness. Various other projects emerged that show a user their browser fingerprint \cite{browser_spy, browserleaks, fp_petportal}, most notably AmIUnique, which aims to implement ``state of the art fingerprinting techniques'' \cite{amiunique} and shares global statistics about a browser's uniqueness.


Beyond that, the research is categorizable into offensive and defensive techniques and large-scale studies. It is essential to consider both sides to understand the whole ecosystem better. 
\subsubsection{Offensive Fingerprinting Techniques}
As stated earlier, a browser is a continually evolving piece of software with a continuously increasing feature set and complexity, resulting in many information sources for fingerprinting scripts. While some information is directly accessible, other can be derived through side-channels.
Since most active fingerprinting scripts rely on JavaScript, \cite{mulazzani2013fast} discovered that the JavaScript engine itself could be used to identify a browser. Similarly, subtle differences and side-channels in the engine can reveal hardware and software details \cite{schwarz2019javascript}. \gls{css} is a vital component for a website's layout. Again, differences between browsers' implementation of certain \gls{css} properties can be used to deduct the browser family \cite{Unger2013}, or installed fonts and more \cite{Takei2015}. \cite{Fifield2015} showed that subtleties in font rendering can be used as well. Browser extensions can increase a fingerprint's uniqueness \cite{Gulyas2018, Starov2017} and can be enumerated through various techniques \cite{Sjsten2017, Starov2019, sanchez_browser_ext}. Likewise, new \gls{html}5 features, such as Audio processing \cite{Englehardt2016, Queiroz2019} or Canvas and WebGL \cite{MS12, Acar2014} aid fingerprinting scripts. The latter was shown to be useful for identifying hardware properties \cite{Cao2017}. However, not all new feature proposals are implemented. After a privacy review and resulting concerns, the battery status \gls{api} was not implemented \cite{battery_status}. It is plausible that other browser features are already or will be utilized for fingerprinting purposes in the future.

\subsubsection{Defensive Fingerprinting Techniques} \label{sec:background:relatedwork:defensive}
Academia not only focuses on offensive techniques but also defensive ones. \cite{Vastel2018} showed that despite a browser's fingerprint changing due to version upgrades, hardware, or software changes, it does not sufficiently prevent tracking. In an experiment using FP-Stalker, they managed to track browsers for more than 54 days on average. There are different approaches to mitigate browser fingerprinting \cite{Laperdrix2020}. One approach is to frequently change the fingerprint by increasing its diversity, e.g., by modifying browser properties to random or pre-defined values \cite{Torres2015,FaizKhademi2015,nikiforakis2014privaricator,Laperdrix2015,Laperdrix2017, Baumann2016}. However, this can lead to the ``Paradox of Fingerprintable Privacy Enhancing Technologies'' \cite{panopticlick}, a term for a situation when mitigating fingerprinting leads to the inverse effect of better fingerprintability. For example, it arises when tools are not consistent or do not cover all occurrences of a property (e.g. changing \texttt{navigator.userAgent} but not the corresponding \texttt{User-Agent} \gls{http} header). Instead of changing properties, \cite{nikiforakis2014privaricator, Laperdrix2017, 235463} and others introduce randomness into return values, e.g., for Canvas, WebGL, Audio, or other hardware-based properties. A different approach is to use the same set of properties across a large set of devices, thus creating a homogenous fingerprint \cite{Laperdrix2020}. It is followed by UniGL \cite{235463}, the Tor Browser \cite{torbrowser_same_fp}, DCB \cite{Baumann2016} and recent browsers, e.g. Firefox \cite{firefox_antifp} or Safari \cite{safari_antifp}. The last resort is to reduce the browser's fingerprinting surface by disabling plugins, features, blocking script execution, or similar measures. For example, the Tor Browser disables WebGL and Canvas by default \cite{torbrowser_same_fp}, but such drastic measures usually come with a usability trade-off that might not be suitable for the average internet user.

\subsubsection{Large-Scale Studies}
While offensive and defense research is crucial to understand the fingerprinting topic, large-scale studies to identify fingerprinting are more relevant to this work. Although the studies performed by \cite{panopticlick} in \citeyear{panopticlick}, \cite{Laperdrix2016} in \citeyear{Laperdrix2016} and \cite{GmezBoix2018} in \citeyear{GmezBoix2018} showed that devices are uniquely identifiable through browser fingerprinting, the data was collected using their own fingerprinting script on a website under their control. This is in contrast with this work's goal to identify fingerprinting scripts, their networks, and actors present on the internet, since it requires analyzing a large set of websites for the presence of fingerprinting scripts. In \citeyear{Acar2013}, \cite{Acar2013} used their FPDetective framework to scan the Alexa Top 1M with a focus on font-based fingerprinting. They found 404 websites and 13 fingerprinters from different providers, such as BlueCava, Perferencement, MaxMind, or others. In the same year, \cite{Nikiforakis2013} analyzed the fingerprinting scripts from BlueCava, Iovation, and ThreatMetrix. According to them, all three providers fingerprinted the browser using JavaScript and Flash, while BlueCava collected the most properties. Their scan of the Alexa Top 10,000 revealed 40 sites using one of the scripts. Using a different set of less popular sites that employ fingerprinting, and website categorization services, they found the majority of categories relating to websites that contain personal or financial information. To their surprise, the two largest categories were ``malicious sites'' and ``spam''. A year later, \cite{Acar2014} crawled the Alexa Top 100,000 websites for Canvas fingerprinting and found 5.5\% employing that technique with 95\% of all scripts belonging to AddThis.com. In total, they found 20 third-party domains with fingerprinting scripts, which they identified by attributing function calls to a script's URL. While 9 of those are reported to be ``in-house'', the remaining 11 fingerprinting scripts are from providers offering them as part of other services. They note that many scripts made use of \emph{FingerprintJS} \cite{fingerprintjs_github}, an open-source fingerprinting library. Two years pass before \cite{Englehardt2016} analyzed fingerprinting scripts on the Alexa Top 1M using the OpenWPM framework. Using this framework, they instrumented selected JavaScript getter \& setter functions and collected the script URL when accessing them. 1.6\% sites are reported to use canvas-based techniques, of which almost all are loaded from 400 third-party domains. Furthermore, they discovered font, WebRTC, audio and battery API fingerprinting activity on a small percentage of the crawled websites. In \citeyear{AlFannah2018}, \cite{AlFannah2018} research showed an increase of first and third-party fingerprinting to approximately 69\%. Contrary to the previous research, their methodology consisted of detecting when browser properties are sent back to a first- or third-party. From the crawled 10,000 websites, 6,876 transmitted fingerprinting information from 1,914 distinct fingerprinting scripts. Of the latter, \texttt{google-analytics.com}, \texttt{doubleclick.net}, \texttt{quantserve.com} and others are among the top third-party domains.

\subsubsection{Positioning The Thesis}
While the research conducted by \cite{Acar2013, Acar2014, Nikiforakis2013} exposed a few actors and websites with fingerprinting scripts, it is quite dated. According to their results, fingerprinting was not commonly used. Similarly, \cite{Englehardt2016}'s study began to instrument JavaScript APIs and to collect script URLs, but their focus lied on specific fingerprinting techniques. Although the alternative approach of detecting browser properties in outgoing traffic by \cite{AlFannah2018} showed that browser fingerprinting is much more common than previously thought, it was limited to only 17 properties. Furthermore, the trend towards more encryption on the internet might hinder this method in the future or lead to less accurate results. As this thesis builds on the work done by colleagues Fietkau and Kybranz, in which they distilled over 100 fingerprinting-related JavaScript functions from various fingerprinting products, it allows for a much broader monitoring scope and more precise fingerprinting script identification. In contrast to previous work, this thesis will analyze the (third-party) script URLs, the fingerprinting scripts' severity, the affected websites, and in manual review the script providers. As stated earlier, the goal is to identify networks and actors who fingerprint users on the internet to evaluate the current state of fingerprinting. Furthermore, we will answer the research questions to better understand it, i.e., to answer if there are actors who aggressively collect browser fingerprints. 


\section{Design} \label{sec:design}
While this chapter introduces the reader to the thesis' studies, major concepts, and work on an abstract level, the next chapter will cover specific implementation details. Since this thesis is a continuation of Kybranz' and Fietkau's work (see \cite{fp_paper,kybranz}), some terminology and concepts are closely related. Therefore, the reader is advised to consult their work for more background or details.

\subsection{Behavior-based Fingerprinting Detection}
With modern JavaScript minification \cite{uglifyjs} and obfuscation \cite{obfuscatorio} techniques trying to minimize a JavaScript script's file size or to hinder static and manual code review, a different analysis method is needed. This thesis defines the behavior-based approach as the means of instrumenting a browser to monitor access to JavaScript properties or observe function calls, thus providing information about access patterns, function arguments, and return values. This is feasible because the browser's JavaScript properties and functions will eventually be called by a (obfuscated) fingerprinting script. Although to the author's knowledge, the term was not coined before, the technique is not new as others have used it before (e.g., \cite{Acar2013, Englehardt2016}). While \citeauthor{Nikiforakis2013} discovered fingerprinting scripts that use Flash or Java functionality \cite{Nikiforakis2013}, both technologies are not actively supported anymore. Adobe announced Flash's end-of-life in 2017 due to \gls{html}5 and other standards taking its place \cite{adobe_flash}. Similarly, Oracle deprecated Java Applets in 2019 \cite{oracle_applets}. Therefore, it can be assumed that fingerprinting primarily focuses on JavaScript rather than deprecated technology. By building the \FPMON~extension, \cite{fp_paper, kybranz} followed a behavior-based approach to detect fingerprinting activity on websites. From the manual analysis of open and closed source fingerprinting libraries, they identified 115 fingerprinting-related properties and functions, which they rate and categorize into 40 \emph{feature groups}. Their categorization is reused and can be found in the appendix in section \qnameref{sec:appendix:featuregroups-ratings}. Each feature group is labeled as either \emph{sensitive} or \emph{aggressive} depending on its potential of misuse and other factors. The more distinct feature groups are observed, the higher the likelihood for potentially privacy-invasive fingerprinting activity. 

\subsection{Script-based Fingerprinting Analysis}\label{sec:design:fp-analysis}
While previous work (e.g., \cite{kybranz}) focuses on the detection of fingerprinting activity on websites, this thesis targets the underlying, individual JavaScript scripts. By monitoring each script's activity, fingerprinting and non-fingerprinting scripts should become distinguishable by their vastly different behavior. Additionally, the collected observations allow more granular further analysis.

\subsubsection{Defining Script Signatures and Scores}\label{sec:design:fp-analysis:scriptsigsscores}
The \emph{script-based fingerprinting signature} is introduced to discern between the two types of scripts. It is the semicolon-separated concatenation of all observations' feature groups ordered by the sID (script-ID) on a per-script basis. Given the example observations in Table \ref{tab:script-signature-observations}, the following two script-based signatures can be computed:
\begin{smallitemize}
	\item {example.com/layout.js}: \verb*|Screen_window;Screen_window;...|
	\item {tracker.com/fp.js}: \verb*|UserAgent;Canvas;WebGL;JS_fonts;...|
\end{smallitemize}
Similarly, a \emph{script's score} is the number of its signature's unique feature groups:
\begin{smallitemize}
	\item {example.com/layout.js}: 1 + ...
	\item {tracker.com/fp.js}: 4 + ...
\end{smallitemize}
\begin{table*}
	\footnotesize
	\centering
	\begin{tabularx}{\linewidth}{rrrXXX}
		\toprule
		gID & sID & oID &Property or function & Script origin & Feature group \\
		\midrule\midrule
		0 & 0 & 0 & navigator.userAgent & tracker.com/fp.js & UserAgent \\
		\midrule
		1 & 0 & 1 & screen.height & example.com/main.js & Screen\_window \\
		\midrule
		2 & 1 & 0 & getImageData() & tracker.com/fp.js & Canvas \\
		\midrule
		3 & 1 & 1 & screen.width & example.com/main.js & Screen\_window \\
		\midrule
		4 & 2 & 0 & drawArrays() & tracker.com/fp.js & WebGL \\
		\midrule
		5 & 3 & 0 & fillText() & tracker.com/fp.js & JS\_fonts \\
		\midrule
		... & ... & ... & ... & ... & ...\\
		\midrule
		\bottomrule
	\end{tabularx}
	\caption{Example property and function observations including the associated feature group and script origin. The global-ID (gID), script-ID (sID) and origin-ID (oID) are essential to preserve the observation order.}
	\label{tab:script-signature-observations}
\end{table*}

\subsubsection{Defining Page Signatures and Scores}\label{sec:design:fp-analysis:pagesignaturesscores}
Our script signatures enable improved \emph{page signatures}. While \cite{kybranz} computes a page signature from feature groups in the order of the observed function calls (see gID in Table \ref{tab:script-signature-observations}), our modifications to \FPMON~change a page signature to the concatenation of all script signatures - ordered by the scripts' \gls{url}. Given the observations from Table \ref{tab:script-signature-observations}, the two example page signatures differ as follows:
\begin{smallitemize}
	\item {\cite{kybranz}:} \verb*|UserAgent;Screen_window;Canvas;Screen_window;WebGL;...|
	\item {Ours:} \verb*|Screen_window;Screen_window;[...];UserAgent;Canvas;WebGL;...|
\end{smallitemize}
The new, script signature-based page signature provides the same information as the old one. However, at the same time, it allows finding script signatures as substrings in page signatures, which is believed to be useful in future work. Since our analysis is agnostic to the order of the script signatures within a page signature, it could also be ordered by the origin-ID (see Table \ref{tab:script-signature-observations}). 

The page score calculation is not affected by this change because the observed events are the same and, therefore, the number of unique feature groups, too. Both examples have a page score of 5.

\begin{figure}[h]
	\centering
	\includegraphics[width=0.9\textwidth]{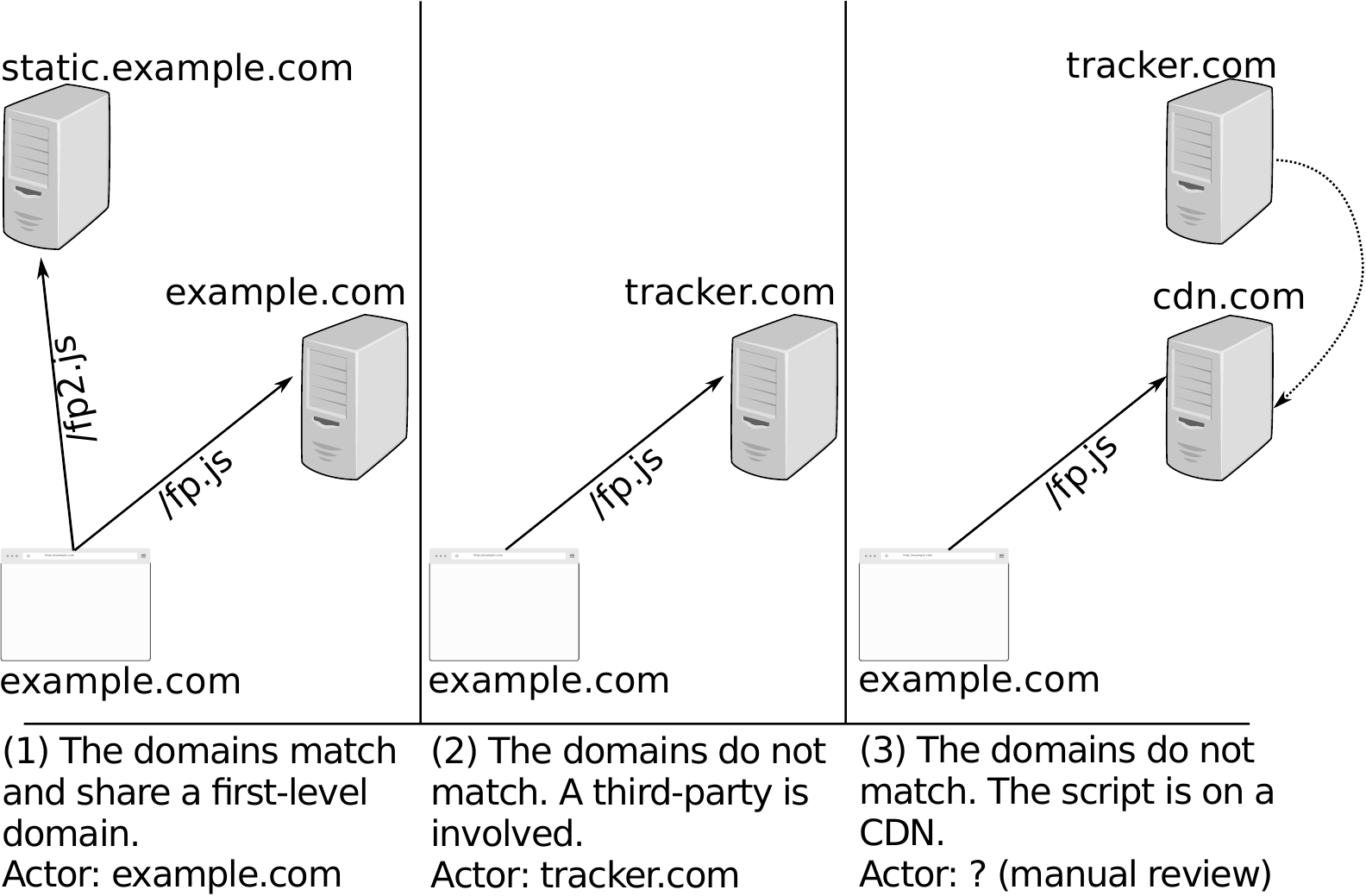}
	\caption{Illustration of the three ways this thesis identifies actors.}
	\label{fig:actor-analysis}
\end{figure}

\subsection{Determining Script Origins and Actors} \label{sec:design:scriptorigins-actors}
The collection of each script's URL - the \emph{script origin} - can provide information about the actor. JavaScript files can be served from the same web server as the website, a \gls{cdn} or a third-party. Depending on the script origin, the attribution to a specific actor (i.e., an organization, a company, or other entity) is of mixed difficulty. An actor can control multiple scripts from different script origins. For example, by serving content from different domains. Associating one or more script origins with one actor is done through manual review. The following subsections explain the three script origin types illustrated in Figure \ref{fig:actor-analysis}.

\subsubsection{Same Origin Actors}
In the case that the script's origin matches the website's domain, it is evident that the script can be attributed to the actor operating the website. This is trivial for cases where \texttt{example.com} is requested, and the script (e.g., \texttt{example.com/script.js}) is served from the same domain. In some cases, only the first-level domain matches when the scripts are served from a subdomain such as \texttt{static.example.com/script.js}. In those cases, the actor is assumed to be the entity operating \texttt{example.com}.

\subsubsection{Third-party Actors}
When the website's domain and the script origin do not match, it is served by a different entity, the so-called third-party. A third-party is considered to be the actor who has full control over the script's contents. Therefore, an actor can change the script's source code at any time. In the context of fingerprinting, this allows the actor to continuously improve the fingerprinting algorithm, its obfuscation or randomization, data collection endpoints, or other properties. Embedding third-party JavaScript code is a pattern that service providers often offer. For example, \cite{google_analytics,google_adsense} and others. However, with third-party domains, the script's behavior can be attributed to the actor having control over the script origin's domain. Given the example in Figure \ref{fig:fp-mechanism}, the actor is \texttt{tracker.com}'s operator.

\subsubsection{CDN-concealed Actors}
We classify \gls{cdn}s as a special type of third-party. \gls{cdn}s usually aim to reduce page load times by providing files (e.g., JavaScript libraries) that are repeatedly included in websites in close proximity to the user, thus benefiting from cache hits and optimized network connectivity. Since a \gls{cdn} can host a vast majority of static resources, it hides the real origin and masks an actor. For example, the popular fingerprintjs2 library can be included from the \gls{cdn} \texttt{JSDELIVR} \cite{jsdelivr_fp2}, thus preventing a direct association with an actor. Although further analysis could reveal the actor if the information is sent to another domain but such an analysis is left to be future work. Although a \gls{cdn} has control over the hosted files, a lot of modern browsers support ``Subresource Integrity'' \cite{subresource_integrity}. It is a browser feature to verify an external resource's integrity before embedding it, thus thwarting malicious changes by a \gls{cdn}. 

\subsection{Conceptualizing the \FPNET~Scanner}
Information about fingerprinting scripts needs to be collected on a large set of websites to evaluate the existence of fingerprinting networks on the internet. Complementary to related and previous work \cite{FaizKhademi2015, Laperdrix2017, Englehardt2016, kybranz}, the most popular websites according to the Alexa traffic ranking \cite{alexa_topsites} will be used. To facilitate the result comparison and as a suitable performance trade-off, only the top 10,000 websites will be analyzed. 

The collection process requires an environment whose functionality and behavior closely matches a real browser. Instrumented headless browsers provide with such requirements while enabling automation. On that basis, \cite{kybranz} implemented a Chrome extension (\FPMON) and scanning tool (\FPCRAWL). Both tools were modified and extended to fit our requirements. We call the resulting scanning system \texttt{Fingerprinting Networks Scanner} or short \FPNET. The post-processing scripts used to analyze the fingerprinting networks and actors are also part of \FPNET. On an abstract level, the following modifications and extensions were realized. The implementation details will be provided in Chapter \ref{sec:implementation}.
\subsubsection{Adapting \FPMON}
\paragraph{Script Monitoring} 
Since the \FPMON~extension initially only monitored a whole page's fingerprinting activity, the monitoring process was modified to work on the more granular script level. This granularity allows observing each script's activity separately, making it possible to compute the corresponding script-based signature and further analyze it.
\paragraph{Script Origin Determination}
To distinguish between two scripts and their activity, functionality was added to determine a script's origin. For each script, its \gls{url} is saved.

\subsubsection{Extending \FPCRAWL}
\paragraph{Browser Parameter Tweaks}
In the assumption that fingerprinting scripts could detect the headless browsers and reduce their activity, environment and browser parameters were adjusted to make it appear more realistic. For example, this includes the virtual screen's size, color depth, dots per inch, or the browser being started in full-screen mode. Furthermore, the browser was instructed to disable web security or sandbox protections and ignore certificate errors to prevent any incompatibilities or bugs with the scanning components. While this leaves room for malicious abuse by fingerprinting scripts, each browser analyzed only one website simultaneously in separate contexts, thus isolating a website from manipulation. 
\paragraph{Runtime Improvements}
Much effort was put into improving the runtime and scanning stability. Changes include improved logging and debug functionality to detect bugs and scanning errors within \FPNET. Additionally, the extended logging features provide information about the amount of successfully analyzed websites. \cite{Englehardt2016,AlFannah2018} reported frequent crashes and instability of selenium. While \cite{kybranz}~worked around them with periodic restarts, \FPNET~replaces the error-prone selenium hub with its custom node management implementation. Furthermore, a custom browser kill script terminates a node's browser process after a predefined time to prevent browser hangs or other stability issues without restarts.
\paragraph{Adding Proxy Support}
For two studies, access to the \gls{http} response headers and file content was required. The scanning system was extended by a \acrfull{mitm} proxy component, similar to previous work \cite{Englehardt2016, Acar2013, Acar2014}. A custom proxy plugin associates each \gls{http} request and response with the scanned website, and saves the network traffic for later analysis. For this approach to work with encrypted connections, the browser had to ignore \gls{tls}/\gls{ssl} errors.

\subsubsection{Consistent Data Processing}
Although the studies' reproducibility depends on some uncontrollable factors, e.g., website's availability or its source code at the time of scanning, this thesis aims to facilitate reproducibility. Therefore, the initial post-processing steps are considered to be part of \FPNET.
 
\paragraph{A Rational Database for Storage}
In order to provide flexibility in analyzing the script activity, the collected observations are parsed and consolidated in a rational database. This step allows using \gls{sql} queries on a defined set of tables to derive answers to research questions or provide the data for further analysis. Furthermore, performance benefits were expected due to parsing and importing all data into the database only once. From there, the data can be accessed in every analysis step.

\paragraph{Analysis Using Post-processing Scripts}
One or more dedicated post-processing scripts are implemented for each study, which queries the database for required information or uses previously processed data. In any case, the set of post-processing scripts should produce the study's results or the data for manual review.
\newpage
\subsection{Uncovering Fingerprinting Networks and Actors}
The collected fingerprinting information is analyzed with regard to the different research questions and the goal of uncovering fingerprinting networks and actors. From the initial questions, five complementary studies were designed.

\subsubsection{Detecting Networks With Script Signatures}\label{sec:design:scriptsignatures}
%
The thesis' main study aims to uncover fingerprinting networks by focusing on the similarity of script signatures. 
\paragraph{Research Questions}
The study aims to answer the following questions:
\begin{smallitemize}
	\item How common and privacy-invasive are browser fingerprinting scripts?
	\item What fingerprinting scripts can be identified?
	\item What are the resulting networks and actors? What size and who are they?
\end{smallitemize}
\paragraph{Building Networks}
For each script, its script signature is computed from the observed behavior. By comparing the signatures, we can compare the scripts despite differences in the script domain, filename or source code. If two or more signatures overlap partially, they are grouped to form a network. Due to their signature behavior, all scripts of a network are likely to share certain functionality. Under the assumption that actors implement their scripts differently, this will naturally separate fingerprinting scripts from each other. A network's shared script signature is used to determine the network's score. 
\paragraph{Identifying Actors}
Since \FPNET~captures the underlying observation's script URL, a list of script domains can be compiled for each network. By counting the script domains or analyzing the score, prominent actors stand out. However, a broad reach does not necessarily imply high-activity fingerprinting scripts. Through a manual review of the networks and their script origin domains, the actors can be identified (see Figure \ref{fig:actor-analysis}). This approach provides insights into who operates fingerprinting scripts on the internet, and what information is collected. It is possible that an actor controls multiple networks.

\subsubsection{Analyzing Scripts for Randomization}\label{sec:design:randomization}
Randomization and obfuscation of a script's origin domain, filename, or file content can impede the static and behavior-based script analysis. Study three examines the following aspects of that topic:
\paragraph{Research Questions}The study aims to answer the following questions:
\begin{smallitemize}
	\item What defensive techniques are utilized by fingerprinting scripts?
	\item Is randomization or obfuscation common?
\end{smallitemize}
\paragraph{Examining Randomization}
The behavior-based approach has its strength in identifying fingerprinting scripts across websites despite randomized script domains or filenames if the behavior does not change. Otherwise, the signatures might not match. Behavior changes can arise from source code changes or altered feature group access patterns. Study 3's two scans in short succession minimize the time window for major adjustments to a website (e.g. rewrite, redesign or script changes), making both scans' scripts comparable for behavior and script origin changes.
\paragraph{Detecting Changed Script Origins}
Analyzing the differences between the scanned pages' scripts and their scripts' URLs will show whether the behavior-based approach has any benefits over a strict URL-based identification method. We evaluate how many URLs have changed but could be re-identified using a matching signature. If re-identification is successful, the behavior-based approach shows its strength and thwarts possible script URL changes, randomization, or obfuscation. 
\paragraph{Detecting Changed Script Signatures}
By matching script origin URLs, the signature accuracy can be evaluated. When the script URLs match, but the script signatures differ, it is either an indication for file content and behavior randomization or weaknesses in \FPMON's observation capabilities.
\paragraph{Evaluating Signature Similarity Matching}
Several string similarity algorithms are evaluated to combat the case when script signatures change. The best performing algorithm is then applied to the script signatures to produce additional script signature matches. This approach could allow for little changes in the script signatures.

\subsubsection{Comparison with a File-Signature Approach}
In study 4, the identification of fingerprinting scripts using file content-based signatures is briefly examined.
\paragraph{Research Questions}The study aims to answer the following questions:
\begin{smallitemize}
	\item Do fingerprinting scripts change or differ?
	\item Can file-signatures support the behavior-based approach or vice versa?
\end{smallitemize}
\paragraph{Imagining a Privacy Enhancement} 
\cite{kybranz} noticed that privacy-enhancing browser extensions depend on URL deny-listing and thus do not prevent fingerprinting from unknown script URLs. Script signatures are unsuitable for a privacy extension, because a signature is computed during  run time. Therefore, a script cannot be blocked before it is executed. At this point, it could already have collected and possibly exfiltrated the fingerprinting information. However, an extension might be developed that blocks outgoing \gls{http} requests using an URL-based deny-list or blocks incoming \gls{http} responses using a file-signature deny-list. 
\paragraph{Defining File Signatures}
In contrast to a behavior-based script signature, a file signature is based on the script's file contents hashed with a suitable algorithm. For this study, the fuzzy hash \texttt{ssdeep} \cite{Kornblum2006} is used. It allows for slight variations in the hashed files, e.g., fingerprinting script configurations. 
\paragraph{Defining File-based Networks}
Based on the file-based signatures, file-based networks can be formed by pair-wise comparison of all signatures. The resulting networks consist of script files whose source code is highly similar. To evaluate the effectiveness of file-based networks, we compare them against signature-based networks through an intermediate \texttt{script domain:filename} mapping of all their scripts. Nevertheless, the file-based approach can be defeated by changes to a file's source code, i.e., changing obfuscation or randomization on each request.

\subsubsection{Examining Security Properties}\label{sec:design:security}
The last study briefly covers web security aspects of the previously identified fingerprinting scripts. It aims to provide insights into the actors' security awareness.
\paragraph{Research Questions}The study aims to answer the following questions:
\begin{smallitemize}
	\item Is transport security used for fingerprinting scripts?
	\item Are web security headers in place?
\end{smallitemize}
\paragraph{Determining Transport Security}
The first part of the study consists of analyzing transport security, more precisely the use of \gls{tls}/\gls{ssl}. Unsecured channels creates opportunities for \gls{mitm} attacks, code execution, or other attack vectors. For example, JavaScript files could be modified to execute malicious code, or the collected fingerprinting information could be obtained or manipulated. All script URLs are checked for \texttt{http://} or \texttt{https://}. Insecure \gls{url}s are checked for redirects to a secure URL. Although redirecting to a secure resource does not prevent the insecure request or response from being attacked, it at least suggests that the resource is available over a secure channel.
\paragraph{Auditing Security Headers}
All scripts' \gls{http} responses are checked for the existence of \gls{http} security headers. The set of analyzed headers includes:
\begin{smallitemize}
	\item \texttt{X-Content-Type-Options}
	\item \texttt{Referrer-Policy}
	\item \texttt{Content-Security-Policy}
	\item \texttt{Strict-Transport-Security} 
\end{smallitemize}
While \texttt{Strict-Transport-Security} is not required to be set on every resource \cite{rfc_6796}, explicitly setting it would ensure that only secured connections are established. Similarly, the \texttt{Content-Security-Policy} header is optional except for JavaScript files executed in the context of workers \cite{csp_workers}. 
\subsection{JavaScript as the Internet's Dependency}\label{sec:design:javascript}
All active fingerprinting activity and behavior-based detection rely on JavaScript being executed in the browser. Therefore, a trivial way to escape active browser fingerprinting would be to disable JavaScript. A survey conducted by Yahoo in 2010 showed only 1\% of users not having JavaScript enabled \cite{yahoo_users}. Today, \cite{w3techs_js} reports more than 96\% of websites using JavaScript as a client-side programming language. When JavaScript is not enabled, a website's client-side interactivity and user experience are impaired. We conduct a study to measure the extent to which JavaScript is required for the intended user experience on the internet:
\paragraph{Research Questions}The study aims to answer the following questions:
\begin{smallitemize}
	\item How relevant is JavaScript for modern websites?
\end{smallitemize}
\paragraph{JavaScript's Impact On User Experience}
JavaScript event handlers enable interactive websites. For example, when the mouse pointer is moved over a \gls{html} element, its \texttt{onmouseover} event is triggered, which could display a tooltip, enlarge an image or otherwise engage with the user interactively. Should JavaScript not be enabled, event handlers cannot trigger, thus not providing the same user experience. Although a \texttt{<noscript>} \gls{html} tag can notify the user about missing JavaScript \cite{w3s_noscript}, it cannot provide the same functionality. However, event handlers are just one aspect of JavaScript's feature set. Since a website might not offer the same user experience or interactivity without JavaScript, the user is coerced into enabling it.
\paragraph{Checking Event Handlers}
Using an early version of the \FPNET~scanner, the Alexa Top 10,000 websites are crawled and checked for 109 event handlers (see appendix \qnameref{sec:appendix:event-handlers}), which were extracted from a real browser. As event handlers can be set in many ways, we analyze three different methods to identify event handlers on a website: 
\begin{smallitemize}
	\item \texttt{onevent} attributes on \gls{html} elements
	\item \texttt{elem.onevent=function()} declarations through JavaScript
	\item \texttt{jQuery}-based event handlers
\end{smallitemize}

\section{Implementation}\label{sec:implementation}
This section aims to outline the most crucial implementation details. As discussed in the previous chapter, the focus lies on changes made to existing tools to answer our research questions.
\subsection{JavaScript Usage On The Internet} \label{sec:implementation:javascript}
To analyze the prevalence of JavaScript on websites, our approach focuses on the existence of event handlers.  
\subsubsection{Compiling a List of Event Handlers}
In order to analyze the event handlers, a list had to be compiled. Since event handlers follow a \texttt{on<eventname>} naming pattern \cite{event_handlers},  the \texttt{window} property was queried for attributes matching the regular expression \texttt{/\^{}on/} as shown in Listing \ref{lst:implementation:event-handler-list}. Running the code in Firefox 82.0 on a Linux computer returns a list of 109 event handlers (see appendix \qnameref{sec:appendix:event-handlers}). This list is not assumed to be exhaustive because event handlers might vary on a device or software basis. For example, touch events \cite{touch_events} were not in the list. Nonetheless, this list constitutes a sufficient basis for the analysis since the focus lies on desktop-based browsers.
\begin{lstlisting}[caption={Compiling a list of event handlers from the \texttt{window} property.}, label={lst:implementation:event-handler-list},language={JavaScript}]
const eventHandlers = [];
for (let eventHandler in window) {
	if (/^on/.test(eventHandler)) {
		eventHandlers[eventHandlers.length] = eventHandler;
	}
}
console.log(eventHandlers)
\end{lstlisting} 
\subsubsection{Detecting Event Handlers on Websites}
Three event handler detection methods were implemented, since event handlers can be set on a \gls{dom} element in multiple ways. Two use native JavaScript functionality \cite{event_handlers}, while the third relies on a library called \texttt{jQuery}. According to \cite{jquery_usage}, ``JQuery [...] simplifies how you traverse HTML documents, handle events, perform animations, [...]'' and more than 83\% of the top 1m websites use it. Once the \gls{dom} finished loading, our detection algorithm iterates over all elements and extracts the event handlers. All occurrences are grouped by the event type and counted.
\paragraph{Checking  on-Attributes}
Elements can set event handlers as \texttt{on-}attributes \cite{event_handlers}. The attribute's value is JavaScript code executed when the event triggers. All \gls{dom} elements' attributes matching the regular expression \texttt{/\^{}on/} were counted.
\paragraph{Checking on-Properties}
Another native JavaScript is to assign functions to \gls{dom} elements' \texttt{on-}attributes in code \cite{event_handlers}. For all \gls{dom} elements, an attribute-based event handler (see \qnameref{sec:appendix:event-handlers}) is counted if it has a function assigned.
\paragraph{Checking jQuery Events}
Listing \ref{lst:implementation:jquery-handlers} shows our implementation for querying jQuery's internal data for defined event handlers. It honors different jQuery versions and certain corner cases as discussed in \cite{jquery_stackoverflow}.
\begin{lstlisting}[caption={Checking an element for event handlers set using jQuery.},label={lst:implementation:jquery-handlers},language={JavaScript}]
jq = window.$ || window.jQuery
if(jq && (jq._data || jq.data)) {
	jq.each((jq._data || jq.data)(jq(element)[0], "events" ), function(e) {
		//counting
	});
}
\end{lstlisting}
\newpage
\subsection{Extending \FPMON}\label{sec:implementation:fpmon}
This section elaborates on the changes to \FPMON. We improve the logging capabilities, enable script-level observations and facilitate manual review of fingerprinting scripts.

\subsubsection{Determining Script Origins}\label{sec:implementation:fpmon-script-sources}
Script URLs are a key part to determine a script's origin and actor. \FPMON's~monitoring functions \texttt{printAccess} and \texttt{detectFingerprinting} call \texttt{constructMetricsArray} for each observed property access or function call. After constructing the metrics-array, a call to \texttt{callOrigin} was implemented, which returns an object with the determined script origin information.

\texttt{callOrigin} uses JavaScript's error handling mechanisms to produce a JavaScript error in a controlled way intentionally. The resulting stack trace provides access to the call stack with its stack frames up to the induced error. Each stack frame holds information about the called function, including the filename, function name, line number, column number, type name, and much more. A script's \gls{url} (\texttt{stack.getFileName()}) was sufficient for the actor analysis. Some stack frames referenced \FPMON~or did not define this property but traversing the call stack eventually revealed the script's origin. Table \ref{tab:call-stack} provides an example call stack.

\begin{table*}
	\centering
	\footnotesize
	\begin{tabularx}{\linewidth}{rXX}
		\toprule
		Index & stack[Index].getFunctionName() & stack[Index].getFileName() \\
		\midrule\midrule
		0 & "callOrigin" & "" \\
		\midrule
		1 & "constructMetricsArray" & "" \\
		\midrule
		2 & "printAccess" & "" \\
		\midrule
		3 & null & "" \\
		\midrule
		4 & "createFingerprint" & "http://tracker.com/fp.js" \\
		\midrule
		\bottomrule
	\end{tabularx}
	\caption{Example call stack that is used to determine an observation's script origin.}
	\label{tab:call-stack}
\end{table*}

\subsubsection{Improving Logging Capabilities}
After the \texttt{callOrigin} function returns, \texttt{constructMetricsArray} further processes the observation for our extended logging purposes. From its function argument \texttt{function\_name}, the corresponding feature group can be derived using the mapping in section \qnameref{sec:appendix:featuregroups-ratings}.

In \texttt{constructMetricsArray}, two objects manage the gathered information: 
\begin{smallitemize}
	\item \texttt{script\_origins} collects all script origins' observations for further processing
	\item \texttt{script\_origins\_internal} stores all script origins' score, unique feature groups and current script-based fingerprinting signature
\end{smallitemize}
The determined script origin is used as the key for both objects. The \texttt{script\_origins} object additionally stores and increments the script origin's index (\emph{origin-ID}, see Table \ref{tab:script-signature-observations}). The \emph{oID} allows to later reconstruct the order in which different scripts were encountered. Similarly, each observation gets assigned a page-based \emph{global ID} (gID) and \emph{script ID} (sID), representing the observation's index in relation to the whole page or the originating script (e.g. Table \ref{tab:script-signature-observations}). Explicitly saving those indexes ensures that the observation order can be reconstructed to avoid analysis errors due JavaScript object key reordering \cite{sc_js_order}. The observation is then added to the internal data structures, which are updated to represent the script's new signature, score and list of unique feature groups.

This approach allows collecting as much information about a website's fingerprint activity on per script basis. A part of this information is displayed in \FPMON's user-facing popup. Furthermore, we store the observations in \gls{json} format to benefit from its higher flexibility in comparison to \gls{csv}.

\begin{figure}[b]
	\centering
	\includegraphics[width=0.6\textwidth]{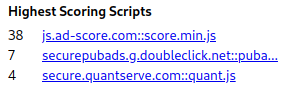}
	\caption{Additional ``Highest Scoring Scripts'' section in \FPMON's popup facilitates manual analysis.}
	\label{fig:fpmon-show-scripts}
\end{figure}

\subsubsection{Facilitating Manual Analysis}
The \FPMON~popup was extended with an additional section to display the most active scripts and their score as Figure \ref{fig:fpmon-show-scripts} shows. The link's content is shortened to \texttt{<domain>::<filename>} for readability but it links to the full \gls{url}. Overall, this feature facilitates manual analysis and helps to empower users against browser fingerprinting by quickly identifying a particularly active script.


\subsection{Developing The \FPNET~Scanner}\label{sec:implementation:fpnet}

The \FPNET~scanner is a tool based on \FPMON~and \FPCRAWL~to analyze a large set of websites for fingerprinting networks in an automated way. This chapter provides the reader with key implementation details about \FPNET~and its internal components.
\subsubsection{Components Overview}
\begin{figure}[h]
	\centering
	\includegraphics[width=0.92\textwidth]{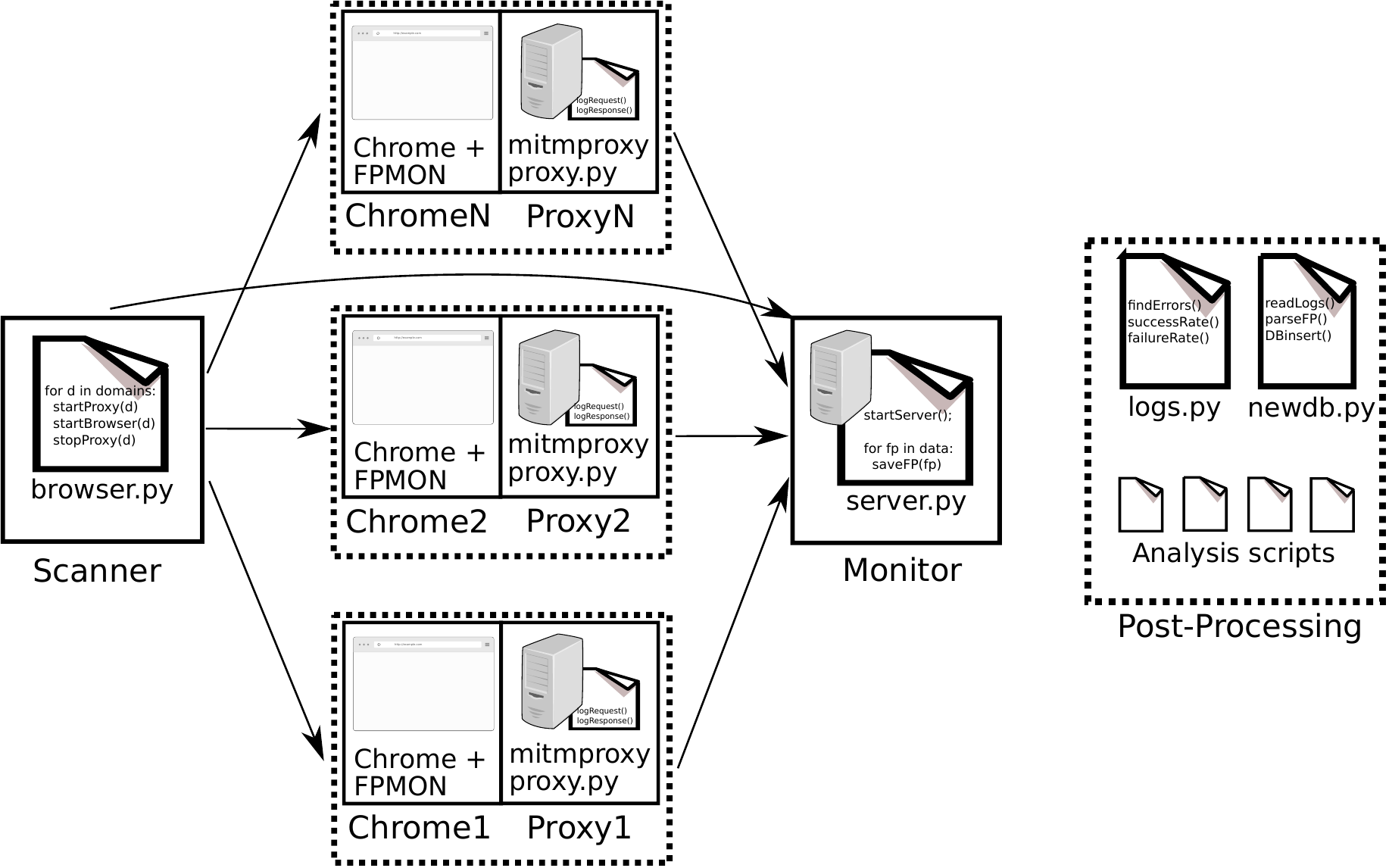}
	\caption{A high-level overview of \FPNET and its components.}
	\label{fig:fpmon-components}
\end{figure}

\FPNET~consists of several components that work together as illustrated in Figure \ref{fig:fpmon-components}. A solid border depicts a component that stands for itself, and a dotted line groups (sub-)components that are closely related. The arrows show how components interact. Post-processing is disconnected from the other components because it only processes data previously collected, but no interaction takes place otherwise.
Each component's purpose will be outlined in the following paragraphs. 
\paragraph{Scanner Component}
The scanner's main script is \texttt{browser.py}, which orchestrates one or more browser \& proxy pairs and manages their environment. It reads a configuration file to determine several scanning parameters, e.g. timeouts, parallelism, or the set of domains to scan. Depending on the configuration, the proxy is disabled or temporary files kept for debugging purposes. The component replaces the unreliable selenium hub by managing all browsers directly. Furthermore, it connects to the monitor component to obtain and send debug information.
\newpage
\paragraph{Browser \& Proxy Component}
This component consists of a headless Chrome browser and an intercepting proxy. For each of the $N$ \texttt{WORKER\_COUNT} threads in the scanner, there is one browser and proxy pair. The headless browser is configured to closely imitate a real one in order to make scanning as reliable as possible. It loads the \FPMON~extension on startup and analyzes the website for fingerprinting activity before sending the results to the monitor. In the end, the browser is gracefully closed or forcefully terminated if it exceeds the defined timeouts. Depending on the \texttt{ENABLE\_PROXY} setting, the browsers use an intercepting proxy. The mitmproxy \cite{mitmproxy} loads the \texttt{proxy.py} module, which captures its browser's \gls{http} traffic and eventually writes it to a compressed \gls{json} file for later analysis.


\paragraph{Monitor Component}
The monitor component consists of a python-based web server, whose main task is to collect the fingerprinting information. It retrieves the data \gls{json} encoded and saves it line-wise in a log file. Furthermore, it supplements \FPNET's debugging features. For one, its \texttt{get\_trace} action returns the unique \texttt{trace\_id} for a domain. Second, it logs the scanner's finish and failure requests. These debug logs later help to determine how many domains were successfully scanned or where errors could have occurred. 
\paragraph{Post-processing Component}
The last component is the set of post-processing scripts. Although those do not interact with the other components during a scan, they are still needed to evaluate the data and uncover fingerprinting networks or actors. The \texttt{logs.py} analyzes \FPNET's log files and debug output to provide information about the scan's failure and success rate. In a second step, the \texttt{newdb.py} script parses the collected fingerprinting information and populates a SQL database. Each analysis script can then query the database for information it requires.

\paragraph{Clarifying The Scanning Process}
For a better understanding how \FPNET~components work together, this paragraph describes individual steps.
{
	\setlength{\itemsep}{0pt}
	\setlength{\parskip}{0pt}
	\setlength{\parsep}{0pt}     
\begin{enumerate}
	\setlength{\itemsep}{0pt}
	\setlength{\parskip}{0pt}
	\setlength{\parsep}{0pt}     
	\item First, the monitor starts a \gls{ssl} enabled web server.
	\item Next, the Chrome \& proxy component starts. The Chrome sub-component starts a service that waits for instructions to handle the browser. The proxy sub-component starts mitmproxy with a custom module.
	\item Now, the scanner component starts and imports the scanning configuration. Separate worker threads are instantiated, and the configured list of domains is read into a queue.
	\item Each worker thread obtains a domain from the queue and processes it:
	\begin{enumerate}[label=\theenumi.\arabic*]
		\setlength{\itemsep}{0pt}
		\setlength{\parskip}{0pt}
		\setlength{\parsep}{0pt}  
		\item A unique \texttt{trace\_id} is requested from the monitor component. It is based on the examined domain and serves tracing and debugging purposes. The monitor keeps a mapping of domains to \texttt{trace\_id}s.
		\item Using the \texttt{trace\_id}, a temporary directory is created, and the \FPMON~extension is copied to it. This unique directory will later contain the browser's profile data.
		\item The extension is edited to include the \texttt{trace\_id} and the domain in specific variables, inter alia, for debugging purposes.
		\item The browser's configuration options are defined. It includes setting the browser's binary to the custom startup script and several command-line arguments, e.g., to start the browser maximized. If the scanning configuration defines \texttt{ENABLE\_PROXY}, the browser is configured to use its accompanying proxy sub-component.
		\item A control connection to the Chrome component is established, and the page load and element access timeouts are configured. The headless browser is launched using the custom startup script. In addition to starting the browser, the script sets a hard timeout that will kill the browser process to prevent browser hangs.
		\item If the scanning configuration enables proxies, the worker now signals the proxy component to begin capturing traffic. The proxy resets itself and begins to save all requests and their responses in memory.
		\item The Chrome component instructs the browser to open the domain. The \FPMON~extension analyzes the website and transmits the fingerprinting information to the monitor component, which saves the data in a central log file.
		\item The worker thread waits a configured amount of time for the extension to successfully finish. If so, the process is considered a success. Otherwise, a failure. This status is sent to the monitor component to indicate that the process has finished. 
		\item The proxy is signaled to stop capturing the network traffic. Upon receiving this signal, the proxy stops recording and saves the recorded data into a compressed \gls{json} file.
		\item The worker instructs the Chrome component to close the browser. Depending on the scanning configuration, it also removes the temporary directory.
	\end{enumerate}
	\item This process is repeated for all domains in the queue.
	\item Once the scanning queue is drained, the automated scanning process ends. The post-processing steps are manual. The two most important steps are the following:
	\begin{enumerate}[label=\theenumi.\arabic*]
		\setlength{\itemsep}{0pt}
		\setlength{\parskip}{0pt}
		\setlength{\parsep}{0pt}  
		\item First, \texttt{logs.py} is used to analyze \FPNET's debug output to determine if any unknown scanning errors occurred and to show the number of successfully analyzed websites.
		\item Second, \texttt{newdb.py} parses the line-based, \gls{json} encoded log entries that \FPNET's monitor created, and inserts the fingerprinting information into a \gls{sql} database.
		\item With the data in the database, each analysis script can access it and query the data it needs.
	\end{enumerate}
\end{enumerate}
}
\subsubsection{Considering Scalability, Stability \& Reliability}
The \FPNET~scanner system was implemented with a focus on scalability, stability, and reliability. This section will highlight key aspects of those efforts.
\paragraph{Taking Advantage of Containerization}
At the time of writing, Docker \cite{docker} is one containerization solution with a wide range available containers, including selenium-based headless Chrome browsers \cite{selenium_chrome}. For the other components, custom containers were built. By containerizing the components, \FPNET~can be used on any system that supports Docker. The only requirement for a host system is providing a shared directory for all containers, which is \texttt{/tmp/chrome/} by default. Furthermore, a \texttt{docker-compose.yml} \cite{docker_compose} file is provided, which facilitates the container orchestration. This containerization concept enables reproducibility and scalability. For the former, because it simplifies the setup, and for the latter because more Chrome \& proxy containers can be started on demand. The special startup script \texttt{start-tmux.sh} starts the required containers in the right order and redirects their output to individual log files.

\paragraph{Replacing The Selenium Hub}
Selenium was reported to be unstable and cause frequent crashes in previous work \cite{Englehardt2016,AlFannah2018}. A core piece of the selenium environment is the \emph{selenium hub}, a central coordinator for all connected selenium-based headless browsers. Similar to previous work, stability issues with selenium occurred. Therefore, the selenium hub was replaced entirely by our scanner component. All selenium Chrome containers offer the same \gls{api} to control the browser as the selenium hub as Listing \ref{lst:implementation:chromecontrol} shows. Therefore, the scanner component implemented the task management and the direct communication to the browser instances itself. With this change, the previously experienced problems disappeared, and no restarts were needed to achieve a 
\begin{lstlisting}[caption={Direct control of the Chrome browser without a central selenium hub.},label={lst:implementation:chromecontrol},language={Python}]
def get_driver(capabilities, chrome_id):
	return webdriver.Remote(command_executor="http://chrome{}:5555/wd/hub".format(chrome_id), desired_capabilities=capabilities)
\end{lstlisting}

\paragraph{Enhancing The Browser Configuration}
Another goal was to make the headless Chrome browsers match their non-headless counterparts as closely as possible to give fingerprinting scripts a lower chance of detecting \FPMON. For that, we changed the following virtual screen properties, which match the author's monitor:
\begin{smallitemize}
	\item Screen width: 1,920px
	\item Screen height: 1,080px
	\item Screen density: 96dpi
	\item Color depth: 24 bit
\end{smallitemize}
Additionally, the browser was configured to start maximized to simulate full-screen use. To supervise the scanning process and facilitate debugging, the \gls{vnc} feature was enabled and the password unset. In order to prevent browser hangs, a hard timeout was realized using a custom startup script (see Listing \ref{lst:implementation:chromestartup}). It starts the Chrome process and kills it after exceeding 90 seconds. The timeout value is the sum of the scanning configuration's \texttt{PAGELOAD\_TIMEOUT} and \texttt{IMPLICITWAIT\_TIMEOUT}. Additionally, it kills the Chrome process when it receives an error or exit signal. Additional command-line arguments for the browser are \texttt{--disable-web-security} and \texttt{--ignore-certificate-errors}, allowing the \FPMON~to send fingerprinting information to the monitor, despite different security contexts. These options, however, increase the browser's attack surface. We minimize the risk of interference or exploitation by running each browser in a separate container, and using a new profile (\texttt{--user-data-dir=/tmp/chrome/<trace\_id>}) for each scanned domain. Since Chrome ignores superfluous arguments, we add \texttt{--domain=<domain>}, \texttt{--traceid=<trace\_id>} and \texttt{--container=chrome<id>} to check for unexpected errors or crashes in the post-processing step.
\begin{lstlisting}[caption={A custom startup script that force-stops the browser process after a defined timeout.},label={lst:implementation:chromestartup},language={Bash}]
	#!/bin/bash
	trap cleanup err exit
	function cleanup() {
		pkill chrome
	}
	timeout -s SIGKILL 90s /opt/google/chrome/chrome "${@:2}"
\end{lstlisting}
\paragraph{Impact on Performance}
Although scanning speed was out of this work's scope due to the focus on accuracy, scalability, stability, and reliability, the following design decisions influenced \FPNET's performance. The \texttt{PAGELOAD\_TIMEOUT} and \texttt{IMPLICITWAIT\_TIMEOUT} were increased to 45 seconds to give fingerprinting scripts enough time to exhibit their behavior. Since \FPMON~sends the data after 20 to 30 seconds, this gives the page 15 to 20 seconds to fully load. The timeouts were chosen arbitrarily but aimed for an accuracy and speed trade-off. With the previously described browser-hang timeout, one analysis finishes after at most 90 seconds. If an error occurs earlier, the scanning component waits for the hard timeout before proceeding. In theory, one selenium-based Chrome instance could open multiple browser tabs to maximize parallelism. Unfortunately, reliability issues occurred, and therefore the scanning process was limited to one tab and domain per Chrome instance. However, the number of Chrome and proxy components is horizontally scalable. On a host system with 24 cores and 40 GB of memory, a parallelism of 20 was achieved for browser-only scans without over-committing resources. Scanning the Alexa Top 10,000 domains took 6 hours. With 15 parallel Chrome and proxy instances, the scan duration was in the range of 9 hours. In theory, the containers could be spanned across multiple systems, allowing higher parallelism and faster scans.
\subsubsection{Proxy Support for Capturing Network Traffic}
\paragraph{Adding new Software}
The file-based and security header analysis imposed the requirement of saving a scanned website's network traffic. Since no programmatic way could be found to export the information directly from the headless browser, a proxy-based solution was chosen. \texttt{mitmproxy} \cite{mitmproxy} had already been used in previous work \cite{Englehardt2016, Acar2013, Acar2014} for similar purposes and its python-based module system allowed us develop custom interception logic. 

\paragraph{Integration Details}
Since multiple Chrome sub-components run in parallel, the traffic had to be separated.  For simplicity, each Chrome was assigned its own proxy. The containerized setup ensured that the \texttt{i-th} proxy only handled \texttt{i-th} browser's traffic. For that reason, Figure \ref{fig:fpmon-components} groups those two components together. Independent of the \texttt{ENABLE\_PROXY} setting, the Chrome browser loads a default page on startup or sends other requests unrelated to our analysis. A signaling mechanism was implemented to filter out such unwanted requests. The scanner component sends the following two types of signal requests directly to the proxy:
\begin{smallitemize}
	\item \texttt{http://startlogging/<domain>} is sent right before the browser is instructed to open the domain.
	\item \texttt{http://stoplogging/<domain>} is sent after \FPMON~finished or an error occurred.
\end{smallitemize}
The proxy monitors all requests for those signaling keywords and when encountering a \texttt{startlogging/<domain>} hostname, it resets its logging state and begins capturing the traffic. All flow-properties, e.g. such as headers, cookies, parameters, are converted to \gls{json} (see Listing \ref{lst:implementation:requestlogging}). Once the proxy receives the stop signal, it saves the whole \texttt{REQUESTS} object into a \gls{json} formatted and GZIP \cite{gzip} compressed file, whose name contains the initially provided domain name. With this method, we obtain a universal, processable log file with a domain's network traffic, including all incoming and outgoing data.
\begin{lstlisting}[caption={Logging the \gls{http} requests and responses based on their flow IDs.},label={lst:implementation:requestlogging},language={Python}]
if not flow.id in REQUESTS:
	REQUESTS[flow.id] = {
		'request': req2dict(flow.request),
		'response': resp2dict(flow.response)
	}
\end{lstlisting}

\paragraph{Impact on Security}
One issue the proxy approach faces is the adoption of encryption communication on the internet. \gls{ssl}/\gls{tls} aim to prevent capturing the plaintext network traffic between two endpoints (e.g. \gls{mitm}). However, the proxy is between the browser and the server, effectively performing a \gls{mitm} attack. To work around this limitation, the proxy issues self-signed \gls{tls}/\gls{ssl} certificates and the browser is instructed to ignore certificate errors.
\newpage
\subsubsection{Processing The Scan Results}
\FPNET's last step is post-processing. While it potentially consists of different scripts to analyze the collected data, this section puts the spotlight on the log analysis and \gls{sql} database.
\paragraph{Analyzing Scan Logs}
After the \FPNET~finished a scan, a log analysis provides information about the scanning process, e.g. occurred errors or the amount of successfully monitored domains. The \texttt{logs.py} script parses the monitor component's and all browser component's log files. The latter are checked for the keywords indicating that a new browser was successfully launched and closed, while the examined domain is extracted from the \texttt{--domain} command-line argument. The resulting list of domains is then compared to the monitor component's log entries, which reveals whether tasks issued by the scanner component correctly reached the Chrome and proxy components. Analyzing the monitor's log file, one learns about:
\begin{smallitemize}
	\item For how many domains \FPMON~successfully transmitted fingerprinting data.
	\item How many domains the scanner component processed successfully.
	\item How many domains failed and what errors occurred, including timeouts, \FPMON~errors, selenium or browser exceptions, and domain differences between \FPNET~components.
\end{smallitemize}
Log analysis proved to be crucial while developing \FPNET~and making it resilient against various errors. For example, in a few cases, the browser or selenium would crash after \FPMON~transmitted the fingerprinting information. In such cases, the error is acceptable, because the monitor successfully recorded the fingerprinting information beforehand.

\paragraph{Storing Data in a SQL Database}\label{sec:impl:fpnet:database}
For the reasons outlined in the previous chapter, the fingerprinting information was parsed from the monitor's log file and transferred into a \gls{sql} database by the \texttt{newdb.py} script. The script uses SQLAlchemy \cite{sqlalchemy} for the following reasons:
\begin{smallitemize}
	\item As an \gls{orm}, SQLAlchemy abstracts away from the \gls{sql} language, allowing to programmatically create, update, modify or query the database and to use the data as objects, facilitating the creation of analysis scripts.
	\item SQLAlchemy supports multiple database backends. Depending on the research environment, the backend can be adjusted to suit a scientist's needs. This thesis defaulted to file-based \texttt{SQLite3} databases for simplicity. 
\end{smallitemize}
Figure \ref{fig:db-schema} (see appendix) represents the database's \gls{erd}. There are four main entities:
\begin{smallitemize}
	\item \texttt{fpcategories} - Represents the 40 feature groups listed in section \qnameref{sec:appendix:featuregroups-ratings}.
	\item \texttt{fpscores} - Consists of the information about a scanned domain and its page-based fingerprinting score.
	\item \texttt{fpscriptorigins} - Contains information about the script origins, i.e. the URLs and script-based fingerprinting scores.
	\item \texttt{fpscriptfunctioncalls} - Contains the individual, observed function call information. 
\end{smallitemize}
The additional association tables realize the entity's relationships to reflect the \gls{json}-structure of the collected data. The \texttt{log\_score} entries in the monitor's log file contain the fingerprinting data transmitted by \FPMON. \texttt{newdb.py} parses that \gls{json} data and inserts it into the aforementioned database tables. To limit the amount of data to analyze, the script has a page score filter. By default, only pages with a score $\geq 10$ are processed. The limit is a trade-off between the amount of data to process and the fingerprinting severity, as privacy-invasive fingerprinters will likely have a substantially higher score. In the last step, the script creates multiple indexes on frequently used columns to speed up query operations.

\paragraph{Analyzing the Raw Data with Custom Scripts}
After a scan's data was pre-processed and transformed into a usable format, it is ready to be analyzed. The separate analysis scripts are written in python and query the database to produce the results or intermediate data for further analysis.

\section{Evaluation} \label{sec:evaluation}
This chapter describes the five studies that were conducted and their results. The first study evaluates the modern internet's dependency on JavaScript. Following that, \FPNET~is used to discover fingerprinting scripts on the Alexa Top 10,000 domains to uncover the underlying fingerprinting networks in study 2. Since the analysis could be affected by randomization, study 3 touches on that subject. Study 4 builds fingerprinting networks based on the scripts' source code as a different approach for comparison. Finally, the fingerprinting scripts undergo a brief analysis from a web security standpoint in study 5.

\subsection*{Technical Setup}
If not stated otherwise, all studies were conducted using an identical setup. The chair provided a virtual server with 24 cores @ 2,660 MHz, 40 GB of \gls{ram}, 200 GB disk space, 1 \gls{ip} address @ 1 Gb/s bandwidth and Debian 10 as the operating system. The latest software versions at the time of writing were used:
\begin{smallitemize}
	\item Docker: 19.03.11
	\item Chrome: selenium/standalone-chrome-debug:3.141.59
	\item Proxy: Mitmproxy 5.1.1
	\item Python inside containers: 3.8.3
	\item Python for analysis scripts: 3.7.3
\end{smallitemize}
By default, the following \FPNET~scanning parameters were set:
\begin{smallitemize}
	\item Domain list: Alexa Top 10,000
	\item Parallelism: 20
	\item Timeouts: 45 seconds page load and implicit wait, 90 seconds browser hangup, 20 seconds \FPMON~analysis timeouts.
	\item Removal of Chrome profiles: Enabled
	\item Proxy: Disabled
\end{smallitemize}
\subsection{Study 1: The Internet's Dependency on JavaScript} \label{sec:evaluation-study-dependency-js}

As explained in section \qnameref{sec:design:javascript}, the basis for active fingerprinting is JavaScript, which could be effectively prevented by disabling JavaScript in the browser. Therefore, this study examines JavaScript's relevance for modern websites since it can be used to build interactive and user-friendly web applications with the goal to improve the user experience. Thus, disabling JavaScript might leave such websites unusable or not fully functional.

\subsubsection{Technical Setup}
Since this study was conducted early into this work, it is based on an initial version of \FPNET, which lacked the external monitor or proxy component, and many of the additional resilience improvements, e.g., custom hub, hard browser hang timeouts, and more. Nevertheless, it produced reasonable results because the event counting script is executed after page-load, and it does not manipulate the page or JavaScript calls in contrast to \FPMON. Therefore, the counting script could return the data directly to the scanning component, where it was collected for further analysis. The parallelism was limited to 4 with a page load timeout of 30 seconds and an additional 15 seconds timeout before the counting script was executed. 
\subsubsection{Methodology}
The Alexa Top 10,000 websites were checked for 109 event handlers using the three different collection methods described in section \qnameref{sec:implementation:javascript}. 
The resulting event handler counts were saved on a per-domain basis and later analyzed. The analysis first determined the number of successfully analyzed websites and the number of sites with at least one event defined. All event handler counts are grouped by their type to get an overview of their frequency and the distribution among all websites. 
\subsubsection{Results}
\begin{figure}[h]
	\centering
	\includegraphics[width=0.98\textwidth]{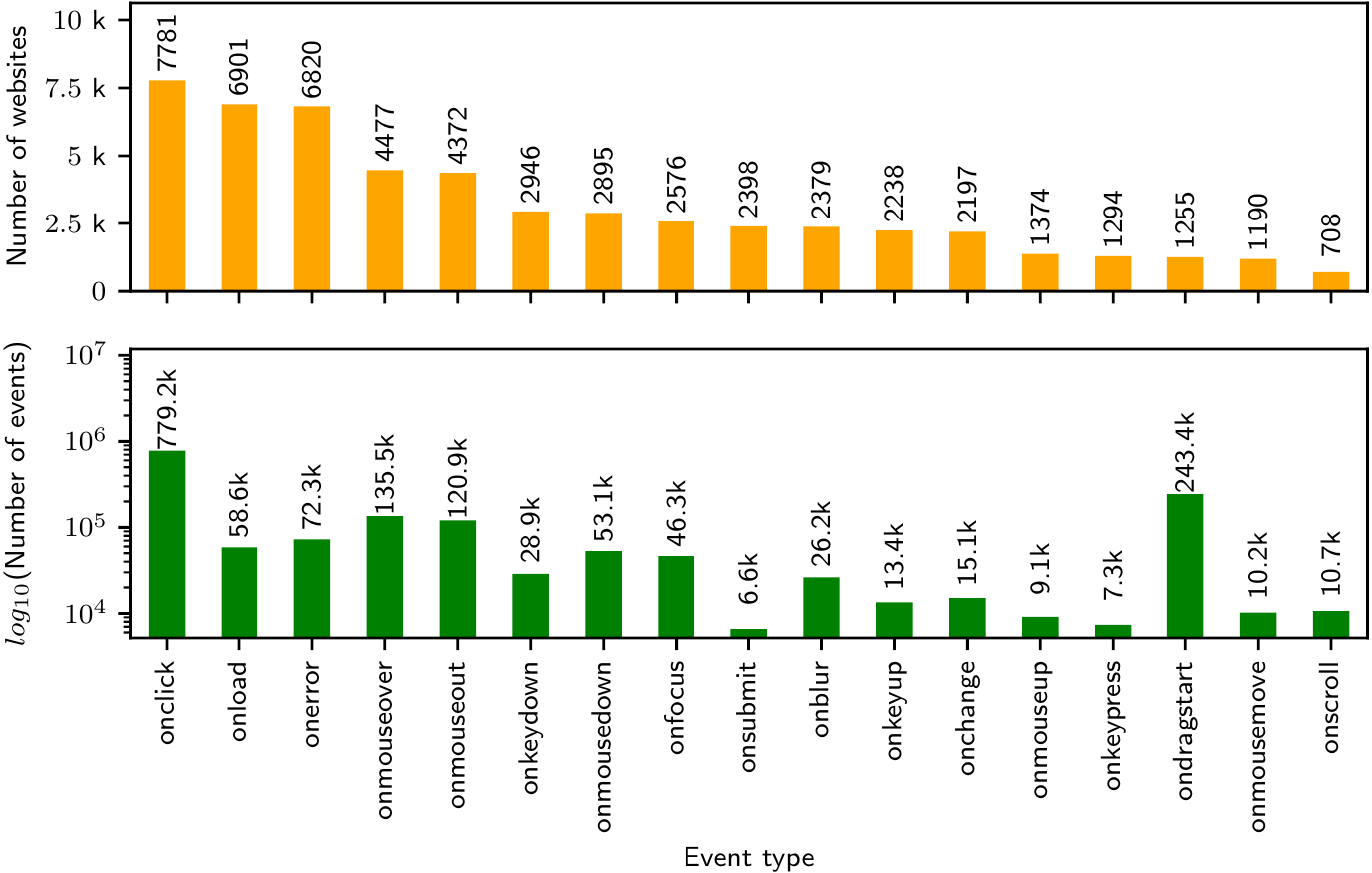}
	\caption{Top 15 event handlers by occurrences and website distribution from an analysis of the Alexa Top 10,000.}
	\label{fig:javascript-event-results}
\end{figure}

From the 10,000 domains, 9,338 were successfully analyzed. 539 pages did not define any event handlers, leaving 87.99\% that make use of at least one event handler. Although our event-based approach poses a lower bound on JavaScript usage, it is only 10\% less than the 96\% reported by \cite{w3techs_js}. All pages define 1,847,445 events in total with the following distribution:
\begin{smallitemize}
	\item \texttt{on-Attribute}: 193,811 events on 5,544 pages
	\item \texttt{on-Property}: 440,256 events on 8,485 pages
	\item \texttt{jQuery}: 1,213,378 events on 5,982 pages
\end{smallitemize}
Slightly less than 50\% of all event handlers are defined using jQuery, although 68\% of pages use that library. With events being a subset of jQuery's functionality, the result is only 15\% lower than the 83\% reported by \cite{jquery_usage} for the whole Alexa Top 1,000,000, which supports the number and acts as a lower estimate. Figure \ref{fig:javascript-event-results} shows the top 15 event handlers by either occurrence or website count. The most commonly occurring and widespread event is \texttt{onclick}. Overall, the top 15 primarily consists of event handlers related to either mouse or keyboard actions. Thus, most events and websites react to the user's input. Since this is not possible when JavaScript is disabled, the usability is reduced on more than 4 out of 5 websites. This makes JavaScript a dependency for modern websites.

\subsection{Study 2: Identifying Fingerprinting Networks using Script Signatures} \label{sec:evaluation-study-script-sigs}
As described in section \qnameref{sec:design:scriptsignatures}, this study applies the behavior-based fingerprinting detection approach to the script-level. The main goal is to explore the landscape of fingerprinting scripts used on the internet and their actors. The idea is to find identical or similar behavior in different scripts across the scanned websites and use the collected data to identify fingerprinting networks. The assumption is that such networks and similar scripts will be traceable back to the actors behind them, thus answering the question about who tracks and collects information about internet users.
\subsubsection{Technical Setup}
With the final and reliable version of \FPNET~the Alexa Top 10,000 was scanned. All domains were requested using \texttt{http://}, but redirects to secure connections were followed. A parallelism of 20 was configured.
\subsubsection{Methodology}
Using \texttt{newdb.py}, the monitored script activity is parsed and stored in the database. As stated in section \qnameref{sec:impl:fpnet:database}, this only processes information from pages with a page score $\geq$ 10 because fingerprinting scripts will likely have a higher activity, and it limits the amount of data that needs to be manually reviewed in later steps. Querying the database, first facts about the collected data can be gathered, such as
\begin{smallitemize}
	\item Number of pages, scripts, or function calls.
	\item Average scores and their standard deviation
	\item Amount of scripts per page and many more.
\end{smallitemize}

In a second step, script-based fingerprinting networks are computed using \texttt{find\_SIGNATURES.py3}. The database is queried for all script information. We check for overlapping script signatures by adapting the algorithm from \cite{kybranz}. When signatures overlap, the scripts have identical behavior to a certain extent. The overlapping signature's length represents the number of identical, consecutive function calls to fingerprinting-related feature groups. Scripts whose signatures overlap form a fingerprinting network. 

The scripts analyzed were filtered with a threshold of at least 6 feature groups to reduce the workload on the following manual analysis and review of the networks and their actors. Again, the reasoning is that more privacy-invasive networks show higher activity and access a wider variety of feature groups. Networks of size $\leq 1$ are omitted since this implies they do not cover more than one website. This, however, makes them irrelevant to finding the largest, most widespread networks and actors. Before saving the fingerprinting networks for later analysis, the list of page domains covered through its scripts is generated for each network. The data provides preliminary results about the fingerprinting networks and their properties, which allows drawing first conclusions about the current state of JavaScript-based, active fingerprinting on the internet.

In the last step, all networks undergo manual review to identify the actor controlling them:
\begin{smallitemize}
	\item The scripts' origin domains are inspected and may lead to an actor.
	\item When no actor can be identified through the script origin domain (e.g., due to \gls{cdn}s or unresolvable domains), a brief search on the internet is performed, including information about the filename and script domain.
\end{smallitemize}
In cases where a network has multiple script domains, the actor's script domain should serve the majority of scripts. This accounts for cases where a page hosts the script themselves instead of including it from a remote domain. If there is no clear majority, no single actor can be assigned to the network, e.g., when multiple pages use an open-source library.

Similar to \cite{Nikiforakis2013}, we try to categorize an actor's domain using TrendMicro's SiteSafety categorization service. 
\newpage
\subsubsection{Results}
\FPNET~successfully scanned 9,180 out of the 10,000 websites without errors and logged 9,322 observations. The difference of 142 stems from failures where \FPMON~did not indicate a successful finish, but an observation was still successfully collected. Overall, the majority of failed observations are from the monitor not receiving an observation request from \FPMON~(507 cases), or the analysis running into a timeout (288 cases). The remaining errors are due to browser hangs which are handled by the custom startup script. From the 9,322 pages, 5,407 are considered for further processing (page score $\geq$ 10), and their data is disseminated into the database. Page scores range from 10 to 38, with an average of 15 and a standard deviation of 5. All pages run 85,618 scripts (42,598 unique script origins) in total, executing 9,419,075 fingerprinting-related function calls. A page has almost 16 script origins on average with a standard deviation of 8. The scripts' scores range from 0 to 38, averaging 3 feature groups with a standard deviation of 3. A script score of 0 is surprising, but some function calls monitored by \FPMON~are not assigned a feature group (e.g., \texttt{getElementsByName()}). Comparing the page and script scores generally shows lower scores for scripts. Since page scores are calculated based on all its scripts, high page scores indicate multiple scripts running in parallel or single, high-activity scripts on that page.

Using the data, we can define three activity ratings for a script, based on the average and the standard deviation as shown in Table \ref{tab:fp-activity-score}.

\begin{table*}
	\centering
	\footnotesize
	\begin{tabularx}{\textwidth}{llX}
		\toprule
		Activity & Score & Interpretation \\
		\midrule\midrule
		Low & score $<$ 3 & The script is likely benign. \\
		\midrule
		Medium & 3 $\leq$ score $\leq$ 6 & The script exhibits limited fingerprinting activity. \\
		\midrule
		High & score $>$ 6 & The script is considered to deliberately fingerprint the user. \\
		\midrule
		\bottomrule
	\end{tabularx}
	\caption{Categorization of scripts into low, medium, high fingerprinting activity based on their score.}
	\label{tab:fp-activity-score}
\end{table*}

\begin{figure}[t]
	\centering
	\includegraphics[width=\linewidth]{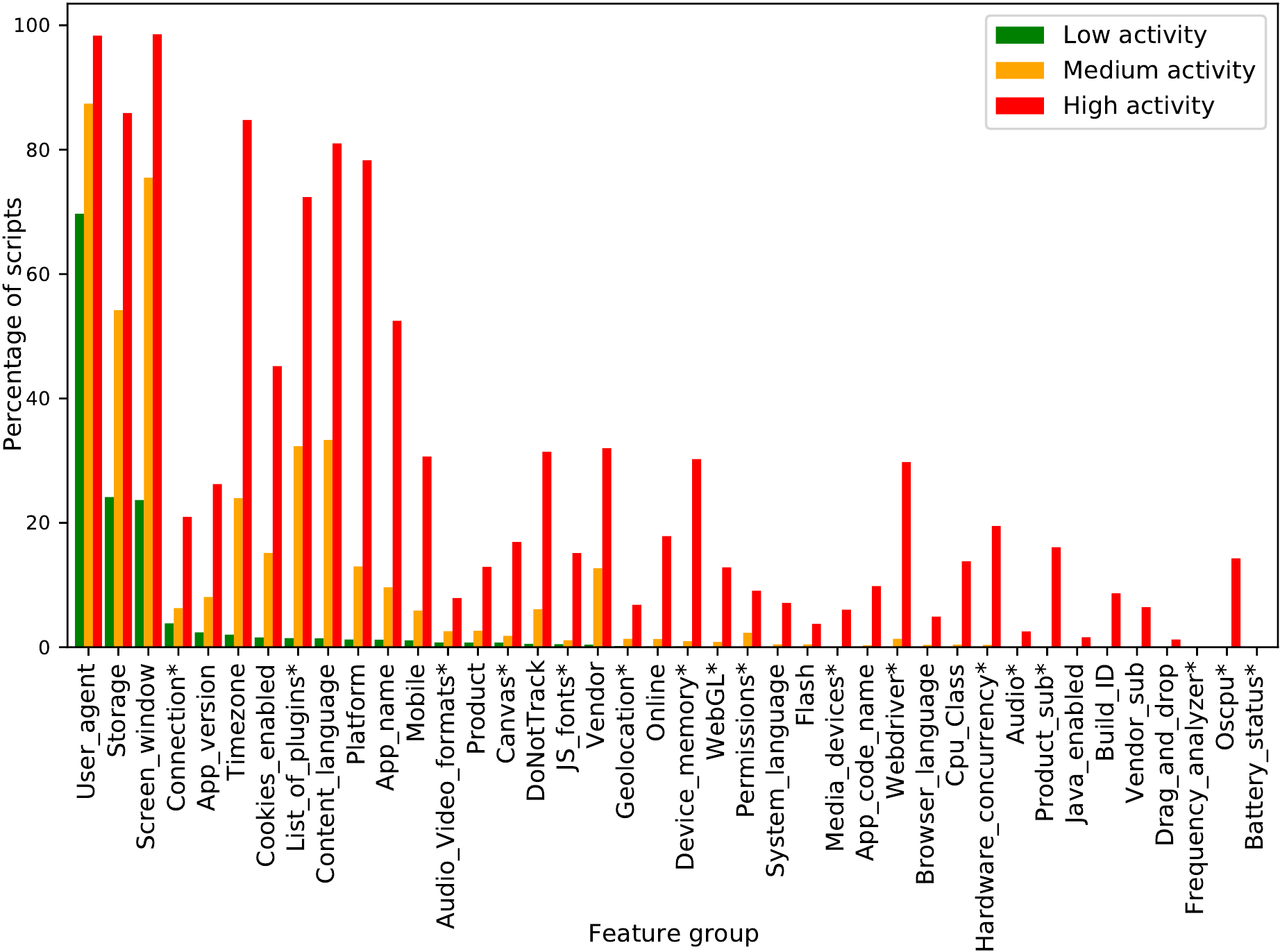}
	\caption{Distribution of feature groups for fingerprinting scripts with low, medium and high activity. * denotes aggressive feature groups.}
	\label{fig:scripts-chart}
\end{figure}

Applying this rating to the 85,618 scripts, we find the majority (59.13\%) to be benign, 32.16\% with medium, and the remaining 8.70\% of scripts with high fingerprinting activity. Figure \ref{fig:scripts-chart} illustrates the distribution of feature groups among scripts with different activity-levels. Fingerprinting scripts with high activity cover a wide range of fingerprintable feature groups, while benign scripts tend to access few non-aggressive feature groups.

After the second analysis step, 379 fingerprinting networks are formed. Their properties vary greatly, as the examples in Table \ref{tab:sample-network-properties} show. The majority of networks have a broad score distribution. The median is 10, and the 75th percentile is 13, which leaves about 25\% of networks with a score up to 36. Similarly, the network sizes' 50th percentile is 3, and the 75th percentile is just 6.5, with only 9 networks covering more than 100 websites. Two lie over the 84th percentile score-wise, and the remaining ones only have a score of 7 or 8. Therefore, most highly active networks have limited coverage. All networks share common behavior defined by the network's signature. The signature's length is a good indicator of how similar the scripts and their behavior are. It ranges from 11 to 4,703 equal, consecutive feature groups with a median of 64. Another valuable property is the number of script domains. The 70th percentile is 1, meaning that all scripts of a network are loaded from the same domain. On the one hand, it proves that identical or similar scripts are successfully grouped into networks, and on the other, it could facilitate the actor identification, since only a single origin must be analyzed. However, there are three networks with close to or above 100 different script domains. These fingerprinting solutions are likely to rely on self-hosting. The column \emph{Files} in Table \ref{tab:sample-network-properties} refers to the number of unique files whose filename and behavior are identical. For example, all scripts from \texttt{doubleblick.net} have the same \texttt{pubads\_impl} prefix, followed by a changing release date. In addition to the filename, the behavioral suffix, consisting of the script's score and signature length, helps identify scripts or networks that change their behavior. Although the \texttt{E-v1.js} files share the same name, their scores range from 7 to 10 and the signature length from 25 to over 2,000. Some fingerprinting scripts are configurable and, therefore, might not produce the same score or signature. Given that the \texttt{doubleclick.net}-based network spans more than 1,300 pages and almost 400 files, identical behavior is observable on multiple pages nonetheless. On the other hand, maxmind.com's \texttt{device.js} proves that the same 269 function calls of a single fingerprinting script can be re-identified across several pages.
\begin{table*}
	\centering
	\footnotesize
	\begin{tabularx}{\textwidth}{rrrrXXX}
		\toprule
		Score & Size & $\text{Sig}_{len}$ & Files & Typ. Names & Script Domain(s) & Page Domain(s) \\
		\midrule\midrule
		8 & 1,343 & 30 & 371 & pubads[...].js, \dots & doubleclick.net & blackdoctor.org, ebay.com, \dots \\
		\midrule
		26 & 232 & 97 & 230 & 1ad6cd50, 5e4f5e70, \dots & dhl.com, dnb.com, \dots & dhl.com, dnb.com, \dots \\
		\midrule
		36 & 5 & 269 & 1 & device.js & maxmind.com & mediafire.com, \dots \\
		\midrule
		7 & 62 & 25 & 40 & E-v1.js, embed\_shepherd-v1.js, \dots & wistia.com, wistia.net & paychex.com, rochester.edu, privy.com, \dots \\
		\midrule
		\dots & \dots & \dots & \dots & \dots & \dots & \dots \\
		\midrule
		\bottomrule
	\end{tabularx}
	\caption{Example fingerprinting networks with their properties.}
	\label{tab:sample-network-properties}
\end{table*}
\begin{figure}[h]
	\centering
	\includegraphics[width=\textwidth]{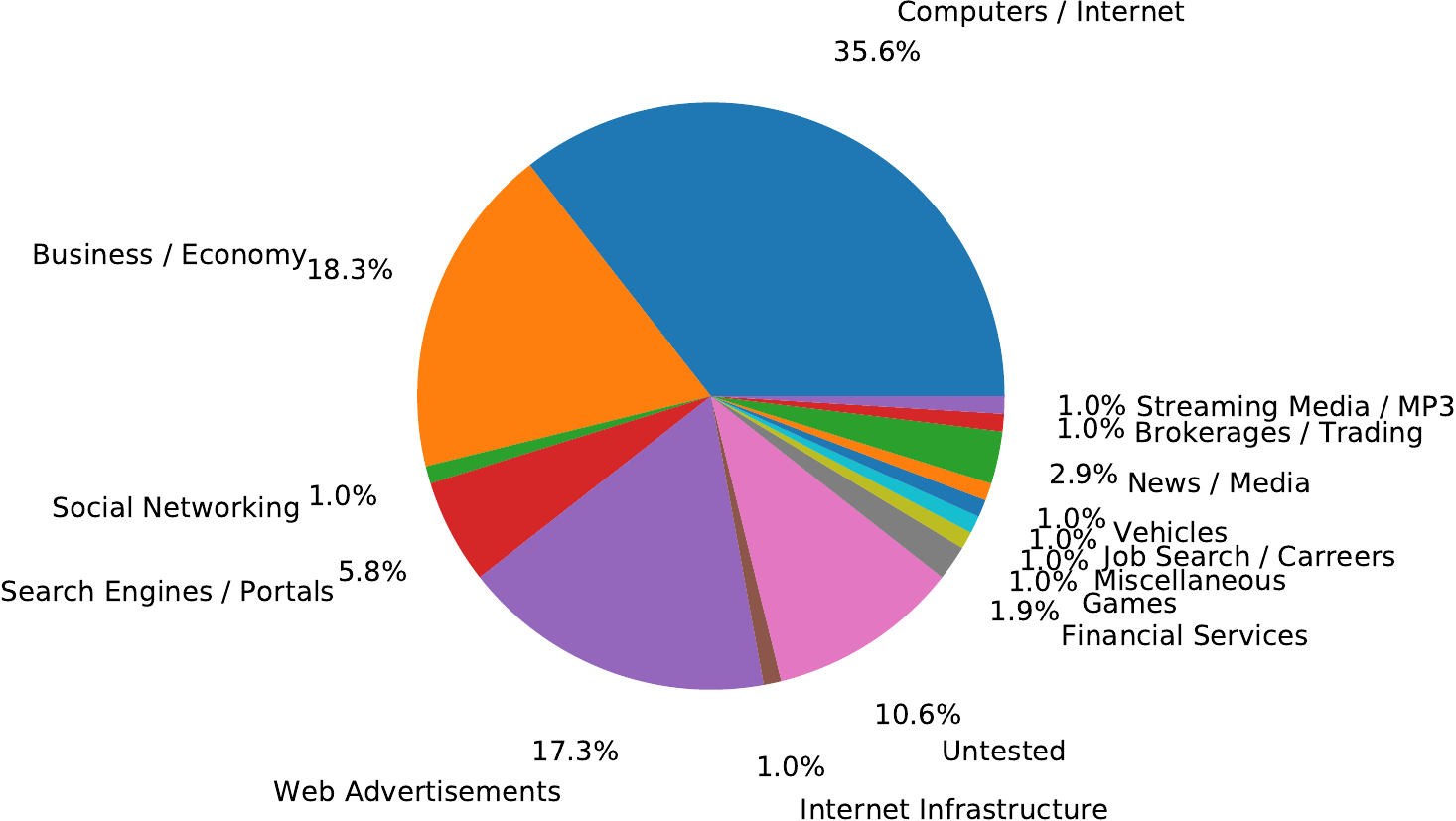}
	\caption{Categorization of the actors' websites according to SiteSafety by TrendMicro}
	\label{fig:actor-categories}
\end{figure}

For 259 out of the 379 networks, we successfully identified an actor. This number excludes 20 rather small networks whose actors operate all network's sites, e.g., a script from \texttt{citi.com} loaded on \texttt{citi.com}, \texttt{citibank.com} and \texttt{citibankonline.com}. In such cases, the actor does not control anything beyond their own pages. For the remaining 100 networks, no single actor could be identified. Of that, 42\% due to being hosted on \gls{cdn}s. For the other 58\%, no clear connection to an entity could be identified. An exception to that is actors who represent a specific technology. Some frameworks that show fingerprinting behavior are open-source and available without cost (i.e FingerprintJS \cite{fingerprintjs_github}, Snowplow \cite{snowplow} or Prebid \cite{prebid_server}. Some businesses offer enhanced versions of these scripts or other related services, e.g., FingerprintJS Pro \cite{fpjs}. The behavior-based approach cannot discern between the open-source or commercial fingerprinting scripts, should the exposed behavior be similar. Therefore, assigning an actor to these networks is difficult since both, commercial and individual actors, are mixed together. To account for the use of these scripts, we merge all these networks into one actor nonetheless.

Before analyzing the resulting landscape of actors, all networks belonging to the same actor are merged. In total, 104 actors are identified. Figure \ref{fig:actor-categories} shows their categorization using TrendMicro's SiteSafety tool. \emph{Untested} represents the uncategorizable technology-based networks or not yet classified domains. Although over a third falls into the \emph{Computers / Internet} category, the other two major categories are \emph{Business / Economy} and \emph{Web Advertisements}. Sectors that usually require a high security standard or user identification are represented as well: \emph{Financial Services}, \emph{Brokerages / Trading}, and \emph{Social Networking}. However, the categorization seems quite broad, as Google Adsense or Yandex Adfox fall into \emph{Search Engines / Portals} instead of the more suitable \emph{Web Advertisements} category. Similarly, Akamai's Bot detection is categorized as \emph{Computers / Internet}.
\begin{table*}
	\centering
	\footnotesize
	\begin{tabularx}{\textwidth}{XXrrr}
		\toprule
		Actor & Category & Networks & Score (Aggr.) & Pages \\
		\midrule\midrule
		Google DoubleClick & Web Advertisements & 19 & 10 (2) & 1,583\\
		\midrule
		Google AdSense & Search Engines & 11 & 8 (1) & 544\\
		\midrule
		Yandex Metrica & Search Engines & 52 & 14 (3) & 367\\
		\midrule
		Akamai & Computers & 2 & 28 (10) & 292 \\
		\midrule
		FingerprintJS & No Category & 9 & 20 (10) & 133 \\
		\midrule
		\bottomrule
	\end{tabularx}
	\caption{Top 5 actors ranked by the amount of pages they can fingerprint.}
	\label{tab:top-actors-affected-domains}
\end{table*}
Depending on the metric, the top-ranking actors change slightly. Table \ref{tab:top-actors-affected-domains} shows the five biggest actors in terms of website coverage. Combining the two Google services, this actor can be found on over 20\% of the scanned websites. In comparison to the other top actors, Google's scripts are not aggressive. Akamai and FingerprintJS use 10 aggressive and more sensitive feature groups to fingerprint browsers on over 4\% of the pages.

Although great coverage is essential to fingerprint as many users as possible, more feature groups provide higher accuracy. The most active fingerprinting scripts come from a different set of actors. Table \ref{tab:top-actors-score} reveals that Maxmind operates one network covering 5 pages on which browsers are fingerprinted with 36 out of 40 possible feature groups. The three following networks span more websites while still accessing more than 50\% of the aggressive feature groups, which were identified to be mainly used for fingerprinting purposes. Our data shows that 25 actors run scripts with a score of 16 or more and at least 3 aggressive feature groups on over 7\% of the Alexa Top 10,000 websites. This includes single network actors such as hCaptcha, Sift, DataDome, and others.

\begin{table*}
	\centering
	\footnotesize
	\begin{tabularx}{\textwidth}{XXrrr}
		\toprule
		Actor & Category & Networks & Score (Aggr.) & Pages \\
		\midrule\midrule
		Maxmind & Computers & 1 & 36 (14) & 5\\
		\midrule
		Moat & Web Advertisements & 5 & 34 (12) & 114\\
		\midrule
		Adsco.re & Web Advertisements & 2 & 29 (10) & 19\\
		\midrule
		Akamai & Computers & 2 & 28 (10) & 292 \\
		\midrule
		ShieldSquare (Perfdrive) & Vehicles & 1 & 23 (6) & 11 \\
		\midrule
		\bottomrule
	\end{tabularx}
	\caption{Top 5 actors covering at least 5 pages ranked by the score.}
	\label{tab:top-actors-score}
\end{table*}

\subsection{Study 3: Examining Scripts for Randomization} \label{sec:evaluation-study-hide-n-seek}
The behavior-based approach requires that script signatures do not significantly change across websites or over time. This study analyzed the data from two consecutive \FPNET~scans to find scripts that try to evade detection by randomization or obfuscation of their fingerprinting signatures or script origins. Furthermore, this could provide insights into fingerprinting scripts' defensive techniques.
\subsubsection{Technical Setup}
For this study, the default setup was used. However, due to the SecT chair requiring additional resources, the VM was slightly down-scaled. Therefore, the parallelism was reduced to 10, which increased the scanning time to over 12 hours. 
\subsubsection{Methodology}
Two consecutive \FPNET~scans were conducted to collect the data for the randomization analysis. Due to the longer scanning time, the two scans were launched 8 hours apart. For both scans, separate databases were created. 

The analysis script \texttt{randomization.py} queries both databases for the page domains and its scripts with script domain, filename, script score, and script signature information. The script domain and filename are concatenated with an underscore to form the script origin property.
Since we can only compare pages that were successfully analyzed in both scans, the intersection of both data sets' page domains is formed. For each comparable page domain, the list of scripts was sorted and the following values computed for both data sets:
\begin{smallitemize}
	\item Number of scripts
	\item Concatenation of all script origins
	\item List of all scripts' signature length
	\item List of all scripts' scores
	\item Based on the list of scores: Maximum, minimum, average, sum
\end{smallitemize}
Using these properties, a page can be checked for changes between the two scans. We define the following three binary values, which compare the aforementioned properties for both data sets:
\begin{smallitemize}
	\item \texttt{is\_equal}: is\_same\_scripts \textit{and} equal list of signature lengths
	\item \texttt{is\_same\_scripts}: Equal number of scripts \textit{and} equal minimum scores \textit{and} equal maximum scores \textit{and} equal sum of scores \textit{and} equal concatenation of script origins
	\item \texttt{is\_same\_score}: Equal number of scripts \textit{and} equal minimum scores \textit{and} equal maximum scores \textit{and} equal list of scores
\end{smallitemize}
If the list of signature lengths, concatenation of script origins, and scores is the same, then \FPMON~monitored the same behavior and the page did not change. \texttt{is\_same\_scripts} can be used to check if script origins changed. A change in a script's score is indicated by \texttt{is\_same\_score}.

\begin{figure}[b]
	\centering
	\begin{tikzpicture}[every tree node/.style={draw,rectangle,align=left},
		level distance=2.25cm,sibling distance=0.2cm,
		edge from parent path={(\tikzparentnode) -- (\tikzchildnode)}]
		\Tree
		[.{Two sets of script origins}
		\edge node[auto=right,pos=.6] {(i) Intersection};
		[.{Identical script origins.\\Comparing script properties}
		\edge node[auto=right,pos=.8] {change};
		[.{Changed behavior} ]
		\edge node[auto=left,pos=.8] {no change};
		[.{Same behavior} ]
		]
		\edge node[auto=left,pos=.6] {(ii) Symm. Difference};
		[.{Distinct script origins.\\Matching based on script signature} 
		\edge node[auto=right,pos=.8] {match};
		[.{Changed origin} ]
		\edge node[auto=left,pos=.8] {no match};
		[.{Changed behavior} ]
		]
		];
	\end{tikzpicture}
	\caption{Steps taken to detect changes in behavior or origins.}
	\label{fig:randomization-steps}
\end{figure}
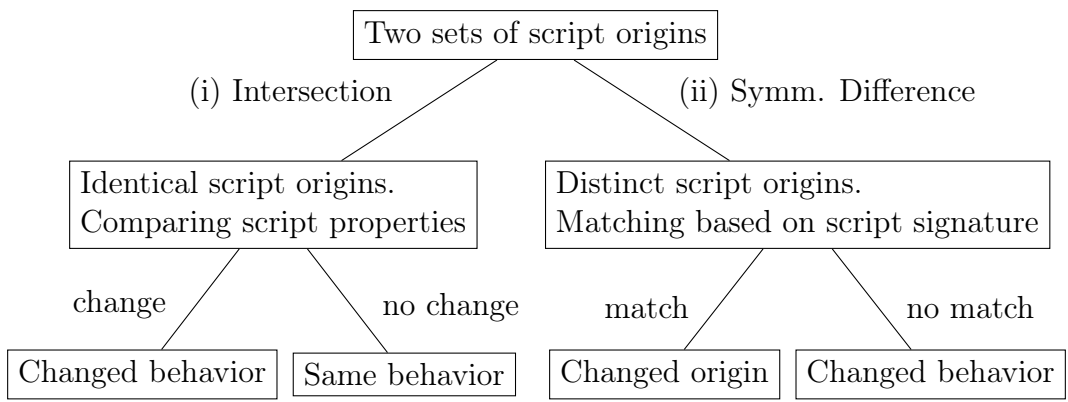

After comparing the pages, we go into more detail and compare each page's script origins. For this comparison, we distinguish scripts whose origins match or do not match as depicted in Figure \ref{fig:randomization-steps}. The former is an intersection, and the latter a symmetric difference of both sets.
\begin{enumerate}[label=(\roman*)]
	\setlength{\itemsep}{0pt}
	\setlength{\parskip}{0pt}
	\setlength{\parsep}{0pt}  
	\item For equal script origins, we analyze the script's properties, e.g., signature, score.
	\item For origins that are neither in both sets, we try to match them based on their signature. All such scripts are compared pair-wise, excluding identity or symmetric pairs. For each pair, the scripts' respective signature is obtained from the data sets, and the script domain and filename are derived from the script origin. The signatures are compared, and if equal, the following three cases arise:
	\begin{smallitemize}
		\item The filenames match, but not the script domains
		\item The filenames do not match, but the script domains match
		\item The neither the filenames nor the script domains match
	\end{smallitemize}
	The fourth case is already covered by the script origin' identity comparison since the filename and script domain are derived from it.
\end{enumerate}
This distinction allows us to draw conclusions about scripts that try to evade the behavior-based analysis by changing their signature, captured by case (i). Case (ii) will prove the behavior-based approach's advantages over a strict \gls{url} based analysis since it reveals cases where the script domain or filename change, but not the signature. Overall, this should answer the initial question of whether randomization or obfuscation techniques occur.

\begin{figure}[h]
	\centering

	\begin{tikzpicture}[every tree node/.style={draw,rectangle,align=left},
		level distance=2.25cm,sibling distance=1cm,
		edge from parent path={(\tikzparentnode) -- (\tikzchildnode)}]
		\Tree
		[.{ssdeep score $\geq$ 95\%,\\equal feature groups,\\$<$ 10\% length difference}
			\edge node[auto=right,pos=.6] {match};
			[.{String-similarity algorithm}
				\edge node[auto=right,pos=.8] {match};
				[.{Correct} ]
				\edge node[auto=left,pos=.8] {no match};
				[.{False-negative} ]
			]
			\edge node[auto=left,pos=.6] {no match};
			[.{String-similarity algorithm} 
				\edge node[auto=right,pos=.8] {match};
				[.{False-positive} ]
				\edge node[auto=left,pos=.8] {no match};
				[.{Correct} ]
			]
		]
	\end{tikzpicture}
	\caption{Decision tree for the string-similarity algorithm comparison.}
\label{fig:stringsim-matching}
\end{figure}
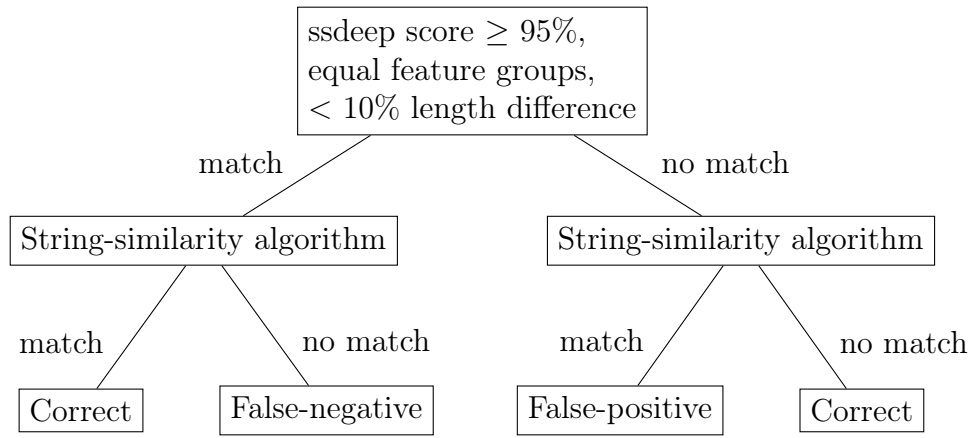

In an attempt to combat signature changes, we explore various string similarity algorithms \cite{strsimpy} to match similar script signatures. We first extend the network building script with ssdeep \cite{Kornblum2006} to find signatures that are more than 95\% similar, have equal feature groups, and at most a 10\% length difference. Using the resulting intermediate data set of 22,166 unique signature match pairs, we evaluate the string similarity algorithms. After identifying the best performing candidate in terms of false-negatives and false-positives (see Figure \ref{fig:stringsim-matching}), we define more criteria to narrow down the error further and to retrieve optimal results after manual review. More matching script signatures provide a higher potential for larger fingerprinting networks.

\subsubsection{Results}
The consecutive scans finished with only 147 page domains not in both data sets. That leaves 5,292 pages for further analysis. Using the initial page comparison, we discover the following numbers:
\begin{smallitemize}
	\item 4,637 pages do not have the same script signatures (not \texttt{is\_equal}).
	\item 3,998 pages do not have the same script origins (not \texttt{is\_same\_scripts}).
	\item 1,823 pages do not have the same script scores (not \texttt{is\_same\_score}).
\end{smallitemize}
Reversely, only 655 pages have exactly the same script origins and same signature lengths. However, as the script origins number suggests, one plausible cause is a change to either the origin's domain or the filename. Since the latter contains all \gls{url} query parameters, subtle differences cause that check to fail. For example, cache busters (e.g., \texttt{?v=<timestamp>}) can be used to trick a browser into reloading a static resource. The third classification excludes the prone script origins from its comparison, thus reporting more than 65\% of pages with exactly the same script scores. Nonetheless, this indicates that the pages' scripts origins frequently change, reinforcing the need to analyze the scripts and their behavior on a script-level.

In total, there are 70,973 scripts whose origins are identical across the two scans. Over 86\% of those have the exact same script score and script signature, thus are re-identifiable using \FPMON. The remaining 9,298 scripts have a different script signature. However, for 91.83\% of the remainder, the script scores stay the same, and except for 2 scripts, the feature groups, too. When the monitored feature groups are equal, the changed script signature indicates that \FPMON~recorded function calls in a slightly diverging order. Possible causes are asynchronous (e.g., \texttt{async}) or timed function calls (e.g. \texttt{setTimeout} or \texttt{setInterval}) executing at varying times, or the scripts deliberately changing the property access patterns. Unfortunately, without a code review and full understanding of a script's functionality, no decisive answer can be given. Despite this, there are 760 scripts where not only the signature but the feature groups change. Analyzing the differences reveals an average of 1.69 feature groups and a standard deviation of 1.55, with the maximum difference being 16 feature groups for certain scripts. Fingerprinting scripts that change the feature groups and script signature are believed to employ anti-detection techniques to some extent. For example, a fingerprinting script could collect all feature groups on the first visit but access fewer feature groups on consecutive page visits while still having enough information to identify the browser uniquely. Again, reviewing and completely analyzing these scripts could bring clarity, but this is left for future work.

When the script origins do not match, our behavior-based approach could still identify the same fingerprinting script using a script signature match, which is not unlikely as the previous analysis suggests. In fact, we re-identify 8,984 scripts where the filename changes, but the script domain and signature are the same. This includes query parameters, e.g. \texttt{?v=482158} vs. \texttt{?v=482181}, but also \gls{url}s that are potentially hard to track on allow-lists as well, e.g. \texttt{ea.com\_710de4e3} vs \texttt{ea.com\_710de559}. Similarly, there are 12 scripts whose script domains change but the signatures and filenames are identical. A prominent example is \texttt{xqheb9yszyrd.com} changing to \texttt{vk77lnizckm6.com}. Last but not least, 817 scripts have a different filename and script domain. However, most of the latter have a script score and signature length of 1, which could accidentally show the same behavior, but not be related at all. Removing these potential false-positives, there are 3 scripts with a score above 3 and significantly longer signatures that can be matched. For example \texttt{missguided.com\_mssgddsdstl.js}, having a score of 25, changes to \texttt{missguided.co.uk\_jywraijzsxptbytq.js} with a different \gls{tld} and randomized filename. The results demonstrate the strengths of a behavior-based fingerprinting detection approach in contrast to just relying on the script URL to identify them. According to \cite{kybranz}, most privacy extensions rely on the latter method to block script execution.

To answer the research questions set in section \qnameref{sec:design:randomization}, the results show that some scripts use filename, script domain, or signature changes as defensive techniques to evade being detected as fingerprinting scripts. We have seen that randomization and obfuscation are quite common as only 12.38\% of pages have absolutely no changes to their scripts.

Using the intermediate matching pairs data set created using ssdeep, four similarity algorithms are identified as being suitable:
\begin{smallitemize}
	\item Cosine similarity
	\item Jaccard index
	\item Sorensen-Dice coefficient
	\item Overlap coefficient
\end{smallitemize}
These do not produce any false-negatives, that is, not finding the similarity in two signatures that match according to the substring analysis and ssdeep (see Figure \ref{fig:stringsim-matching}). Cosine similarity has the highest false-positive rate, that is, finding the similarity in two signatures that are not matched by ssdeep. These false-positives are potential candidates to enlarge existing networks.
In order to increase the accuracy and facilitate the manual review, we define the following criteria that similar signatures must fulfill:
\begin{smallitemize}
	\item Length over 20
	\item Algorithm's string similarity is above 95\%
	\item Equal signature lengths and at most 1 different feature group
	\item Equal number of feature groups and at most 1 different feature group
	\item At most 5\% of the shorter signature in length difference and at most 1 different feature group
\end{smallitemize}
With this, we find up to 250 signature matches using Cosine similarity ($N=40$), which were manually verified and indeed considered to be similar. Comparing the fingerprinting networks for scripts with a score $\geq$ 25, these additional matches enlarge at least two fingerprinting networks. Depending on one's accuracy requirements, the above criteria could be adjusted. Anyhow, the string similarity analysis shows that signature randomization does not evade a behavior-based analysis, at least to some degree. 

\subsection{Study 4: Comparing Script- and File-Signature Networks} \label{sec:evaluation-study-file-matching}
The fourth study explores the relationship between a script's signature and content. According to \cite{kybranz}, privacy extensions are the most effective against fingerprinting scripts that are on an extension's deny-list. However, as we noticed in the previous study, some fingerprinting scripts change their script domain or filename to evade such deny-lists. Using a signature-based detection does not seem feasible since the signature can only be calculated during run time. This limitation prevents detecting and blocking the script execution preemptively. Therefore, in an attempt to strengthen a user's privacy, this study tries to build file-based fingerprinting networks, whose files could be compared and blocked through the returning \gls{http} request.
\subsubsection{Technical Setup}
Analyzing the file contents requires capturing network traffic so that a file's source code can be associated with the script origin. Therefore, this study's setup enabled \texttt{ENABLED\_PROXY} and reduced the overall parallelism down to 15 to accommodate the additional proxy containers. With that configuration, a new \FPNET~scan was launched, which took about 9.5 hours to finish. The extended scanning time likely stems from the proxies processing all \gls{http} requests sequentially. 
\subsubsection{Methodology}
Analogous to the other studies, we generate an SQL database from the \FPNET~logs and query it for the list of page domains, their fingerprinting scripts, and script domains.
The process of building file-based networks consists of several steps, each implemented in a separate script.

At first, \texttt{files\_extractor.py} checks if the proxy successfully captured network traffic for a page domain. Page domains for which no proxy logs exist are not considered for further analysis. For all scripts of a page, the traffic log is searched for a matching \gls{http} request and response. If a match is found, the script's source code is extracted into a file whose filename encodes its properties: \texttt{<page domain>\_<script domain>\_<script score>\_<filename>}. Unfortunately, the underlying file system (ext4) only supports filenames shorter than 256 characters. Thus, longer files were exempt from further processing. Matching the network traffic based on the fingerprinting script's domain and filename limited the analysis to dedicated files because JavaScript contents from \texttt{<script></script>} tags could not reliably be identified and extracted.

The next step is performed by \texttt{files\_matcher\_pooled.py}, which uses a process pool to distribute the following analysis onto 40 \gls{cpu} cores. It reads the contents of all previously extracted script files and computes the ssdeep \cite{Kornblum2006} hash for each one. \emph{ssdeep} is a fuzzy hashing algorithm that weakens a hash algorithm's avalanche effect, thus providing a similarity score for files that are just slightly different. Manual review of the fingerprinting scripts throughout the other studies suggested that some scripts only vary slightly. All resulting ssdeep hashes are compared pairwise while skipping identity and symmetric pairs for performance reasons. When ssdeep reports a similarity $\geq$ 95\% (referred to as \emph{match score}), the two underlying fingerprinting scripts are considered to be highly related and similar. All successful matches are exported to a \gls{csv} file for the next analysis step, including information about both scripts' filename, domain, score, page domains, and match-score. At this point, the data allows us to draw first insights about what kind of files are similar.

In the third step, we use \texttt{files\_analyzer.py} and the previous matches to build file-based networks. For that, we assume transitivity, e.g., two matches (a,b) and (b,c) implying similarity of (a,c). With a graph library, the matches form a set of distinct clusters, of which each represents a group of matching pairs. Each cluster's entries are then mapped back to their corresponding ssdeep hashes and their original files. We call the clusters of similar files \emph{file-based networks}, and after filtering out networks of size $\leq 1$, the results are saved to a separate \gls{csv} file.

The last analysis step consists of \texttt{files\_networks.py}, which gathers further information about the different file-based networks and their properties, e.g., scores, sizes, files, and more. 
Additionally, a comparison of script-based and file-based networks is achieved through a mapping to \texttt{<script domain>:<filename>} and computing the resulting sets' intersection. To facilitate manual review of the results, we limit the script-based networks to a minimum score over 15.

\subsubsection{Results}
The initial preparation step finds proxy traffic for 4,488 domains, which leads to 20,638 extracted script files. The ssdeep matching process results in 9,748,440 similar script pairs. From those, over half a million (about 5\%) scripts have different filenames but can be matched across multiple websites. 30\% of those matches have different script origin domains, which gives the file-based approach an advantage in detecting fingerprinting scripts over a deny-list. Any improvement in identifying and blocking a fingerprinting script is a potential privacy gain for internet users. On the other hand, there is a case of 25 scripts found on the same page domain, but with different filenames and script domains, where the latter is the result of \gls{http} redirects and the former likely randomization as discovered in the previous study.

The matching pairs form 2,926 distinct clusters, and after conversion to file-based networks, 1,534 networks with more than one script remain. In total, the networks cover 11,704 fingerprinting scripts. We categorize all file-based networks using the following boolean criteria:
\begin{smallitemize}
	\item Same match-score
	\item Same script-score
	\item Same script domain
	\item Same script filename
\end{smallitemize}
Manually reviewing the 16 resulting categories for prominent properties, we discover the following insights: The largest category with 506 file-based networks has a majority of 233 networks with an average script score of 1. Another 242 networks have an average score below 8, leaving only a few highly active networks. However, all files per network have the same script domain, script filename, script score, and matching-score, making it likely that each network represents a unique actor whose scripts do not change. Another observation is that for many networks, the filename changes come from cache busters, version numbers, client identifiers, and slight name variations. These modifications are likely to require changes to deny-list-based privacy solutions, although their file contents are the same. There are multiple categories where the script domain changes, either due to the use of \gls{cdn}s or due to self-hosted fingerprinting scripts. In some cases, these scripts' filenames change as well. As discussed in section \qnameref{sec:design:scriptorigins-actors}, determining the actor of such scripts is difficult, and deny-lists cannot be used without the risk of blocking unrelated resources. One network that stood out was \texttt{hCaptcha}, which seems to change the accessed feature groups, as the observed script score changes. Overall, the observations overlap with the randomization study's results using the signature-based approach. Furthermore, the examined samples are consistent across the various networks in terms of matching similar files, even though the files have slightly changes.
\begin{table*}
	\centering
	\footnotesize
	\begin{tabularx}{\linewidth}{rrrrrX}
		\toprule
		Score & \# SB & \# FB & {SB$\cap$FB} & SB$\cup$FB & \texttt{<script domain>:<filename>} \\
		\midrule\midrule
		32 & 2 & 1 & 1 & 2 & exponential.com:tags.js \\
		\midrule 
		27 & 141 & 8 & 7 & 142 & easyjet.com:37f42850 \\
		\midrule 
		25 & 16 & 19 & 14 & 21 & athome.co.jp:eadjaxlayqcmrfpn.js \\
		\midrule 
		22 & 2 & 7 & 2 & 7 & skyscanner.ru:init.js \\
		\midrule 
		21 & 3 & 2 & 1 & 4 & px-cloud.net:main.min.js \\
		\midrule 
		21 & 2 & 7 & 1 & 8 & zazzle.com:init.js \\
		\midrule 
		20 & 4 & 4 & 4 & 4 & sift.com:s.js \\
		\midrule 
		20 & 2 & 3 & 2 & 3 & yabidos.com:flimpobj.js?cb=[...] \\
		\midrule 
		20 & 11 & 13 & 9 & 15 & adform.net:?pm=1511358\&[...] \\
		\midrule 
		19 & 8 & 2 & 1 & 9 & dns-shop.ru:log-action.js \\
		\midrule 
		19 & 2 & 4 & 1 & 5 & hclips.com:barbar4.12.2.8ba[...].js \\
		\midrule 
		19 & 2 & 2 & 2 & 2 & datadome.co:tags.js \\
		\midrule 
		18 & 88 & 160 & 82 & 166 & celine.com:1e846d2919242fbe7[...] \\
		\midrule 
		18 & 3 & 3 & 2 & 4 & detik.net.id:thetracker-cnn-v3.min.js \\
		\midrule 
		18 & 2 & 160 & 2 & 160 & hilton.com:2fe9b03fa8242608be[...] \\
		\midrule 
		18 & 18 & 160 & 18 & 160 & jetstar.com:02e931a6e189a5dded[...] \\
		\midrule
		\bottomrule
	\end{tabularx}
	\caption{Comparison of signature-based (SB) networks versus file-based (FB) networks, including the number of script files resulting from an intersection or union of both network types, and a script example.}
	\label{tab:signature-vs-file-based-networks}
\end{table*}
To answer the second question of whether the file-based and signature-based detection approaches can work together, we compared both network types. Table \ref{tab:signature-vs-file-based-networks} shows the 16 networks whose scripts overlap between a signature-based and network-based network. Both analysis approaches identify two identical, highly active networks by the actors DataDome and Sift. The remaining overlaps are of partial nature only but reveal an important relationship: Both approaches can be combined to form larger networks. For example, there is a signature-based network of size 2 that matches with a file-based network of size 160. Vice versa, there is a signature-based network of size 141 which overlaps with a file-based network of size 8. Other networks are more balanced and do not show such high differences. Nonetheless, the results promote to combine both approaches in order to increase the networks' coverage:
\begin{smallenumerate}
	\item Identify fingerprinting scripts using a behavior-based approach, such as \FPMON.
	\item Use fuzzy hashing and a suitable extension to block the previously identified fingerprinting scripts or similar ones.
\end{smallenumerate}
Furthermore, having all similar files of a signature-based network and vice versa provides the ability to further analyze behavior, content, networks, or actors. For example, the file-based network of size 160 covers three different signature-based networks, providing the additional insight that these networks are potentially based on the same script code or operated by the same actor.

\subsection{Study 5: A Web Security Analysis} \label{sec:evaluation-study-security}
The last study gauges if fingerprinting scripts are following basic web security principles as outlined in section \qnameref{sec:design:security}. In particular, we analyze whether the files use transport security and common \gls{http} security headers.
\subsubsection{Technical Setup}
Since all \gls{http} requests and responses with their headers are recorded when \texttt{ENABLE\_PROXY} is set, this study can reuse the previous study's data. Other than a separate analysis script, no technical setup was needed. 
\subsubsection{Methodology}
At first, \texttt{security\_analyzer.py} compiles a list of domains and their scripts with script domain, script filename, and script score. Since only domains with a network traffic log can be analyzed, the analysis tool checks for the existence of a domain's traffic log file. If no log file is present, the domain is skipped. Otherwise, all captured \gls{http} requests' host-header and filename are matched with the fingerprinting scripts' domain and filenames. Since \FPMON's monitored script origins were requested and loaded by the browser, their corresponding request and response must be found in the captured network traffic. However, it can only match scripts with an external \gls{url}, since JavaScripts directly embedded into the page would not have a distinct filename. From these traffic and fingerprinting script matches, three groups are formed as follows:
\begin{smallitemize}
	\item If the request's \gls{url} begins with \texttt{https://}, it is categorized \texttt{secure}.
	\item If the \gls{url} begins with \texttt{http://}, two cases are distinguished:
	\begin{smallitemize}
		\item If the response's status code indicates a \gls{http} redirect (301 or 302) and the destination \gls{url} begins with \texttt{https://}, then the group is \texttt{redirects}.
		\item Otherwise, it is \texttt{insecure}.
	\end{smallitemize}
\end{smallitemize}
Having all fingerprinting scripts in these groups answers the transport security question and becomes the basis for further analysis. Each group has their scripts' scores grouped and counted for an overview of each group's severity. Similarly, all three group's \gls{http} responses are analyzed for the web security headers listed in \qnameref{sec:design:security}, and their occurrences are counted, thus answering the question about their usage.

In addition to the initially asked questions, the three groups allow deducing information about the actors by intersecting (denoted by \texttt{\&}) the sets of script domains:
\begin{smallitemize}
	\item \texttt{Insecure \& Secure}
	\item \texttt{Redirects \& Secure}
	\item \texttt{Redirects \& Insecure}
\end{smallitemize}
Code reviews for JavaScript-based vulnerabilities like cross-site-scripting or open-redirects are left for future work due to most fingerprinting scripts being minified or obfuscated.

\subsubsection{Results}

We find 4,490 domains that have matching network traffic and undergo further analysis. From those, 57,414 fingerprinting script and traffic request matches are found. With 56,351 matches (98.15\%) and 27,240 unique request \gls{url}s, the \texttt{secure} group indicates that most fingerprinting scripts are loaded over a secure channel. However, there are 999 fingerprinting scripts with 807 unique \gls{url}s that are insecurely transferred. 64 requests to 45 distinct \gls{url}s receive a redirect to a secure \gls{url}. Although the upgrade to a secure connection is well intended, it does not prevent network-level attacks (e.g., \gls{mitm}) because the initial request and response are still transferred insecurely. 
\begin{figure}[b]
	\centering
	\includegraphics[width=0.90\textwidth]{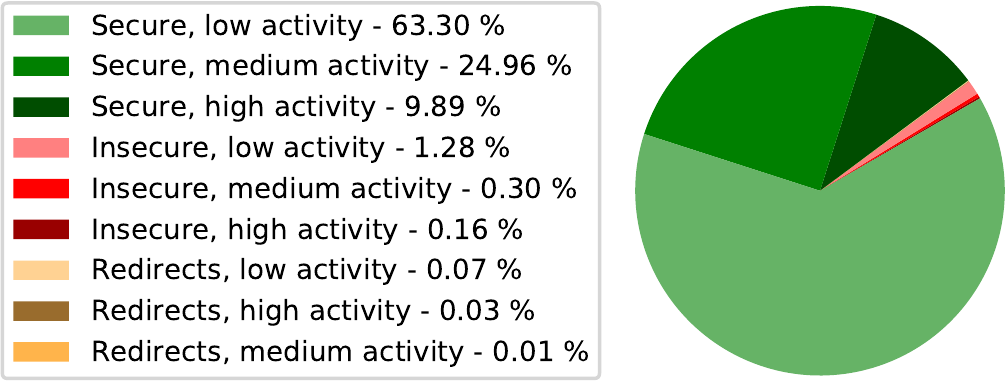}
	\caption{Script classification into high, medium and low activity for all fingerprinting script and \gls{http} request matches per group in study 5.}
	\label{fig:security-group-ratings}
\end{figure}

\begin{figure}[t]
	\centering
	\includegraphics[width=0.935\textwidth]{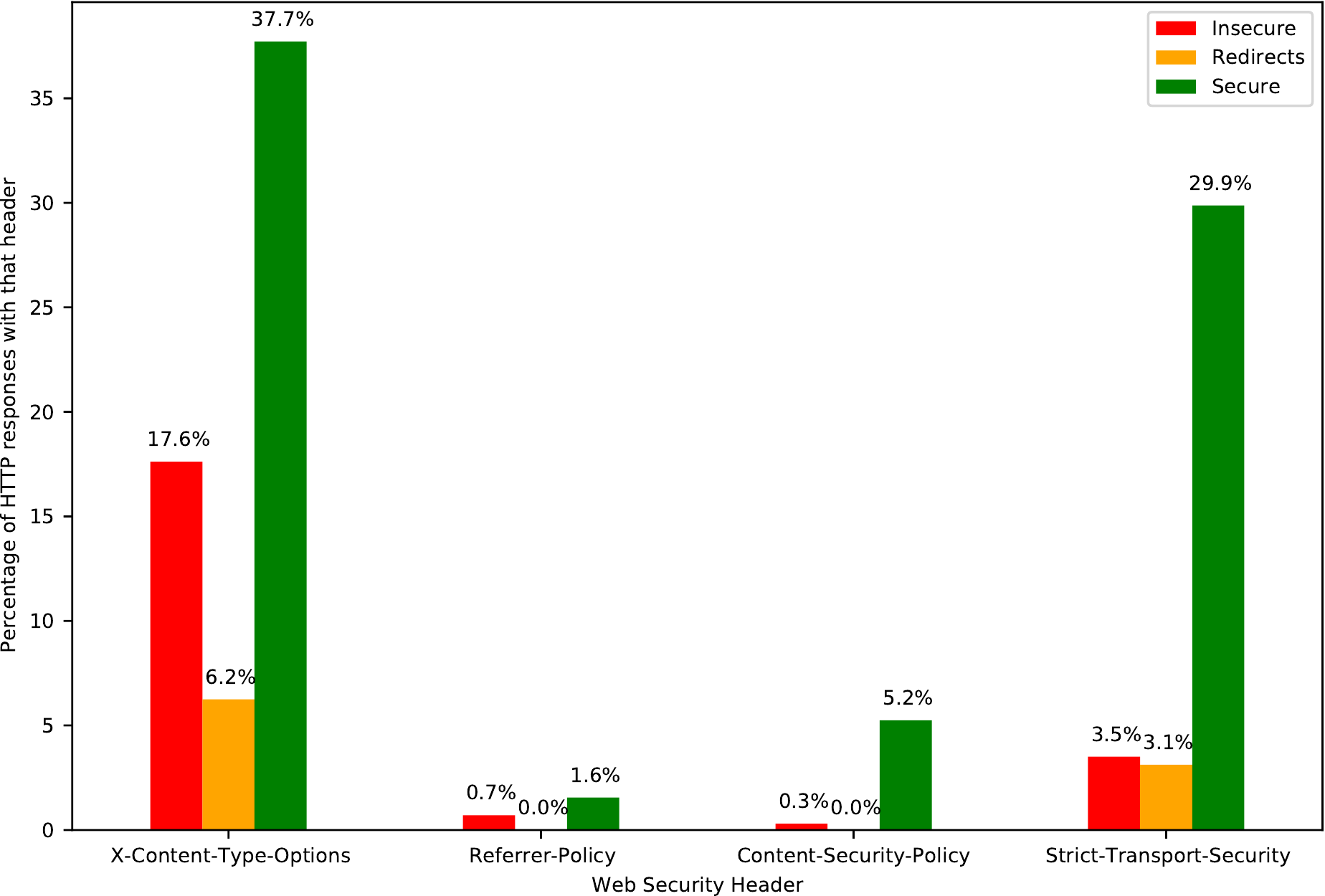}
	\caption{Distribution of web security headers for secure, insecure and redirect script \gls{url}s}
	\label{fig:security-headers}
\end{figure}

Figure \ref{fig:security-group-ratings} shows the three groups and their scripts' scores categorized into high, medium, and low activity. Although the insecure and redirect groups have an insignificant share, each group has medium and high activity scripts, which leaves the source code or data at risk. The header analysis shows a similar trend. Almost one-third of the files in the secure group set the Strict-Transport-Security and X-Content-Type-Options header to force browsers requesting it securely, as Figure \ref{fig:security-headers} depicts. Only 5\% set the Content-Security-Policy header, which is optional for scripts that do not use workers. The Referer-Policy header lacks popularity with at most 1.6\% (secure group). Remarkably, \gls{http} responses from the insecure and redirects group set the Strict-Transport-Security header, thus upgrading to secure connections for future requests until the header expires.
Comparing the three groups, we discover that 134 script origin domains serve their scripts over both secure and insecure transport channels. Furthermore, 36 script domains serve their script securely and upgrade insecure requests using a redirect response, but this only mitigates attacks after a successful upgrade to a secure connection. Only one domain is in the \texttt{insecure} and \texttt{redirects} group. It is plausible that only some resources return a redirect, but not all. Ultimately, the responsibility of embedding a script correctly (secure vs. insecure) is left to the website developer or administrator.

The results show that only a slim percentage of fingerprinting scripts is requested over insecure protocols. Nonetheless, some of them have a medium or high score. Furthermore, we discover that security-conscious fingerprinting script providers secure the transport and application layer using web security headers. Certain headers reach around 30\% or more, which provides additional security, but leaves a majority of script files and connections with room for improvement. However, to mitigate risks from insecurely loaded active content, modern browsers prevent the content's execution when the page was requested securely \cite{mixed_content}.

\section{Discussion} \label{sec:discussion}
This section aims to discuss and put the previous section's results into perspective.
\subsection{Comparing The Results to Related Work}
To bring our results into context, we compare them to related work. Since our data only focuses on the Alexa Top 10,000 and pages with a page score $\geq$ 10, the results must be considered a lower bound. Eight years ago, \cite{Acar2013} scanned the Alexa Top 1,000,000 with a focus on font fingerprinting. They found 404 pages that employ font-based fingerprinting, whereas our results find a 200\% increase to over 1,300 pages within the Alexa Top 10,000. They identified MaxMind as one actor, which we discovered to run fingerprinting scripts with the highest score. In the same year, \cite{Nikiforakis2013} analyzed the fingerprinting scripts from BlueCava, Iovation, and ThreatMetrix in regard to browser-related properties (e.g., plugins, cookies, timezone, useragent, screen resolution, fonts) and discovered 40 pages using them within the Alexa Top 10,000. Today, Iovation's scripts can be found on 15 pages using up to 13 feature groups in our data set.
In \citeyear{Acar2014}, \cite{Acar2014} discovered roughly 500 pages within the Alexa Top 10,000 that use Canvas fingerprinting techniques. Our scans reveal at least 786 pages using the Canvas feature group - a 50\% increase over the last 7 years. Again, actors like \texttt{Addthis.com}, \texttt{admicro1.vnmedia.com} or \texttt{rackcdn.com} are still active and host fingerprinting scripts. In comparison to \cite{Englehardt2016} in \citeyear{Englehardt2016}, our results indicate a significant increase in Canvas fingerprinting by almost 100\%. Furthermore, the ``infrequent'' usage of Audio features reported by \cite{Englehardt2016} grew to 1.63\% according to our data set. Google and its subsidiaries were identified as the most prevalent actor by \cite{AlFannah2018} in \citeyear{AlFannah2018}. Additionally, they discover Yandex to collect the most data. Our analysis finds both actors' networks covering the most websites (see Table \ref{tab:top-actors-affected-domains}). In fact, we can confirm \texttt{Quantserve.com}, \texttt{Rubiconproject.com}, \texttt{omtrdc.net}, \texttt{openx.net} as domains related to fingerprinting. \cite{AlFannah2018} concluded that almost 69\% of websites conduct fingerprinting. Our result has a natural lower bound of 54.07\% since we limited the set of analyzed pages. However, in that data set we find 4288 pages with a high activity script ($>$ 6 feature groups), which we classify as fingerprinting scripts, and 5287 pages with at least medium activity ($\geq$ 3 but $\leq 6$ feature groups). Although, we did not consider the transmission of collected properties or fingerprints in contrast to \cite{AlFannah2018}, the security analysis study suggests that \gls{ssl}/\gls{tls} adoption in third party JavaScript fingerprinting scripts increased to 98\%. That result is significantly higher than the previously reported numbers by \cite{AlFannah2018,Englehardt2016}. 

\subsection{Inspecting Identified Fingerprinting Networks \& Actors}
Overall the comparison indicates that fingerprinting is becoming more prevalent on the internet, and more users are exposed to it. As outlined in Chapter \ref{sec:background-related-work}, fingerprinting can serve a variety of purposes: Financial services might use it to improve a customer's security and prevent fraud, but advertising businesses might rely on fingerprinting to improve their tracking abilities. While \FPNET~and \FPMON~can identify scripts with fingerprinting behavior, both tools cannot determine the use-case. However, by grouping scripts' similar behavior into networks and pinpointing the actors behind them, we try to answer that question. In contrast to previous work, which based its analysis on single or small sets of features, our approach classifies scripts based on their detected behavior. Figure \ref{fig:scripts-chart} shows that high activity scripts do not limit themselves to only one fingerprinting technique but tend to leverage various browser functionality to achieve their goal.

\subsubsection{Exposing Fingerprinting Networks On The Internet}
Using the behavior-based approach, we identified hundreds of fingerprinting networks of various sizes and properties in study 2. The networks' median signature length is 64. Thus, half of the identified networks consist of fingerprinting scripts that access at least 64 feature groups in the same order. Therefore, we consider this to be a strong and reliable indicator that these networks are formed by similar scripts. However, the number of networks and their signature length is a lower bound since study 3 showed that randomization can influence a signature. Nonetheless, the networks' properties are surprising. 75\% span less than 7 websites, and only 9 networks have a coverage of 100 or more. An actor's reach might exceed that if they control multiple networks. Overall, the relationship between network size and score seems to follow an anti-proportional trend: Larger networks tend to have lower scores and smaller ones have higher scores. Still, all analyzed networks consist of high activity scripts (score $>$ 6) and, therefore, can gather enough entropy for accurate fingerprints. A plausible explanation for this relationship could be the fingerprinting use-cases: Smaller but more aggressive networks likely belong to actors that offer specialized products which require high accuracy or security, e.g., identification, bot detection, fraud detection. These networks are small since not every website needs such guarantees about its users on the home page. On the other hand, larger and less aggressive networks operate in business areas where a slightly reduced fingerprinting precision might be tolerable, e.g., tracking, advertising, or others.


Another notable result stems from networks' script domains. A handful of networks have the number of script domains close to the number of covered websites, leading to an almost one-to-one relation between website and script domain. A brief, manual review of several networks showed that both domains are indeed mostly identical. Therefore, we call such networks \emph{self-hosted} following Figure \ref{fig:actor-analysis}. Depending on the network, collected fingerprinting data is not sent to a third-party but to the website's domain. Although this does not legitimate consent-less fingerprinting, it can be considered more privacy-friendly if the data is only processed by one entity. This requires running suitable software (e.g Matomo \cite{matomo} for analytics, Revive \cite{revive} for advertising). Otherwise, data could be shared with third-parties through server-side backend communication, but this cannot be determined with our tools. However, self-hosted networks can belong to actors nonetheless. For example, Akamai (see Table \ref{tab:top-actors-score}) controls a self-hosted network, but is still a third-party. Unless the script filenames share a pattern, such networks might circumvent deny-list-based privacy extensions since all script domains are different. The Akamai network could be an explanation for \cite{kybranz}'s observation that privacy extensions are not always effective because this network has random filenames and different domains.

In contrast to self-hosted networks, single \emph{third-party} networks were also identified. These kinds of networks usually have only one script domain unless an actor uses several similar script domains (e.g., same first-level domain, but different top-level domains, abbreviations, or slight variations). A small set of script domains facilitates identifying an actor, and the fingerprinting data is likely sent back to the third-party for processing. Third-party networks with high coverage and scores are especially insightful from a privacy standpoint. With such networks, a single actor could collect highly accurate fingerprints for large numbers of users across many websites, enabling user identification or tracking and other use-cases. As Table \ref{tab:top-actors-score} shows, most high-score actors have rather small networks. However, some single third-party networks are based on \gls{cdn}s, which makes them hard to analyze, as depicted in Figure \ref{fig:actor-analysis} because they hide the actor's script domains and improve the network's script delivery. About 50 networks were not reliably identifiable due to \gls{cdn}s. For networks with multiple script domains, an actor could be revealed if not all scripts were requested from a \gls{cdn}.

Studies 3 and 4 provided valuable insights into the behavior-based approach's reliability and the networks' defensive measures.
The comparison of two consecutive scans showed that more than 75\% of the successfully analyzed pages do not have an identical set of script origins. However, a brief analysis of the differences provides a plausible explanation: Cache busters, timestamps, or slightly varied \gls{url} query parameters cause a mismatch. This error propagates into the script signature comparison since the list of signatures is ordered by the scripts' URLs. So a change in a script origin might alter the lists' order. Another source of error could be \FPMON~if it records slightly different script signatures, leading to a unsuccessful script list comparison due to different script scores. However, that kind of error should be revealed in the script analysis. In contrast, only 35\% of pages differ when the script origins are not part of the comparison, e.g., only checking the scripts' scores. Therefore, we believe that although most script origins change over time, their behavior or activity does not. The file-based comparison supports that with the fact that over 86\% of more than 70,000 scripts have the same script score and signature. In fact, the results also show that several scripts either deliberately change their signature or \FPMON~fails to monitor them properly. Unfortunately, distinguishing between both cases would require a more in-depth analysis of the scripts or their behavior, which is future work. Nonetheless, we succeeded in recovering several changed signature matches using string similarity algorithms. On the other hand, the behavior-based approach helped us re-identify almost 9,000 scripts whose filename, domain, or both change. From the examples provided in study 3, we discover a few cases where fingerprinting scripts deliberately try to evade detection or privacy deny-lists by changing or randomizing the script domain or filename. This observation is important because it shows that a minority of networks and actors use defensive mechanisms. Without such defenses, a fingerprinting-based product might not be able to collect the necessary data and therefore not properly produce its value.

In study 4, we used the script files' contents to build file-based networks. The results show similar changes in filenames or script domains. Comparing file-based and signature-based networks (see Table \ref{tab:signature-vs-file-based-networks}), we discover a weakness of the behavior-based approach: Some signature-based networks are smaller in size than their file-based equivalent. So although the file contents are quite similar, their behavior is not necessarily. Several factors might have influenced this result: First, the assumption of transitivity in the file-based matches might not hold. For example, the script-matches (a,c) in a file-based network might not be similar according to ssdeep, although (a,b) and (b,c) are. Second, variations in script behavior or execution due to script changes or configuration changes. The latter was observed in brief reviews of the file-based networks, where script files are almost identical except for website-specific configuration parameters (e.g., a website's \gls{url}). 
Despite possible weaknesses , the combination of the behavior-based and file-based approach could help to build an extension to increase users' privacy and better detect fingerprinting scripts. 

\subsubsection{Unmasking Actors Controlling Fingerprinting Networks}

As the results of study 2 reveal, we uncovered more than 100 distinct actors, often controlling several networks. To our knowledge, this is the first extensive list (see Table \ref{tab:actor-list}) of fingerprinting related entities. In contrast to related work \cite{nikiforakis2014privaricator, Nikiforakis2013, Acar2013, Baumann2016}, which often discovered and analyzed \texttt{BlueCava} as an actor, the company does not appear in our results. Similarly, \texttt{addthis.com} and others are not in our list of actors. An explanation might be changing technology or privacy policies. \cite{FaizKhademi2015} noticed that addthis.com stopped using canvas fingerprinting in 2014. In fact, some fingerprinting scripts might only get active once \gls{gdpr} consent is obtained. Other scripts and actors might use fewer feature groups and therefore were not included in the final analysis. This work did not investigate where the collected fingerprinting material is sent to, but a brief analysis was conducted for the paper \cite{fp_paper}, in which the authors showed that most actors are sending data back to their third-party servers. Only Akamai and FingerprintJS send it to the website's domain. \cite{Torres2015} performed a comparable analysis and discovered that fingerprinting hashes and gathered properties are sent to the collecting server. In particular, Iovation and FingerprintJS are among the actors in our data set.

It is not surprising that a significant share of the actors falls into the following categories according to TrendMicro's SiteSafety (see Figure \ref{fig:actor-analysis}): \emph{web advertisements}, \emph{search engines}, \emph{business / economy} or \emph{computers / internet}. While \emph{computer / internet} is a too broad generalization, the other categories appear to fit the described purposes of either providing security or analytics and advertising (see section \qnameref{sec:background-related-work:fp-purposes}). Tables \ref{tab:top-actors-affected-domains} and \ref{tab:top-actors-score} indicate a rather anti-proportional trend between score and coverage, which supports the hypothesis that each fingerprinting purpose has varying requirements in terms of accuracy or collected data.

Google DoubleClick and Google AdSense are operated by the company Google. Together, they are the actors with the broadest coverage but with a relatively low score of 10. Controversially, Google itself has policies against fingerprinting in Google Analytics \cite{google_analytics_policy} and discourages the use of ``fingerprinting for identification'' by adopting \gls{tcf} v2.0 \cite{google_adsense_tcf} for AdSense. It is questionable why some of Google's products access fingerprinting-related browser properties, including two aggressive ones, nonetheless. Since Google operates in the advertisement and analytics business, their wide coverage might allow them to require fewer properties to track users. In fact, AdSense uses one sensitive and one aggressive feature group less than DoubleClick. The 2nd largest actor is Yandex Metrica, which is a web analytics platform describing itself as ``From traffic trends to mouse movements - get a comprehensive understanding of your online audience and drive business growth'' and ``[...] easy-to-use attribution models, filter out spam, and control bot traffic'' \cite{yandex_metrica}. The scripts used by Yandex Metrica access 3 aggressive feature groups and 11 sensitive ones. Clearly, collecting many browser properties helps to achieve the aforementioned goals. Tracking website visitors utilizing browser fingerprinting could be beneficial when cookies cannot be used, or the user switches browsers or surfs in private mode. 
FingerprintJS claims to ``stop fraud, spam, and account takeovers [...]'' on their website \cite{fpjs}, which caters more to the security fingerprinting purpose. In fact, their product focuses on ``fraud prevention'' or ``account sharing'' \cite{fpjs}. Their documentation mentions a ``Server \gls{api}'' \cite{fpjs_doc}, which can be used to retrieve a visitor's browser details. Furthermore, it proves that products relying on backend communication exist, and even data transmitted to the same website might be processed by a third-party actor. 


One actor has a top rank measured in both score and affected websites: Akamai. Its product ``Bot Manager'' uses browser fingerprinting to ``identify automated or headless browsers and detect anomalies that indicate an automated bot'' \cite{akam_botmgr}. With 10 aggressive and 18 sensitive feature groups, it is not the highest-scoring actor offering fraud or bot detection products. Maxmind uses its highly aggressive device tracking technology to ``capture more data and catch more fraud'' \cite{maxmind}. Similarly, Adsco.re classifies traffic to detect bots, scrapers, or ``Low Quality Human Traffic''~\cite{adscore}. The company behind ShieldSquare titles itself a ``leader in bot management'' and promises to defend against ``Account Takeover'', ``API Abuse'', ``Price Scraping'' and other attack vectors \cite{radware}. A press release about ``new device fingerprinting technology'' dates back to 2015 \cite{radware_press}, but certainly, their technology has evolved since then. Their fingerprinting script accesses 6 aggressive and 17 sensitive feature groups. The only actor not focusing on security aspects, but on advertising and analytics in the top 5 by score, is Moat \cite{moat}. They do so using 34 feature groups on over 100 websites (see Table \ref{tab:top-actors-score}). On the one hand, accurate fingerprinting could help prevent advertisement fraud (i.e., unintentional clicks or bot traffic). On the other hand, the collected data could be valuable for other purposes of Oracle's Data Cloud, e.g., serving as training sets for machine learning or showing personalized advertisements.

Concluding, the analysis of the actors' fingerprinting use-cases reinforces the observed division into security (fraud protection, bot detection, and other) or tracking (advertisements, analytics) purposes. Indeed, the latter aims for greater coverage than accuracy since tracking an individual across a large set of websites potentially improves the advertising profile. In contrast, security products have lower coverage but try to distinguish illegitimate from legitimate users or traffic more accurately.  


\newpage
\subsection{Privacy Implications of Fingerprinting}
We identified a large set of fingerprinting networks and the actors behind them on the most popular 10,000 websites. Study 1 confirmed that disabling JavaScript reduces a website's usability. Thus, one cannot simply disable JavaScript to avoid active fingerprinting. Furthermore, research by \cite{kybranz} showed that privacy-enhancing browser plugins do not completely eliminate fingerprinting. According to \cite{gdpr_paper}, fingerprinting data is regulated by the \gls{gdpr}, so users should need to consent before being fingerprinting. In reality, all scans conducted in the studies did not provide any consent at all. In fact, we believe in detecting larger networks or overall higher fingerprinting activity if consent would be given. Nonetheless, our results match the observation by \cite{Trevisan2019} that not all websites are fully \gls{gdpr} compliant yet. 

In other words, the identified actors do not honor a user's privacy. As fingerprinting is practically invisible to users, they are certainly not aware of their privacy being violated. The larger an actor's coverage and score, the better an actor can re-identify a user on multiple websites, despite having cookies disabled, privacy plugins enabled or using a browser's privacy mode. One reason for the increased use of fingerprinting technology might come from the regulation around cookie usage (see \cite{Trevisan2019}). Websites use so-called cookie banners (see Figure \ref{fig:cookie_banner}) to gather consent for setting cookies, which are not necessarily needed for the website's operation, e.g., ones used for tracking or advertising. Therefore, advertisers lose out on potential data or revenue by not being able to track or profile users that deny such cookies. Fingerprinting circumvents this by not requiring persistent storage for providing data or a high-entropy identifier. Study 5 shows that \gls{tls}/\gls{ssl} is being adopted by fingerprinting actors, as more than 98\% of scripts are served over a secure connection. Since some browsers prevent \emph{mixed content} \cite{mixed_content}, actors are forced to adopt secure transport protocols if they want their scripts executed on websites that are served over them. The security protection includes outgoing connections. Thus, one can assume that collected fingerprinting information is sent over a secure channel, too. This is an important development for users' privacy, as it prevents other attackers from obtaining sensitive information (e.g., using \gls{mitm}).

Nonetheless, allowing the absolute minimum of cookies in cookie-consent banners or disabling them altogether might leave users with a false sense of privacy. It is therefore critical to educate users about new techniques, such as active fingerprinting. Stateless fingerprinting can compensate for missing information from stateful fingerprinting when the device properties are unique enough. In an attempt to combat this trend, some browsers began to implement anti-fingerprinting techniques. For example, Apple's Safari or Mozilla's Firefox try to reduce the amount of data or entropy that can be collected \cite{safari_antifp, antifp_mozilla}. Although users could install privacy-enhancing browser extensions, study 3 showed that deny-list approaches can fail due to domains or filenames being randomized or fingerprinting scripts being served from hard-to-block domains (e.g., self-hosted or \gls{cdn}s). Even extensions that temporarily or selectively block scripts from executing automatically at page load, their usage usually comes with usability constraints. Furthermore, the fingerprinting code might not be distinguishable from regular website code, e.g., due to minification or obfuscation. Academia has not been inactive in researching defensive techniques as the related work indicates (see section \qnameref{sec:background-related-work:related-work}). We believe that this research topic will gain importance and attention in the future. Especially since fingerprinting is a dual-use technology, recognizing a fingerprinter's purpose is not trivial because most browser properties exist to improve the web browsing experience. For example, a website's layout can be changed depending on the screen width or height. Similarly, WebGL and other properties can be used to implement complex web applications, e.g., games or dashboards, and are therefore not easily distinguishable from being used for fingerprinting.

\subsection{Assessing Faced Limitations}
The thesis faced some technical and non-technical limitations that might have influenced the studies' outcomes or posed a threat to the validity.

\subsubsection{Technical Objections}
Although \FPNET~was built to be as resilient to errors as possible, unhandled exceptions or corner cases might still exist, which could prevent a successful collection of fingerprinting information.
However, during the \FPNET~scans only a few controlled errors occurred. For example, in study 2 only 0.025\% domains failed with errors related to the automated browser instrumentation (e.g., Java exceptions), and no unknown failures occurred. Single errors did not impact the remaining scanning process since each scan is performed in a new browser instance, and the modified startup script resets defunct browser instances after a timeout. 

The chosen timeouts and parallelism could affect the results. Too short timeouts might prevent a page to fully load and execute the fingerprinting script. The timeouts chosen for this thesis' studies were slightly longer than used in \cite{kybranz}: With 45 seconds for a page to load and another 45 seconds to collect the fingerprint, we gave the page almost twice as much time than needed according to \cite{httparchive}. Too high parallelism could influence the pages' load time or analysis due to exhaustion of the shared resources, e.g., disk or network \gls{io}. To our knowledge, no resource exhaustion occurred during the scans since the provided virtual machine had enough spare resources and neither \gls{cpu}, \gls{ram}, network bandwidth nor disk \gls{io} were utilized 100\%. However, \FPNET's parallelism is configurable and can be adapted to a less resourceful environment, allowing slower or less resource-intensive scans to be performed. Furthermore, \FPNET~provides each scanned page an isolated environment with a clean, fresh browser profile to prevent interference.

Since domain names have no \gls{http} protocol associated, \FPNET~requests \texttt{http://<domain>}, because this does not limit the analysis to \gls{ssl}/\gls{tls} secured websites. Websites that support both usually upgrade the connection from insecure to secure using a \gls{http} redirect. When no upgrade is possible, and the website is not served over \texttt{http://}, \FPNET~fails to establish a connection, leading to a timeout. Therefore, choosing \texttt{http://} as the primary connection protocol could be one cause for the 288 timeouts during study 2's scan.

For the proxy component to capture the network traffic, it must lever out the \gls{ssl}/\gls{tls} encryption by masquerading the certificate. Although the Chrome browser was instructed to ignore \gls{ssl}/\gls{tls} related errors, connection issues cannot be fully eliminated. For example, \gls{hpkp} was invented to thwart \gls{mitm} attacks by checking the certificate's public key against a header \cite{hpkp}. To eliminate other potential issues, mitmproxy's certificate-authority could be imported into Chrome's certificate store. However, due to the increased complexity, the \texttt{--ignore-certificate-errors} solution was favored, and the improvement was left for future work. For the scans performed in study 4, 1.75\% of pages could not be analyzed due to certificate and proxy-related errors.

All scans were executed from one virtual machine with one \gls{ip} address from the chair's address pool 130.149.230.0/24. In theory, the recurring requests from one single \gls{ip} address or its origin from a university network could be detected, and the fingerprinting activity limited. This would require passive fingerprinting on the script origin's server. However, several devices or users sharing a single \gls{ip} address (e.g., in \gls{nat} networks) would be affected, too, which is detrimental to a fingerprinting script's goal of maximizing the coverage and accuracy. Nonetheless, the risk of missing highly sophisticated fingerprinting networks or actors was accepted in favor of a consistent setup. As outlined in future work, this problem could be solved with a community-based approach.

Another technical limitation is the missing interaction with a website. The headless browsers did not interact with a page like a human would, e.g., there was no mouse or keyboard input. Fingerprinting scripts could detect the lack thereof and reduce their fingerprinting activity accordingly. Mouse movements are supported by selenium, but an algorithm for natural mouse behavior would need to be implemented and is left for future work.

\subsubsection{Improving Detection Accuracy}


All scans were limited to the analysis of a domain's landing page. It is possible that some fingerprinting scripts are not executed on the landing page but on other \mbox{(sub-)pages}, which were not analyzed. For example, a shopping website might only need to inspect the user's browser during checkout before finishing the purchase to prevent fraud. 

\FPNET~can only uncover fingerprinting networks and actors for fingerprinting scripts identified by \FPMON. \cite{kybranz, fp_paper} analyzed and distilled function names or properties into a feature group mapping as listed in Table \ref{tab:feature-groups}. Fingerprinting scripts that use other techniques or functions will not be processed by \FPNET. Therefore, \FPNET~might not uncover novel fingerprinting techniques. However, \FPNET~and \FPMON~can be extended with new feature groups or properties if necessary.

Furthermore, \FPNET~did not implement consenting to \gls{gdpr} popups. Therefore, fingerprinting scripts that run after consent was provided could not be detected. In fact, our results highlight cases in which users are potentially tracked despite not consenting to any data processing. Therefore, we expect higher fingerprint activity and larger fingerprinting networks if consent would be provided.

In comparison to \cite{kybranz}, this work analysis' focuses on single scripts. The signatures and scores are calculated for each script on a page. As noted in study 2, this leads to situations where a page's score is higher than its scripts' scores. Furthermore, networks are formed from high activity scripts (score $>$ 6) only. This bears the chance that \FPNET~or the post-processing step missed scripts which reduced their score by distributing their fingerprinting behavior across several scripts. Depending on the score $S$ and the amount $N$ of sub-scripts, each sub-script could have a score $\frac{S}{N}$ $<$ 3, but still gather enough fingerprinting information in total. However, such an approach would require $N$ \gls{http} requests to load the sub-scripts, which increases a page's load time and, therefore, is detrimental to fast and efficient fingerprinting. In fact, Maxmind's top-ranking script would require $N \geq 18$ to get a score of 36 categorized as low activity.

A limitation that was evaluated in study 3 is randomization. \FPNET~post processing is based on the assumption that a script's execution is deterministic and the collected script signatures do not change. Otherwise, two scripts might not be identified as similar, thus preventing finding a network's actual size. Asynchronous loading of scripts is not an issue for \FPNET~since each script is monitored independently. However, asynchronous or parallel execution, e.g., \texttt{setTimeout}, \texttt{setInterval} or other functions which introduce variable delays, can influence the order in which feature groups are monitored by \FPMON. Study 3 succeeded in matching slightly varying script signatures using fuzzy hashing and string similarity algorithms. Although \FPNET~successfully uncovered networks and actors of high fingerprinting activity, designing script similarity algorithms that do not rely on a strict order of feature groups could further increase the identified networks' sizes.

Study 4 faced limitations due to the exclusion of non-external scripts and the file system's filename length limit. These limitations merely reduce the number of scripts processed in the study. Therefore, the amount of file-based networks or their properties are potentially lower than in reality. The filename length limit was unforeseen, but reformatting the server's file system was unfeasible. Including non-external scripts into the analysis could be achieved by matching scripts and the proxy traffic with more properties than just \gls{url}, e.g., monitored function names. 

\subsubsection{Challenges of Attribution}
Attributing a network to an actor became a challenge for several networks in study 2. Although single script domain networks are potentially easier to attribute to an entity, some first-level domains did not resolve nor provided other information about an actor. Similarly, several networks had more than one script origin domain, and in multiple cases identifying a single actor was not possible. Especially \gls{cdn}s and networks of open-source or self-hosted solutions pose a limitation to the actor analysis. Therefore, such networks were not further considered. The resulting actors in study 2 control the networks with high certainty. However, leaving out networks for the actor analysis only lowers the identified actors' coverage or score. In essence, the results are a lower bound, and presumably, more actors exist, or their actual score and size are higher.

\subsection{Future Work} \label{sec:discussion-future-work}
While conducting the research, several new questions and ideas arose around the browser fingerprinting topic.

\subsubsection{Remediating Limitations}
The \FPNET~system could benefit from future work on the limitations outlined in the previous section. However, not only technical improvements are beneficial. Designing and testing new methods to uncover fingerprinting networks and actors is an attractive area of research. 

Human-like keyboard and mouse input could further move the headless Chrome browsers closer to reality. With such improvements, fingerprinting behavior changes after consenting to \gls{gdpr} popups could be evaluated.  

\subsubsection{Improving Fingerprinting Detection}
We used behavior-based signatures and proxy-based file-signatures to detect fingerprinting activity. The script origin was determined solely from its script domain and filename. However, the exception stack trace (see section \qnameref{sec:implementation:fpmon-script-sources}) provides many other properties (e.g., line numbers, function names or others) that could be used to identify fingerprinting networks better or gather more insights about them. At best, this could lead to alternative ways of determining a script signature or behavior similarities that are harder to evade using randomization or obfuscation.

Using the proxy-enabled \FPNET~system, future work could apply \cite{AlFannah2018}'s technique of analyzing outgoing \gls{http} requests for fingerprinting data. Such an approach could answer what information is collected and how or where it is transferred. \cite{Torres2015}'s analysis of several fingerprinting products shows that not only the fingerprint's hash is sent back to the server, but the gathered information as well. Furthermore, this might help to discern the networks that use self-hosted or open-source fingerprinting scripts. 

\subsubsection{Advanced Analysis Methods}
This work's analysis was limited to the Alexa Top 10,000 domains. In future work, larger data sets such as the Alexa Top 100,000 or 1,000,000 could capture an even larger landscape, uncovering more fingerprinting networks or so far unknown actors. 

In order to collect a greater variety of fingerprinting script behavior and circumvent location-related configurations (e.g., fingerprinting scripts only executing for visitors of a specific country), a collaborative community approach could be a solution. Internet users could be asked to load a modified \FPMON~extension while browsing the internet. It would report the fingerprinting signatures and filenames to a central data collection endpoint. We believe that with such an approach, more fingerprinting scripts and subtle differences could be identified. Furthermore, this would eliminate the virtualized environment, and the real user's interaction could capture a fingerprinting script's true behavior, should more sophisticated fingerprinting scripts exist. 

Furthermore, increased analysis efforts could help understand how fingerprinting scripts work (e.g., reverse engineering the obfuscated files). Such an analysis might find specific behavior patterns or identify data transfer methods. For example, a specific order or details about function calls to perform Audio or Canvas fingerprinting might be revealed, leading to a higher identification or script matching accuracy, which could benefit the re-identification of slightly randomized or obfuscated fingerprinting scripts. 

Our actor identification was based on the scripts' domain. Future work could investigate and evaluate other ways of identifying an actor, e.g., for files hosted on \gls{cdn}s. A more in-depth analysis of the actors, their products, and websites their scripts run on could provide insights into the market and create awareness of what sectors use fingerprinting technology. 

\subsubsection{Developing Privacy Extensions}
Two privacy extensions could be evaluated in future work. For one, as indicated in study 4, an extension could be developed which checks all \gls{http} responses' file content against a deny-list of known fingerprinting scripts' file-signatures. In addition to existing domain or filename-based privacy extensions, such an extension could further improve a user's privacy.

In a similar approach, an extension could be developed that prevents the outgoing transmission of the collected fingerprinting information or the calculated fingerprint itself. When this data does not reach the network or actor, the fingerprinting script provides little to no value. Although this might come with its caveats (e.g., a server-side action requiring the successful transmission of a fingerprint), this could be another approach to an anti-fingerprinting privacy extension, similar to \cite{AlFannah2018}'s tool \texttt{FingerprintAlert} and its blocking feature.

\section{Conclusion} \label{sec:conclusion}
This thesis set out to uncover and analyze fingerprinting networks and their actors using a behavior-based approach. While previous work focused on the effectiveness of fingerprinting, offensive or defensive fingerprinting techniques as outlined in section \qnameref{sec:background-related-work}, this work captured the status quo of fingerprinting networks and their actors on the Alexa Top 10,000. Further, we explored the importance of JavaScript for modern websites, fingerprinting scripts' defensive techniques, and web security practices. 

In order to answer our research questions, we implemented an extensible fingerprinting network scanning system (\FPNET) and conducted five complementary studies. First, we showed that at least 87\% of websites rely on JavaScript for an optimal user experience, leaving internet users prone to active fingerprinting. In contrast to previous work, \FPNET's focus on single JavaScript files enabled the analysis and the identification of fingerprinting scripts in the following studies. While our behavior-based approach could not determine a script's intent of accessing specific browser properties, we classified over 8.50\% of all scripts as fingerprinting scripts based on their high activity. Manual analysis revealed over 100 actors controlling more than 250 fingerprinting networks. With a reach of up to 1,500 websites or a score of up to 36 feature groups, we uncovered actors with varying coverage and fingerprinting intensity. We argued that an actor's properties align with the fingerprinting purpose, e.g., higher accuracy for security or greater coverage for tracking. Furthermore, we discovered some fingerprinting scripts randomizing their script origin or behavior in an attempt to evade detection. While deny-list approaches might be prone to such changes, our behavior-based approach successfully re-identifies several fingerprinting scripts. Finally, our results show over 98\% of the scripts loaded over secure connections, while standard web security headers are only present on around one-third of them. From a security perspective, this is a positive trend, although there is room for improvement.

Using these results, we succeeded in answering our research questions and providing a big picture of the current state of fingerprinting on the internet. Overall, our numbers show a steep increase in various fingerprinting techniques and broad adoption of fingerprinting in general compared to previous work. Therefore, we predict a continuous rise in popularity, leading to the development of new fingerprinting methods. Unfortunately, it remains a dual-use technology, which can enhance security or reduce privacy. Since fingerprinting happens unnoticeably in the background, it is not trivial for a user to recognize a fingerprinting script's intent. Additionally, it is not transparent to a user what data is accessed or where and by whom it is processed — our results and modifications to \FPMON~lay the groundwork for changing this. While current regulation enables users to opt-out of cookies used for privacy-invasive purposes, no such possibility exists for fingerprinting. In fact, \FPNET~did not consent to any data processing during the scans but was fingerprinted nonetheless. Therefore, we argue that our results are a lower bound, and users have no chance to protect their privacy sufficiently against fingerprinting. With respect to the randomization study, even popular privacy extensions might not be effective against script origin changes. In order to create awareness and educate the public about this topic, we call out actors that do not honor the users' privacy by publishing our extensive list. Although the list is not exhaustive due to the limitations faced, we are confident that \FPNET~and our studies enable further research as outlined in section \qnameref{sec:discussion-future-work}.

\setcounter{secnumdepth}{-1}
\section{References}
\begingroup 
\renewcommand{\section}[2]{}
\printbibliography[title={}]
\endgroup

\setcounter{secnumdepth}{-1}
\section{Appendix} \label{sec:appendix}


\subsection*{Feature Groups and Ratings}\label{sec:appendix:featuregroups-ratings}
The following table contains the 40 feature groups (of which 17 are rated aggressive), and their associated properties or functions monitored using \FPMON. Feature groups rated \emph{aggressive} are considered more privacy-invasive and more commonly misused for fingerprinting, while the user experience benefits from access to \emph{sensitive} \cite{fp_paper}. 
{
	
\begin{footnotesize}
	\begin{longtabu} to \textwidth {llL}
		\toprule
		Feature group & Rating & Properties or functions\\
		\midrule\midrule
		App\_code\_name & sensitive & navigator.appCodeName\\
		\midrule
		App\_name & sensitive & navigator.appName\\
		\midrule
		App\_version & sensitive & navigator.appVersion\\
		\midrule
		Audio & aggressive & createAnalyser(), createOscillator(), createGain(), createScriptProcessor(), createDynamicsCompressor()\\
		\midrule
		Audio\_Video\_formats & aggressive & canPlayType()\\
		\midrule
		Battery\_status & aggressive & navigator.getBattery\\
		\midrule
		Browser\_language & sensitive & navigator.browserLanguage\\
		\midrule
		Build\_ID & sensitive & navigator.buildID\\
		\midrule
		Canvas & aggressive & getImageData(), getLineDash(), measureText(), isPointInPath()\\
		\midrule
		Connection & aggressive & navigator.connection\\
		\midrule
		Content\_language & sensitive & navigator.languages, navigator.userLanguage, navigator.language\\
		\midrule
		Cookies\_enabled & sensitive & navigator.cookieEnabled\\
		\midrule
		Cpu\_Class & sensitive & navigator.cpuClass\\
		\midrule
		Device\_memory & aggressive & navigator.deviceMemory\\
		\midrule
		DoNotTrack & sensitive & navigator.doNotTrack, navigator.msDoNotTrack\\
		\midrule
		Drag\_and\_drop & sensitive & navigator.dragDrop\\
		\midrule
		Flash & sensitive & window.swfobject\\
		\midrule
		Frequency\_analyzer & aggressive & getFloatFrequencyData(), getByteFrequencyData(), getFloatTimeDomainData(), getByteTimeDomainData()\\
		\midrule
		Geolocation & aggressive & navigator.geolocation\\
		\midrule
		Hardware\_concurrency & aggressive & navigator.hardwareConcurrency\\
		\midrule
		Java\_endabled & sensitive & navigator.enabled\\
		\midrule
		JS\_fonts & aggressive & fill(), fillText()\\
		\midrule
		List\_of\_plugins & aggressive & navigator.plugins\\
		\midrule
		Media\_devices & aggressive & navigator.mediaDevices\\
		\midrule
		Mobile & sensitive & window.ondeviceproximity, window.onuserproximity, window.DeviceOrientationEvent, window.DeviceMotionEvent, navigator.maxTouchPoints, navigator.msMaxTouchPoints, navigator.touch\\
		\midrule
		Online & sensitive & navigator.onLine\\
		\midrule
		Oscpu & aggressive & navigator.oscpu\\
		\midrule
		Permissions & aggressive & navigator.permissions\\
		\midrule
		Platform & sensitive & navigator.platform\\
		\midrule
		Product & sensitive & navigator.product\\
		\midrule
		Product\_sub & aggressive & navigator.productSub\\
		\midrule
		Screen\_window & sensitive & window.devicePixelRatio, window.innerWidth, window.innerHeight, window.emit, window.outerWidth, window.outerHeight, screen.colorDepth, screen.width, screen.availWidth, screen.availHeight, screen.pixelDepth, screen.height, screen.availTop, screen.availLeft, screen.deviceXDPI, screen.logicalXDPI, screen.fontSmoothingEnabled, screen.screenInfo, navigator.orientation\\
		\midrule
		Storage & sensitive & window.sessionStorage, window.localStorage, window.indexedDB, window.openDatabase, navigator.webkitTemporaryStorage, navigator.webkitPersistentStorage, navigator.openDatabase, navigator.localStorage\\
		\midrule
		System\_language & sensitive & navigator.systemLanguage\\
		\midrule
		Timezone & sensitive & getTimezoneOffset(), window.Intl\\
		\midrule
		User\_agent & sensitive & navigator.userAgent\\
		\midrule
		Vendor & sensitive & navigator.vendor\\
		\midrule
		Vendor\_sub & sensitive & navigator.vendorSub\\
		\midrule
		Webdriver & aggressive & window.webdriver, navigator.webdriver\\
		\midrule
		WebGL & aggressive & getParameter(), getSupportedExtensions(), getContextAttributes(), getShaderPrecisionFormat(), getExtension(), readPixels(), getUniformLocation(), getAttribLocation(), clearColor(), enable(), depthFunc(), clear(), createBuffer(), bindBuffer(), bufferData(), createProgram(), createShader(), shaderSource(), compileShader(), attachShader(), linkProgram(), useProgram(), drawArrays()\\
		\midrule
		\bottomrule
		\caption{Observable feature groups, their rating and associated properties or functions.}
		\label{tab:feature-groups}
	\end{longtabu}

\end{footnotesize}
}
\newpage
\subsection*{List of Event Handlers}\label{sec:appendix:event-handlers}
Using the JavaScript code from Listing \ref{lst:implementation:event-handler-list}, the following 109 event handlers are retrieved:
{
	\begin{footnotesize}
	\begin{longtabu} to \textwidth {LLL}
		\toprule
		\midrule
onabort & onabsolutedeviceorientation & onafterprint\\ \midrule
onanimationcancel & onanimationend & onanimationiteration\\ \midrule
onanimationstart & onauxclick & onbeforeprint\\ \midrule
onbeforeunload & onblur & oncanplay\\ \midrule
oncanplaythrough & onchange & onclick\\ \midrule
onclose & oncontextmenu & oncuechange\\ \midrule
ondblclick & ondevicelight & ondevicemotion\\ \midrule
ondeviceorientation & ondeviceproximity & ondrag\\ \midrule
ondragend & ondragenter & ondragexit\\ \midrule
ondragleave & ondragover & ondragstart\\ \midrule
ondrop & ondurationchange & onemptied\\ \midrule
onended & onerror & onfocus\\ \midrule
onformdata & ongotpointercapture & onhashchange\\ \midrule
oninput & oninvalid & onkeydown\\ \midrule
onkeypress & onkeyup & onlanguagechange\\ \midrule
onload & onloadeddata & onloadedmetadata\\ \midrule
onloadend & onloadstart & onlostpointercapture\\ \midrule
onmessage & onmessageerror & onmousedown\\ \midrule
onmouseenter & onmouseleave & onmousemove\\ \midrule
onmouseout & onmouseover & onmouseup\\ \midrule
onmozfullscreenchange & onmozfullscreenerror & onoffline\\ \midrule
ononline & onpagehide & onpageshow\\ \midrule
onpause & onplay & onplaying\\ \midrule
onpointercancel & onpointerdown & onpointerenter\\ \midrule
onpointerleave & onpointermove & onpointerout\\ \midrule
onpointerover & onpointerup & onpopstate\\ \midrule
onprogress & onratechange & onrejectionhandled\\ \midrule
onreset & onresize & onscroll\\ \midrule
onseeked & onseeking & onselect\\ \midrule
onselectstart & onshow & onstalled\\ \midrule
onstorage & onsubmit & onsuspend\\ \midrule
ontimeupdate & ontoggle & ontransitioncancel\\ \midrule
ontransitionend & ontransitionrun & ontransitionstart\\ \midrule
onunhandledrejection & onunload & onuserproximity\\ \midrule
onvolumechange & onwaiting & onwebkitanimationend\\ \midrule
onwebkitanimationiteration & onwebkitanimationstart & onwebkittransitionend\\ \midrule
onwheel\\ \midrule
		\bottomrule
		\caption{List of event handlers defined on a web browser's \texttt{window} property.}
		\label{tab:event-handlers}
	\end{longtabu}

\end{footnotesize}
}

\newpage
\subsection*{List of Identified Actors}\label{sec:appendix:identified-actors}
{
\begin{footnotesize}
\begin{longtabu} to \textwidth {Xrrr}
	\toprule
	Actor & Networks & Score (Aggr.) & Covered websites\\
	\midrule\midrule
	
	google doubleclick & 19 & 10 (2) & 1,583 \\
	\midrule
	google adsense & 11 & 8 (1) & 544 \\
	\midrule
	yandex metrica & 52 & 14 (3) & 367 \\
	\midrule
	akamai.com & 2 & 28 (10) & 292 \\
	\midrule
	fingerprintjs2 & 9 & 20 (10) & 133 \\
	\midrule
	yandex adfox & 3 & 7 (1) & 125 \\
	\midrule
	google.com & 3 & 8 (3) & 122 \\
	\midrule
	hubspot.com & 1 & 8 (0) & 116 \\
	\midrule
	moat.com & 5 & 34 (12) & 114 \\
	\midrule
	snowplow & 2 & 8 (1) & 89 \\
	\midrule
	drift.com & 2 & 8 (1) & 66 \\
	\midrule
	wistia.com & 1 & 7 (1) & 62 \\
	\midrule
	crazyegg.com & 1 & 8 (3) & 56 \\
	\midrule
	top100.ru & 1 & 9 (1) & 55 \\
	\midrule
	prebid.org & 12 & 10 (1) & 53 \\
	\midrule
	siftscience (sift.com) & 1 & 20 (6) & 45 \\
	\midrule
	piano.io & 1 & 9 (1) & 43 \\
	\midrule
	google ima sdk & 3 & 13 (5) & 32 \\
	\midrule
	adform.com & 2 & 20 (3) & 32 \\
	\midrule
	riskified.com & 1 & 11 (5) & 31 \\
	\midrule
	self-hosted webtrekk & 2 & 7 (1) & 28 \\
	\midrule
	incapsula.com & 1 & 10 (4) & 24 \\
	\midrule
	hcaptcha.com & 1 & 21 (5) & 23 \\
	\midrule
	zopim.com & 1 & 7 (1) & 22 \\
	\midrule
	tealium.com & 7 & 14 (3) & 21 \\
	\midrule
	decibelinsight (decibel.com) & 3 & 7 (0) & 21 \\
	\midrule
	akamai boomerang & 6 & 8 (2) & 19 \\
	\midrule
	adsco.re & 2 & 29 (10) & 19 \\
	\midrule
	cxsense (piano.io) & 6 & 7 (1) & 17 \\
	\midrule
	datadome.co & 1 & 20 (11) & 16 \\
	\midrule
	iovation.com & 3 & 13 (3) & 15 \\
	\midrule
	crisp.chat & 1 & 7 (0) & 15 \\
	\midrule
	vidazoo.com & 3 & 12 (3) & 14 \\
	\midrule
	gumgum.com & 1 & 12 (3) & 14 \\
	\midrule
	ptengine.com & 1 & 9 (1) & 13 \\
	\midrule
	openx.net & 2 & 7 (1) & 13 \\
	\midrule
	perfdrive (radwarebotmanager.com) & 1 & 23 (6) & 11 \\
	\midrule
	appsflyer.com & 1 & 11 (2) & 11 \\
	\midrule
	zaloapp.com & 1 & 12 (6) & 10 \\
	\midrule
	yabidos.com & 1 & 20 (6) & 10 \\
	\midrule
	vdo.ai & 3 & 8 (1) & 10 \\
	\midrule
	pushly.com & 2 & 9 (1) & 10 \\
	\midrule
	tawk.to & 1 & 8 (2) & 9 \\
	\midrule
	sas.com & 1 & 7 (1) & 9 \\
	\midrule
	webtrends.com & 2 & 7 (1) & 9 \\
	\midrule
	yotpo.com & 1 & 7 (1) & 8 \\
	\midrule
	wunderkind.co & 3 & 12 (3) & 8 \\
	\midrule
	vwo.com & 1 & 7 (0) & 7 \\
	\midrule
	spiceworks.com & 1 & 14 (4) & 7 \\
	\midrule
	petametrics.com & 3 & 7 (0) & 7 \\
	\midrule
	brightedge.com & 1 & 20 (9) & 7 \\
	\midrule
	braze.com & 3 & 7 (0) & 7 \\
	\midrule
	bebi.com & 2 & 8 (1) & 7 \\
	\midrule
	adskeeper.com & 3 & 9 (5) & 7 \\
	\midrule
	forseepower.com & 2 & 12 (3) & 6 \\
	\midrule
	webtrekk.com & 1 & 8 (1) & 5 \\
	\midrule
	webterren.com & 1 & 8 (1) & 5 \\
	\midrule
	subscribers.com & 1 & 9 (0) & 5 \\
	\midrule
	maxmind.com & 1 & 36 (14) & 5 \\
	\midrule
	mathereconomics.com & 2 & 9 (1) & 5 \\
	\midrule
	detik.com & 1 & 18 (9) & 5 \\
	\midrule
	acquia.com & 1 & 7 (0) & 5 \\
	\midrule
	viafoura.com & 2 & 8 (0) & 4 \\
	\midrule
	valuecommerce.co.jp & 1 & 20 (4) & 4 \\
	\midrule
	trackjs.com & 1 & 7 (1) & 4 \\
	\midrule
	riotgames.com & 1 & 10 (4) & 4 \\
	\midrule
	pixlee.com & 1 & 18 (9) & 4 \\
	\midrule
	mgid.com & 2 & 9 (5) & 4 \\
	\midrule
	jwplayer.com & 2 & 8 (1) & 4 \\
	\midrule
	forter.com & 1 & 12 (6) & 4 \\
	\midrule
	amazon ads & 1 & 7 (0) & 4 \\
	\midrule
	acecounter.com & 1 & 8 (1) & 4 \\
	\midrule
	underdogmedia.com & 1 & 7 (0) & 3 \\
	\midrule
	teads.com & 1 & 7 (2) & 3 \\
	\midrule
	razorpay.com & 1 & 10 (2) & 3 \\
	\midrule
	pushe.co & 1 & 18 (8) & 3 \\
	\midrule
	onead (guoshipartners.com) & 1 & 18 (6) & 3 \\
	\midrule
	loop11.com & 1 & 16 (7) & 3 \\
	\midrule
	geetest.com & 1 & 19 (5) & 3 \\
	\midrule
	cellit.io & 1 & 8 (1) & 3 \\
	\midrule
	carbonplatform.com & 1 & 8 (1) & 3 \\
	\midrule
	brightcove.com & 1 & 8 (1) & 3 \\
	\midrule
	breaktime.com.tw & 1 & 18 (8) & 3 \\
	\midrule
	blueconic.com & 1 & 9 (1) & 3 \\
	\midrule
	bitmedia.at & 1 & 12 (4) & 3 \\
	\midrule
	abtshield.com & 1 & 11 (4) & 3 \\
	\midrule
	yandex.ru & 1 & 10 (6) & 2 \\
	\midrule
	perimeterx.com & 1 & 19 (4) & 2 \\
	\midrule
	particularaudience.com & 1 & 8 (1) & 2 \\
	\midrule
	oracle.com & 1 & 7 (0) & 2 \\
	\midrule
	native.ai & 1 & 9 (0) & 2 \\
	\midrule
	nanigans.com & 1 & 33 (11) & 2 \\
	\midrule
	mql5.com & 1 & 16 (7) & 2 \\
	\midrule
	meetrics.com & 1 & 7 (3) & 2 \\
	\midrule
	logrocket.com & 1 & 8 (0) & 2 \\
	\midrule
	kustomer.com & 1 & 7 (2) & 2 \\
	\midrule
	gatedcontent.com & 1 & 11 (1) & 2 \\
	\midrule
	fetnet.net & 1 & 7 (0) & 2 \\
	\midrule
	eastmoney.com & 1 & 16 (7) & 2 \\
	\midrule
	dcmn.com & 1 & 24 (7) & 2 \\
	\midrule
	coingecko.com & 1 & 7 (1) & 2 \\
	\midrule
	auone.jp & 1 & 7 (0) & 2 \\
	\midrule
	adobe.com & 1 & 7 (1) & 2 \\
	\midrule
	self-hosted genesys & 1 & 7 (0) & 2 \\
	\midrule
	
	\bottomrule
	\caption{Identified actors, their number of networks, score and covered websites.}
	\label{tab:actor-list}
\end{longtabu}

\end{footnotesize}
}
\newpage
\subsection*{Source Code}
The source code for \FPNET~and the analysis scripts is available in a public repository: \href{https://github.com/gehaxelt/MasterThesis-FPNET/}{https://github.com/gehaxelt/MasterThesis-FPNET/}

\subsubsection*{Database Schema}\label{sec:appendix:database-schema}
\begin{figure}[h]
	\hspace*{-1cm}
	\includegraphics[width=1.2\textwidth]{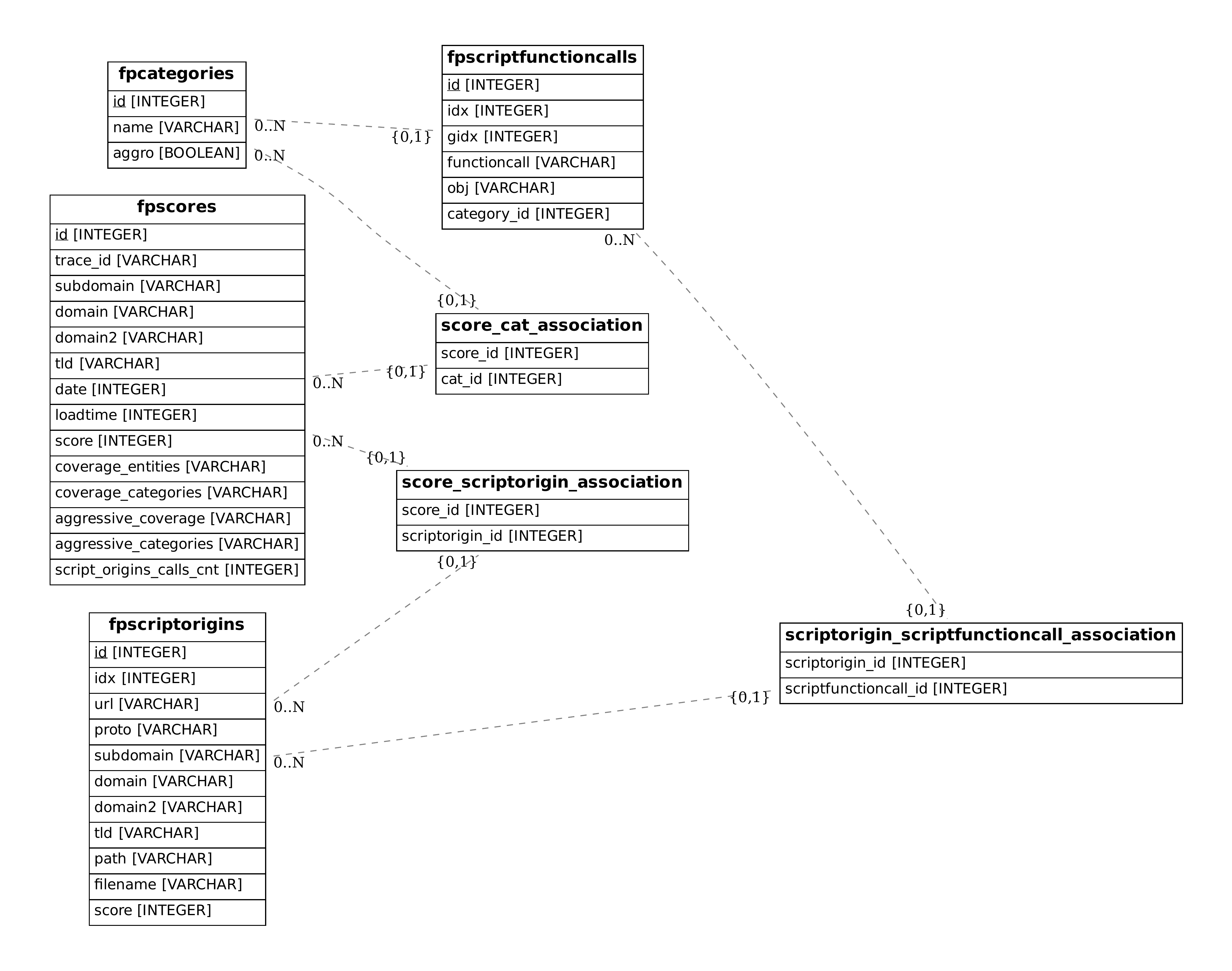}
	\caption{Entity Relationship Diagram of the database schema used in \FPNET's post-processing step.}
	\label{fig:db-schema}
\end{figure}

\end{document}